\documentclass[10pt,english,british,traditabstract,longauth,nonamebreak]{aa}
\usepackage[T1]{fontenc}
\usepackage[utf8]{inputenc}
\usepackage[a4paper]{geometry}
\usepackage{color}
\usepackage{babel}
\usepackage{verbatim}
\usepackage{float}
\usepackage{url}
\usepackage{amstext}
\usepackage{fixltx2e}
\usepackage{graphicx}
\usepackage[authoryear]{natbib}
\usepackage[unicode=true,pdfusetitle,
 bookmarks=true,bookmarksnumbered=false,bookmarksopen=false,
 breaklinks=false,pdfborder={0 0 1},colorlinks=false] %,backref=section
 {hyperref}
\hypersetup{
 breaklinks,colorlinks,citecolor=blue}

\makeatletter
\pdfoutput=1
\bibpunct{(}{)}{;}{a}{}{,}
\usepackage{txfonts}
\usepackage{color}
\usepackage[mathscr]{eucal}
\usepackage{array}
\usepackage{multirow}
\usepackage{morefloats}
\newcommand{\rev}[1]{#1}  %% for final production
\newcommand{\revv}[1]{#1}  %%  for final production

\newcommand{\IMO}{\texttt{IMO}}

\newcommand{\spice}{\texttt{spice}}

\def\MHz{\ifmmode $MHz$\else \,MHz\fi}
\def\Hz{\ifmmode $Hz$\else \,Hz\fi}

\def\setsymbol#1#2{\expandafter\def\csname #1\endcsname{#2}}
\def\getsymbol#1{\csname #1\endcsname}

\def\Planck{\textit{Planck}}

\def\HeJT{$^4$He-JT}

\def\all2013resultspapers{\nocite{planck2013-p01, planck2013-p02, planck2013-p02a, planck2013-p02d, planck2013-p02b, planck2013-p03, planck2013-p03c, planck2013-p03f, planck2013-p03d, planck2013-p03e, planck2013-p01a, planck2013-p06, planck2013-p03a, planck2013-pip88, planck2013-p08, planck2013-p11, planck2013-p12, planck2013-p13, planck2013-p14, planck2013-p15, planck2013-p05b, planck2013-p17, planck2013-p09, planck2013-p09a, planck2013-p20, planck2013-p19, planck2013-pipaberration, planck2013-p05, planck2013-p05a, planck2013-pip56, planck2013-p06b}}

\newbox\tablebox    \newdimen\tablewidth
\def\leaderfil{\leaders\hbox to 5pt{\hss.\hss}\hfil}
\def\endPlancktable{\tablewidth=\columnwidth 
    $$\hss\copy\tablebox\hss$$
    \vskip-\lastskip\vskip -2pt}
\def\endPlancktablewide{\tablewidth=\textwidth 
    $$\hss\copy\tablebox\hss$$
    \vskip-\lastskip\vskip -2pt}
\def\tablenote#1 #2\par{\begingroup \parindent=0.8em
    \abovedisplayshortskip=0pt\belowdisplayshortskip=0pt
    \noindent
    $$\hss\vbox{\hsize\tablewidth \hangindent=\parindent \hangafter=1 \noindent
    \hbox to \parindent{$^#1$\hss}\strut#2\strut\par}\hss$$
    \endgroup}
\def\doubleline{\vskip 3pt\hrule \vskip 1.5pt \hrule \vskip 5pt}

\def\L2{\ifmmode L_2\else $L_2$\fi}

\def\DeltaT{\ifmmode \Delta T\else $\Delta T$\fi}
\def\deltat{\ifmmode \Delta t\else $\Delta t$\fi}
\def\fknee{\ifmmode f_{\rm knee}\else $f_{\rm knee}$\fi}
\def\Fmax{\ifmmode F_{\rm max}\else $F_{\rm max}$\fi}
\def\solar{\ifmmode{\rm M}_{\mathord\odot}\else${\rm M}_{\mathord\odot}$\fi}
\def\Msolar{\ifmmode{\rm M}_{\mathord\odot}\else${\rm M}_{\mathord\odot}$\fi}
\def\Lsolar{\ifmmode{\rm L}_{\mathord\odot}\else${\rm L}_{\mathord\odot}$\fi}
\def\inv{\ifmmode^{-1}\else$^{-1}$\fi}
\def\mo{\ifmmode^{-1}\else$^{-1}$\fi}
\def\sup#1{\ifmmode ^{\rm #1}\else $^{\rm #1}$\fi}
\def\expo#1{\ifmmode \times 10^{#1}\else $\times 10^{#1}$\fi}
\def\,{\thinspace}
\def\lsim{\mathrel{\raise .4ex\hbox{\rlap{$<$}\lower 1.2ex\hbox{$\sim$}}}}
\def\gsim{\mathrel{\raise .4ex\hbox{\rlap{$>$}\lower 1.2ex\hbox{$\sim$}}}}

\def\simprop{\mathrel{\raise .4ex\hbox{\rlap{$\propto$}\lower 1.2ex\hbox{$\sim$}}}}
\def\deg{\ifmmode^\circ\else$^\circ$\fi}
\def\pdeg{\ifmmode $\setbox0=\hbox{$^{\circ}$}\rlap{\hskip.11\wd0 .}$^{\circ}
          \else \setbox0=\hbox{$^{\circ}$}\rlap{\hskip.11\wd0 .}$^{\circ}$\fi}
\def\arcs{\ifmmode {^{\scriptstyle\prime\prime}}
          \else $^{\scriptstyle\prime\prime}$\fi}
\def\arcm{\ifmmode {^{\scriptstyle\prime}}
          \else $^{\scriptstyle\prime}$\fi}
\newdimen\sa  \newdimen\sb
\def\parcs{\sa=.07em \sb=.03em
     \ifmmode \hbox{\rlap{.}}^{\scriptstyle\prime\kern -\sb\prime}\hbox{\kern -\sa}
     \else \rlap{.}$^{\scriptstyle\prime\kern -\sb\prime}$\kern -\sa\fi}
\def\parcm{\sa=.08em \sb=.03em
     \ifmmode \hbox{\rlap{.}\kern\sa}^{\scriptstyle\prime}\hbox{\kern-\sb}
     \else \rlap{.}\kern\sa$^{\scriptstyle\prime}$\kern-\sb\fi}
\def\ra[#1 #2 #3.#4]{#1\sup{h}#2\sup{m}#3\sup{s}\llap.#4}
\def\dec[#1 #2 #3.#4]{#1\deg#2\arcm#3\arcs\llap.#4}
\def\deco[#1 #2 #3]{#1\deg#2\arcm#3\arcs}
\def\rra[#1 #2]{#1\sup{h}#2\sup{m}}

\def\dots{\relax\ifmmode \ldots\else $\ldots$\fi}
\def\WHzsr{\ifmmode $W\,Hz\mo\,sr\mo$\else W\,Hz\mo\,sr\mo\fi}
\def\mHz{\ifmmode $\,mHz$\else \,mHz\fi}
\def\GHz{\ifmmode $\,GHz$\else \,GHz\fi}
\def\mKs{\ifmmode $\,mK\,s$^{1/2}\else \,mK\,s$^{1/2}$\fi}
\def\muKs{\ifmmode \,\mu$K\,s$^{1/2}\else \,$\mu$K\,s$^{1/2}$\fi}
\def\muKRJs{\ifmmode \,\mu$K$_{\rm RJ}$\,s$^{1/2}\else \,$\mu$K$_{\rm RJ}$\,s$^{1/2}$\fi}
\def\muKHz{\ifmmode \,\mu$K\,Hz$^{-1/2}\else \,$\mu$K\,Hz$^{-1/2}$\fi}
\def\MJysr{\ifmmode \,$MJy\,sr\mo$\else \,MJy\,sr\mo\fi}
\def\MJysrmK{\ifmmode \,$MJy\,sr\mo$\,mK$_{\rm CMB}\mo\else \,MJy\,sr\mo\,mK$_{\rm CMB}\mo$\fi}
\def\microns{\ifmmode \,\mu$m$\else \,$\mu$m\fi}

\def\muK{\ifmmode \,\mu$K$\else \,$\mu$\hbox{K}\fi}
\def\microK{\ifmmode \,\mu$K$\else \,$\mu$\hbox{K}\fi}
\def\muW{\ifmmode \,\mu$W$\else \,$\mu$\hbox{W}\fi}
\def\kms{\ifmmode $\,km\,s$^{-1}\else \,km\,s$^{-1}$\fi}
\def\kmsMpc{\ifmmode $\,\kms\,Mpc\mo$\else \,\kms\,Mpc\mo\fi}
\setsymbol{LFI:center:frequency:70GHz:units}{70.3\,GHz}
\setsymbol{LFI:center:frequency:44GHz:units}{44.1\,GHz}
\setsymbol{LFI:center:frequency:30GHz:units}{28.5\,GHz}

\setsymbol{LFI:center:frequency:70GHz}{70.3}
\setsymbol{LFI:center:frequency:44GHz}{44.1}
\setsymbol{LFI:center:frequency:30GHz}{28.5}

\setsymbol{LFI:center:frequency:LFI18:Rad:M:units}{71.7\GHz}
\setsymbol{LFI:center:frequency:LFI19:Rad:M:units}{67.5\GHz}
\setsymbol{LFI:center:frequency:LFI20:Rad:M:units}{69.2\GHz}
\setsymbol{LFI:center:frequency:LFI21:Rad:M:units}{70.4\GHz}
\setsymbol{LFI:center:frequency:LFI22:Rad:M:units}{71.5\GHz}
\setsymbol{LFI:center:frequency:LFI23:Rad:M:units}{70.8\GHz}
\setsymbol{LFI:center:frequency:LFI24:Rad:M:units}{44.4\GHz}
\setsymbol{LFI:center:frequency:LFI25:Rad:M:units}{44.0\GHz}
\setsymbol{LFI:center:frequency:LFI26:Rad:M:units}{43.9\GHz}
\setsymbol{LFI:center:frequency:LFI27:Rad:M:units}{28.3\GHz}
\setsymbol{LFI:center:frequency:LFI28:Rad:M:units}{28.8\GHz}
\setsymbol{LFI:center:frequency:LFI18:Rad:S:units}{70.1\GHz}
\setsymbol{LFI:center:frequency:LFI19:Rad:S:units}{69.6\GHz}
\setsymbol{LFI:center:frequency:LFI20:Rad:S:units}{69.5\GHz}
\setsymbol{LFI:center:frequency:LFI21:Rad:S:units}{69.5\GHz}
\setsymbol{LFI:center:frequency:LFI22:Rad:S:units}{72.8\GHz}
\setsymbol{LFI:center:frequency:LFI23:Rad:S:units}{71.3\GHz}
\setsymbol{LFI:center:frequency:LFI24:Rad:S:units}{44.1\GHz}
\setsymbol{LFI:center:frequency:LFI25:Rad:S:units}{44.1\GHz}
\setsymbol{LFI:center:frequency:LFI26:Rad:S:units}{44.1\GHz}
\setsymbol{LFI:center:frequency:LFI27:Rad:S:units}{28.5\GHz}
\setsymbol{LFI:center:frequency:LFI28:Rad:S:units}{28.2\GHz}

\setsymbol{LFI:center:frequency:LFI18:Rad:M}{71.7}
\setsymbol{LFI:center:frequency:LFI19:Rad:M}{67.5}
\setsymbol{LFI:center:frequency:LFI20:Rad:M}{69.2}
\setsymbol{LFI:center:frequency:LFI21:Rad:M}{70.4}
\setsymbol{LFI:center:frequency:LFI22:Rad:M}{71.5}
\setsymbol{LFI:center:frequency:LFI23:Rad:M}{70.8}
\setsymbol{LFI:center:frequency:LFI24:Rad:M}{44.4}
\setsymbol{LFI:center:frequency:LFI25:Rad:M}{44.0}
\setsymbol{LFI:center:frequency:LFI26:Rad:M}{43.9}
\setsymbol{LFI:center:frequency:LFI27:Rad:M}{28.3}
\setsymbol{LFI:center:frequency:LFI28:Rad:M}{28.8}
\setsymbol{LFI:center:frequency:LFI18:Rad:S}{70.1}
\setsymbol{LFI:center:frequency:LFI19:Rad:S}{69.6}
\setsymbol{LFI:center:frequency:LFI20:Rad:S}{69.5}
\setsymbol{LFI:center:frequency:LFI21:Rad:S}{69.5}
\setsymbol{LFI:center:frequency:LFI22:Rad:S}{72.8}
\setsymbol{LFI:center:frequency:LFI23:Rad:S}{71.3}
\setsymbol{LFI:center:frequency:LFI24:Rad:S}{44.1}
\setsymbol{LFI:center:frequency:LFI25:Rad:S}{44.1}
\setsymbol{LFI:center:frequency:LFI26:Rad:S}{44.1}
\setsymbol{LFI:center:frequency:LFI27:Rad:S}{28.5}
\setsymbol{LFI:center:frequency:LFI28:Rad:S}{28.2}

\setsymbol{LFI:white:noise:sensitivity:70GHz:units}{134.7\muKs}
\setsymbol{LFI:white:noise:sensitivity:44GHz:units}{164.7\muKs}
\setsymbol{LFI:white:noise:sensitivity:30GHz:units}{143.4\muKs}

\setsymbol{LFI:white:noise:sensitivity:70GHz}{134.7}
\setsymbol{LFI:white:noise:sensitivity:44GHz}{164.7}
\setsymbol{LFI:white:noise:sensitivity:30GHz}{143.4}

\setsymbol{LFI:white:noise:sensitivity:LFI18:Rad:M:units}{512.0\muKs}
\setsymbol{LFI:white:noise:sensitivity:LFI19:Rad:M:units}{581.4\muKs}
\setsymbol{LFI:white:noise:sensitivity:LFI20:Rad:M:units}{590.8\muKs}
\setsymbol{LFI:white:noise:sensitivity:LFI21:Rad:M:units}{455.2\muKs}
\setsymbol{LFI:white:noise:sensitivity:LFI22:Rad:M:units}{492.0\muKs}
\setsymbol{LFI:white:noise:sensitivity:LFI23:Rad:M:units}{507.7\muKs}
\setsymbol{LFI:white:noise:sensitivity:LFI24:Rad:M:units}{462.2\muKs}
\setsymbol{LFI:white:noise:sensitivity:LFI25:Rad:M:units}{413.6\muKs}
\setsymbol{LFI:white:noise:sensitivity:LFI26:Rad:M:units}{478.6\muKs}
\setsymbol{LFI:white:noise:sensitivity:LFI27:Rad:M:units}{277.7\muKs}
\setsymbol{LFI:white:noise:sensitivity:LFI28:Rad:M:units}{312.3\muKs}
\setsymbol{LFI:white:noise:sensitivity:LFI18:Rad:S:units}{465.7\muKs}
\setsymbol{LFI:white:noise:sensitivity:LFI19:Rad:S:units}{555.6\muKs}
\setsymbol{LFI:white:noise:sensitivity:LFI20:Rad:S:units}{623.2\muKs}
\setsymbol{LFI:white:noise:sensitivity:LFI21:Rad:S:units}{564.1\muKs}
\setsymbol{LFI:white:noise:sensitivity:LFI22:Rad:S:units}{534.4\muKs}
\setsymbol{LFI:white:noise:sensitivity:LFI23:Rad:S:units}{542.4\muKs}
\setsymbol{LFI:white:noise:sensitivity:LFI24:Rad:S:units}{399.2\muKs}
\setsymbol{LFI:white:noise:sensitivity:LFI25:Rad:S:units}{392.6\muKs}
\setsymbol{LFI:white:noise:sensitivity:LFI26:Rad:S:units}{418.6\muKs}
\setsymbol{LFI:white:noise:sensitivity:LFI27:Rad:S:units}{302.9\muKs}
\setsymbol{LFI:white:noise:sensitivity:LFI28:Rad:S:units}{285.3\muKs}

\setsymbol{LFI:white:noise:sensitivity:LFI18:Rad:M}{512.0}
\setsymbol{LFI:white:noise:sensitivity:LFI19:Rad:M}{581.4}
\setsymbol{LFI:white:noise:sensitivity:LFI20:Rad:M}{590.8}
\setsymbol{LFI:white:noise:sensitivity:LFI21:Rad:M}{455.2}
\setsymbol{LFI:white:noise:sensitivity:LFI22:Rad:M}{492.0}
\setsymbol{LFI:white:noise:sensitivity:LFI23:Rad:M}{507.7}
\setsymbol{LFI:white:noise:sensitivity:LFI24:Rad:M}{462.2}
\setsymbol{LFI:white:noise:sensitivity:LFI25:Rad:M}{413.6}
\setsymbol{LFI:white:noise:sensitivity:LFI26:Rad:M}{478.6}
\setsymbol{LFI:white:noise:sensitivity:LFI27:Rad:M}{277.7}
\setsymbol{LFI:white:noise:sensitivity:LFI28:Rad:M}{312.3}
\setsymbol{LFI:white:noise:sensitivity:LFI18:Rad:S}{465.7}
\setsymbol{LFI:white:noise:sensitivity:LFI19:Rad:S}{555.6}
\setsymbol{LFI:white:noise:sensitivity:LFI20:Rad:S}{623.2}
\setsymbol{LFI:white:noise:sensitivity:LFI21:Rad:S}{564.1}
\setsymbol{LFI:white:noise:sensitivity:LFI22:Rad:S}{534.4}
\setsymbol{LFI:white:noise:sensitivity:LFI23:Rad:S}{542.4}
\setsymbol{LFI:white:noise:sensitivity:LFI24:Rad:S}{399.2}
\setsymbol{LFI:white:noise:sensitivity:LFI25:Rad:S}{392.6}
\setsymbol{LFI:white:noise:sensitivity:LFI26:Rad:S}{418.6}
\setsymbol{LFI:white:noise:sensitivity:LFI27:Rad:S}{302.9}
\setsymbol{LFI:white:noise:sensitivity:LFI28:Rad:S}{285.3}

\setsymbol{LFI:knee:frequency:70GHz:units}{29.5\mHz}
\setsymbol{LFI:knee:frequency:44GHz:units}{56.2\mHz}
\setsymbol{LFI:knee:frequency:30GHz:units}{113.7\mHz}

\setsymbol{LFI:knee:frequency:70GHz}{29.5}
\setsymbol{LFI:knee:frequency:44GHz}{56.2}
\setsymbol{LFI:knee:frequency:30GHz}{113.7}

\setsymbol{LFI:knee:frequency:LFI18:Rad:M:units}{16.3\mHz}
\setsymbol{LFI:knee:frequency:LFI19:Rad:M:units}{15.1\mHz}
\setsymbol{LFI:knee:frequency:LFI20:Rad:M:units}{18.7\mHz}
\setsymbol{LFI:knee:frequency:LFI21:Rad:M:units}{37.2\mHz}
\setsymbol{LFI:knee:frequency:LFI22:Rad:M:units}{12.7\mHz}
\setsymbol{LFI:knee:frequency:LFI23:Rad:M:units}{34.6\mHz}
\setsymbol{LFI:knee:frequency:LFI24:Rad:M:units}{46.2\mHz}
\setsymbol{LFI:knee:frequency:LFI25:Rad:M:units}{24.9\mHz}
\setsymbol{LFI:knee:frequency:LFI26:Rad:M:units}{67.6\mHz}
\setsymbol{LFI:knee:frequency:LFI27:Rad:M:units}{187.4\mHz}
\setsymbol{LFI:knee:frequency:LFI28:Rad:M:units}{122.2\mHz}
\setsymbol{LFI:knee:frequency:LFI18:Rad:S:units}{17.7\mHz}
\setsymbol{LFI:knee:frequency:LFI19:Rad:S:units}{22.0\mHz}
\setsymbol{LFI:knee:frequency:LFI20:Rad:S:units}{8.7\mHz}
\setsymbol{LFI:knee:frequency:LFI21:Rad:S:units}{25.9\mHz}
\setsymbol{LFI:knee:frequency:LFI22:Rad:S:units}{15.8\mHz}
\setsymbol{LFI:knee:frequency:LFI23:Rad:S:units}{129.8\mHz}
\setsymbol{LFI:knee:frequency:LFI24:Rad:S:units}{100.9\mHz}
\setsymbol{LFI:knee:frequency:LFI25:Rad:S:units}{38.9\mHz}
\setsymbol{LFI:knee:frequency:LFI26:Rad:S:units}{58.9\mHz}
\setsymbol{LFI:knee:frequency:LFI27:Rad:S:units}{104.4\mHz}
\setsymbol{LFI:knee:frequency:LFI28:Rad:S:units}{40.7\mHz}

\setsymbol{LFI:knee:frequency:LFI18:Rad:M}{16.3}
\setsymbol{LFI:knee:frequency:LFI19:Rad:M}{15.1}
\setsymbol{LFI:knee:frequency:LFI20:Rad:M}{18.7}
\setsymbol{LFI:knee:frequency:LFI21:Rad:M}{37.2}
\setsymbol{LFI:knee:frequency:LFI22:Rad:M}{12.7}
\setsymbol{LFI:knee:frequency:LFI23:Rad:M}{34.6}
\setsymbol{LFI:knee:frequency:LFI24:Rad:M}{46.2}
\setsymbol{LFI:knee:frequency:LFI25:Rad:M}{24.9}
\setsymbol{LFI:knee:frequency:LFI26:Rad:M}{67.6}
\setsymbol{LFI:knee:frequency:LFI27:Rad:M}{187.4}
\setsymbol{LFI:knee:frequency:LFI28:Rad:M}{122.2}
\setsymbol{LFI:knee:frequency:LFI18:Rad:S}{17.7}
\setsymbol{LFI:knee:frequency:LFI19:Rad:S}{22.0}
\setsymbol{LFI:knee:frequency:LFI20:Rad:S}{8.7}
\setsymbol{LFI:knee:frequency:LFI21:Rad:S}{25.9}
\setsymbol{LFI:knee:frequency:LFI22:Rad:S}{15.8}
\setsymbol{LFI:knee:frequency:LFI23:Rad:S}{129.8}
\setsymbol{LFI:knee:frequency:LFI24:Rad:S}{100.9}
\setsymbol{LFI:knee:frequency:LFI25:Rad:S}{38.9}
\setsymbol{LFI:knee:frequency:LFI26:Rad:S}{58.9}
\setsymbol{LFI:knee:frequency:LFI27:Rad:S}{104.4}
\setsymbol{LFI:knee:frequency:LFI28:Rad:S}{40.7}

\setsymbol{LFI:slope:70GHz:units}{$-1.03$\mHz}
\setsymbol{LFI:slope:44GHz:units}{$-0.89$\mHz}
\setsymbol{LFI:slope:30GHz:units}{$-0.87$\mHz}

\setsymbol{LFI:slope:70GHz}{$-1.03$}
\setsymbol{LFI:slope:44GHz}{$-0.89$}
\setsymbol{LFI:slope:30GHz}{$-0.87$}

\setsymbol{LFI:slope:LFI18:Rad:M:units}{$-1.04$\mHz}
\setsymbol{LFI:slope:LFI19:Rad:M:units}{$-1.09$\mHz}
\setsymbol{LFI:slope:LFI20:Rad:M:units}{$-0.69$\mHz}
\setsymbol{LFI:slope:LFI21:Rad:M:units}{$-1.56$\mHz}
\setsymbol{LFI:slope:LFI22:Rad:M:units}{$-1.01$\mHz}
\setsymbol{LFI:slope:LFI23:Rad:M:units}{$-0.96$\mHz}
\setsymbol{LFI:slope:LFI24:Rad:M:units}{$-0.83$\mHz}
\setsymbol{LFI:slope:LFI25:Rad:M:units}{$-0.91$\mHz}
\setsymbol{LFI:slope:LFI26:Rad:M:units}{$-0.95$\mHz}
\setsymbol{LFI:slope:LFI27:Rad:M:units}{$-0.87$\mHz}
\setsymbol{LFI:slope:LFI28:Rad:M:units}{$-0.88$\mHz}
\setsymbol{LFI:slope:LFI18:Rad:S:units}{$-1.15$\mHz}
\setsymbol{LFI:slope:LFI19:Rad:S:units}{$-1.00$\mHz}
\setsymbol{LFI:slope:LFI20:Rad:S:units}{$-0.95$\mHz}
\setsymbol{LFI:slope:LFI21:Rad:S:units}{$-0.92$\mHz}
\setsymbol{LFI:slope:LFI22:Rad:S:units}{$-1.01$\mHz}
\setsymbol{LFI:slope:LFI23:Rad:S:units}{$-0.95$\mHz}
\setsymbol{LFI:slope:LFI24:Rad:S:units}{$-0.73$\mHz}
\setsymbol{LFI:slope:LFI25:Rad:S:units}{$-1.16$\mHz}
\setsymbol{LFI:slope:LFI26:Rad:S:units}{$-0.79$\mHz}
\setsymbol{LFI:slope:LFI27:Rad:S:units}{$-0.82$\mHz}
\setsymbol{LFI:slope:LFI28:Rad:S:units}{$-0.91$\mHz}

\setsymbol{LFI:slope:LFI18:Rad:M}{$-1.04$}
\setsymbol{LFI:slope:LFI19:Rad:M}{$-1.09$}
\setsymbol{LFI:slope:LFI20:Rad:M}{$-0.69$}
\setsymbol{LFI:slope:LFI21:Rad:M}{$-1.56$}
\setsymbol{LFI:slope:LFI22:Rad:M}{$-1.01$}
\setsymbol{LFI:slope:LFI23:Rad:M}{$-0.96$}
\setsymbol{LFI:slope:LFI24:Rad:M}{$-0.83$}
\setsymbol{LFI:slope:LFI25:Rad:M}{$-0.91$}
\setsymbol{LFI:slope:LFI26:Rad:M}{$-0.95$}
\setsymbol{LFI:slope:LFI27:Rad:M}{$-0.87$}
\setsymbol{LFI:slope:LFI28:Rad:M}{$-0.88$}
\setsymbol{LFI:slope:LFI18:Rad:S}{$-1.15$}
\setsymbol{LFI:slope:LFI19:Rad:S}{$-1.00$}
\setsymbol{LFI:slope:LFI20:Rad:S}{$-0.95$}
\setsymbol{LFI:slope:LFI21:Rad:S}{$-0.92$}
\setsymbol{LFI:slope:LFI22:Rad:S}{$-1.01$}
\setsymbol{LFI:slope:LFI23:Rad:S}{$-0.95$}
\setsymbol{LFI:slope:LFI24:Rad:S}{$-0.73$}
\setsymbol{LFI:slope:LFI25:Rad:S}{$-1.16$}
\setsymbol{LFI:slope:LFI26:Rad:S}{$-0.79$}
\setsymbol{LFI:slope:LFI27:Rad:S}{$-0.82$}
\setsymbol{LFI:slope:LFI28:Rad:S}{$-0.91$}

\setsymbol{LFI:FWHM:70GHz:units}{13\parcm01}
\setsymbol{LFI:FWHM:44GHz:units}{27\parcm92}
\setsymbol{LFI:FWHM:30GHz:units}{32\parcm65}

\setsymbol{LFI:FWHM:70GHz}{13.01}
\setsymbol{LFI:FWHM:44GHz}{27.92}
\setsymbol{LFI:FWHM:30GHz}{32.65}

\setsymbol{LFI:FWHM:LFI18:units}{13\parcm39}
\setsymbol{LFI:FWHM:LFI19:units}{13\parcm01}
\setsymbol{LFI:FWHM:LFI20:units}{12\parcm75}
\setsymbol{LFI:FWHM:LFI21:units}{12\parcm74}
\setsymbol{LFI:FWHM:LFI22:units}{12\parcm87}
\setsymbol{LFI:FWHM:LFI23:units}{13\parcm27}
\setsymbol{LFI:FWHM:LFI24:units}{22\parcm98}
\setsymbol{LFI:FWHM:LFI25:units}{30\parcm46}
\setsymbol{LFI:FWHM:LFI26:units}{30\parcm31}
\setsymbol{LFI:FWHM:LFI27:units}{32\parcm65}
\setsymbol{LFI:FWHM:LFI28:units}{32\parcm66}

\setsymbol{LFI:FWHM:LFI18}{13.39}
\setsymbol{LFI:FWHM:LFI19}{13.01}
\setsymbol{LFI:FWHM:LFI20}{12.75}
\setsymbol{LFI:FWHM:LFI21}{12.74}
\setsymbol{LFI:FWHM:LFI22}{12.87}
\setsymbol{LFI:FWHM:LFI23}{13.27}
\setsymbol{LFI:FWHM:LFI24}{22.98}
\setsymbol{LFI:FWHM:LFI25}{30.46}
\setsymbol{LFI:FWHM:LFI26}{30.31}
\setsymbol{LFI:FWHM:LFI27}{32.65}
\setsymbol{LFI:FWHM:LFI28}{32.66}

\setsymbol{LFI:FWHM:uncertainty:LFI18:units}{0.170\arcm}
\setsymbol{LFI:FWHM:uncertainty:LFI19:units}{0.174\arcm}
\setsymbol{LFI:FWHM:uncertainty:LFI20:units}{0.170\arcm}
\setsymbol{LFI:FWHM:uncertainty:LFI21:units}{0.156\arcm}
\setsymbol{LFI:FWHM:uncertainty:LFI22:units}{0.164\arcm}
\setsymbol{LFI:FWHM:uncertainty:LFI23:units}{0.171\arcm}
\setsymbol{LFI:FWHM:uncertainty:LFI24:units}{0.652\arcm}
\setsymbol{LFI:FWHM:uncertainty:LFI25:units}{1.075\arcm}
\setsymbol{LFI:FWHM:uncertainty:LFI26:units}{1.131\arcm}
\setsymbol{LFI:FWHM:uncertainty:LFI27:units}{1.266\arcm}
\setsymbol{LFI:FWHM:uncertainty:LFI28:units}{1.287\arcm}

\setsymbol{LFI:FWHM:uncertainty:LFI18}{0.170}
\setsymbol{LFI:FWHM:uncertainty:LFI19}{0.174}
\setsymbol{LFI:FWHM:uncertainty:LFI20}{0.170}
\setsymbol{LFI:FWHM:uncertainty:LFI21}{0.156}
\setsymbol{LFI:FWHM:uncertainty:LFI22}{0.164}
\setsymbol{LFI:FWHM:uncertainty:LFI23}{0.171}
\setsymbol{LFI:FWHM:uncertainty:LFI24}{0.652}
\setsymbol{LFI:FWHM:uncertainty:LFI25}{1.075}
\setsymbol{LFI:FWHM:uncertainty:LFI26}{1.131}
\setsymbol{LFI:FWHM:uncertainty:LFI27}{1.266}
\setsymbol{LFI:FWHM:uncertainty:LFI28}{1.287}

\setsymbol{HFI:center:frequency:100GHz:units}{100\,GHz}
\setsymbol{HFI:center:frequency:143GHz:units}{143\,GHz}
\setsymbol{HFI:center:frequency:217GHz:units}{217\,GHz}
\setsymbol{HFI:center:frequency:353GHz:units}{353\,GHz}
\setsymbol{HFI:center:frequency:545GHz:units}{545\,GHz}
\setsymbol{HFI:center:frequency:857GHz:units}{857\,GHz}

\setsymbol{HFI:center:frequency:100GHz}{100}
\setsymbol{HFI:center:frequency:143GHz}{143}
\setsymbol{HFI:center:frequency:217GHz}{217}
\setsymbol{HFI:center:frequency:353GHz}{353}
\setsymbol{HFI:center:frequency:545GHz}{545}
\setsymbol{HFI:center:frequency:857GHz}{857}

\setsymbol{HFI:Ndetectors:100GHz}{8}
\setsymbol{HFI:Ndetectors:143GHz}{11}
\setsymbol{HFI:Ndetectors:217GHz}{12}
\setsymbol{HFI:Ndetectors:353GHz}{12}
\setsymbol{HFI:Ndetectors:545GHz}{3}
\setsymbol{HFI:Ndetectors:857GHz}{4}

\setsymbol{HFI:FWHM:Maps:100GHz:units}{9\parcm88}
\setsymbol{HFI:FWHM:Maps:143GHz:units}{7\parcm18}
\setsymbol{HFI:FWHM:Maps:217GHz:units}{4\parcm87}
\setsymbol{HFI:FWHM:Maps:353GHz:units}{4\parcm65}
\setsymbol{HFI:FWHM:Maps:545GHz:units}{4\parcm72}
\setsymbol{HFI:FWHM:Maps:857GHz:units}{4\parcm39}
\setsymbol{HFI:FWHM:Maps:100GHz}{9.88}
\setsymbol{HFI:FWHM:Maps:143GHz}{7.18}
\setsymbol{HFI:FWHM:Maps:217GHz}{4.87}
\setsymbol{HFI:FWHM:Maps:353GHz}{4.65}
\setsymbol{HFI:FWHM:Maps:545GHz}{4.72}
\setsymbol{HFI:FWHM:Maps:857GHz}{4.39}

\setsymbol{HFI:beam:ellipticity:Maps:100GHz}{1.15}
\setsymbol{HFI:beam:ellipticity:Maps:143GHz}{1.01}
\setsymbol{HFI:beam:ellipticity:Maps:217GHz}{1.06}
\setsymbol{HFI:beam:ellipticity:Maps:353GHz}{1.05}
\setsymbol{HFI:beam:ellipticity:Maps:545GHz}{1.14}
\setsymbol{HFI:beam:ellipticity:Maps:857GHz}{1.19}

\setsymbol{HFI:FWHM:Mars:100GHz:units}{9\parcm37}
\setsymbol{HFI:FWHM:Mars:143GHz:units}{7\parcm04}
\setsymbol{HFI:FWHM:Mars:217GHz:units}{4\parcm68}
\setsymbol{HFI:FWHM:Mars:353GHz:units}{4\parcm43}
\setsymbol{HFI:FWHM:Mars:545GHz:units}{3\parcm80}
\setsymbol{HFI:FWHM:Mars:857GHz:units}{3\parcm67}

\setsymbol{HFI:FWHM:Mars:100GHz}{9.37}
\setsymbol{HFI:FWHM:Mars:143GHz}{7.04}
\setsymbol{HFI:FWHM:Mars:217GHz}{4.68}
\setsymbol{HFI:FWHM:Mars:353GHz}{4.43}
\setsymbol{HFI:FWHM:Mars:545GHz}{3.80}
\setsymbol{HFI:FWHM:Mars:857GHz}{3.67}

\setsymbol{HFI:beam:ellipticity:Mars:100GHz}{1.18}
\setsymbol{HFI:beam:ellipticity:Mars:143GHz}{1.03}
\setsymbol{HFI:beam:ellipticity:Mars:217GHz}{1.14}
\setsymbol{HFI:beam:ellipticity:Mars:353GHz}{1.09}
\setsymbol{HFI:beam:ellipticity:Mars:545GHz}{1.25}
\setsymbol{HFI:beam:ellipticity:Mars:857GHz}{1.03}

\setsymbol{HFI:CMB:relative:calibration:100GHz}{$\lsim 1\%$}
\setsymbol{HFI:CMB:relative:calibration:143GHz}{$\lsim 1\%$}
\setsymbol{HFI:CMB:relative:calibration:217GHz}{$\lsim 1\%$}
\setsymbol{HFI:CMB:relative:calibration:353GHz}{$\lsim 1\%$}
\setsymbol{HFI:CMB:relative:calibration:545GHz}{}
\setsymbol{HFI:CMB:relative:calibration:857GHz}{}

\setsymbol{HFI:CMB:absolute:calibration:100GHz}{$\lsim 2\%$}
\setsymbol{HFI:CMB:absolute:calibration:143GHz}{$\lsim 2\%$}
\setsymbol{HFI:CMB:absolute:calibration:217GHz}{$\lsim 2\%$}
\setsymbol{HFI:CMB:absolute:calibration:353GHz}{$\lsim 2\%$}
\setsymbol{HFI:CMB:absolute:calibration:545GHz}{}
\setsymbol{HFI:CMB:absolute:calibration:857GHz}{}

\setsymbol{HFI:FIRAS:gain:calibration:accuracy:statistical:100GHz}{}
\setsymbol{HFI:FIRAS:gain:calibration:accuracy:statistical:143GHz}{}
\setsymbol{HFI:FIRAS:gain:calibration:accuracy:statistical:217GHz}{}
\setsymbol{HFI:FIRAS:gain:calibration:accuracy:statistical:353GHz}{2.5\%}
\setsymbol{HFI:FIRAS:gain:calibration:accuracy:statistical:545GHz}{1\%}
\setsymbol{HFI:FIRAS:gain:calibration:accuracy:statistical:857GHz}{0.5\%}

\setsymbol{HFI:FIRAS:gain:calibration:accuracy:systematic:100GHz}{}
\setsymbol{HFI:FIRAS:gain:calibration:accuracy:systematic:143GHz}{}
\setsymbol{HFI:FIRAS:gain:calibration:accuracy:systematic:217GHz}{}
\setsymbol{HFI:FIRAS:gain:calibration:accuracy:systematic:353GHz}{}
\setsymbol{HFI:FIRAS:gain:calibration:accuracy:systematic:545GHz}{7\%}
\setsymbol{HFI:FIRAS:gain:calibration:accuracy:systematic:857GHz}{7\%}

\setsymbol{HFI:FIRAS:zero:point:accuracy:100GHz:units}{0.8\MJysr}
\setsymbol{HFI:FIRAS:zero:point:accuracy:143GHz:units}{}
\setsymbol{HFI:FIRAS:zero:point:accuracy:217GHz:units}{}
\setsymbol{HFI:FIRAS:zero:point:accuracy:353GHz:units}{1.4\MJysr}
\setsymbol{HFI:FIRAS:zero:point:accuracy:545GHz:units}{2.2\MJysr}
\setsymbol{HFI:FIRAS:zero:point:accuracy:857GHz:units}{1.7\MJysr}

\setsymbol{HFI:FIRAS:zero:point:accuracy:100GHz}{0.8}
\setsymbol{HFI:FIRAS:zero:point:accuracy:143GHz}{}
\setsymbol{HFI:FIRAS:zero:point:accuracy:217GHz}{}
\setsymbol{HFI:FIRAS:zero:point:accuracy:353GHz}{1.4}
\setsymbol{HFI:FIRAS:zero:point:accuracy:545GHz}{2.2}
\setsymbol{HFI:FIRAS:zero:point:accuracy:857GHz}{1.7}

\setsymbol{HFI:unit:conversion:100GHz:units}{0.2415\MJysrmK}
\setsymbol{HFI:unit:conversion:143GHz:units}{0.3694\MJysrmK}
\setsymbol{HFI:unit:conversion:217GHz:units}{0.4811\MJysrmK}
\setsymbol{HFI:unit:conversion:353GHz:units}{0.2883\MJysrmK}
\setsymbol{HFI:unit:conversion:545GHz:units}{0.05826\MJysrmK}
\setsymbol{HFI:unit:conversion:857GHz:units}{0.002238\MJysrmK}

\setsymbol{HFI:unit:conversion:100GHz}{0.2415}
\setsymbol{HFI:unit:conversion:143GHz}{0.3694}
\setsymbol{HFI:unit:conversion:217GHz}{0.4811}
\setsymbol{HFI:unit:conversion:353GHz}{0.2883}
\setsymbol{HFI:unit:conversion:545GHz}{0.05826}
\setsymbol{HFI:unit:conversion:857GHz}{0.002238}

\setsymbol{HFI:colour:correction:alpha=-2:V1.01:100GHz}{0.9893}
\setsymbol{HFI:colour:correction:alpha=-2:V1.01:143GHz}{0.9759}
\setsymbol{HFI:colour:correction:alpha=-2:V1.01:217GHz}{1.0007}
\setsymbol{HFI:colour:correction:alpha=-2:V1.01:353GHz}{1.0028}
\setsymbol{HFI:colour:correction:alpha=-2:V1.01:545GHz}{1.0019}
\setsymbol{HFI:colour:correction:alpha=-2:V1.01:857GHz}{0.9889}

\setsymbol{HFI:colour:correction:alpha=0:V1.01:100GHz}{1.0008}
\setsymbol{HFI:colour:correction:alpha=0:V1.01:143GHz}{1.0148}
\setsymbol{HFI:colour:correction:alpha=0:V1.01:217GHz}{0.9909}
\setsymbol{HFI:colour:correction:alpha=0:V1.01:353GHz}{0.9888}
\setsymbol{HFI:colour:correction:alpha=0:V1.01:545GHz}{0.9878}
\setsymbol{HFI:colour:correction:alpha=0:V1.01:857GHz}{1.0014}

\providecommand{\sorthelp}[1]{}

\let\simlt=\lsim
\let\simgt=\gsim

\def\all2103resultspapers{\nocite{planck2013-p01, planck2013-p02, planck2013-p02a, planck2013-p02d, planck2013-p02b, planck2013-p03, planck2013-p03c, planck2013-p03f, planck2013-p03d, planck2013-p03e, planck2013-p01a, planck2013-p06, planck2013-p03a, planck2013-pip88, planck2013-p08, planck2013-p11, planck2013-p12, planck2013-p13, planck2013-p14, planck2013-p15, planck2013-p05b, planck2013-p17, planck2013-p09, planck2013-p09a, planck2013-p20, planck2013-p19, planck2013-pipaberration, planck2013-p05, planck2013-p05a, planck2013-pip56, planck2013-p06b}}

 \def\leaderfil{\leaders\hbox to 5pt{\hss.\hss}\hfil}
\makeatother

\begin{document}
\titlerunning{HFI data processing}

\authorrunning{Planck Collaboration}   %% NOT italics in "Planck Collaboration"!

%This author list corresponds to \title{Author list for SVN P03\_HFI\_Processing, Proj. Ref. 1\_7: The High Frequency Instrument Data Processing}
%Prepared by R. Leonardi (rleonardi@sciops.esa.int), ESAC/ESA
%This version is from Thu Sep 05 13:37:08 2013 CET
%\subtitle{There are 242 co-authors in this list}
\author{\small
Planck Collaboration:
P.~A.~R.~Ade\inst{86}
\and
N.~Aghanim\inst{59}
\and
C.~Armitage-Caplan\inst{90}
\and
M.~Arnaud\inst{73}
\and
M.~Ashdown\inst{70, 6}
\and
F.~Atrio-Barandela\inst{19}
\and
J.~Aumont\inst{59}
\and
C.~Baccigalupi\inst{85}
\and
A.~J.~Banday\inst{93, 10}
\and
R.~B.~Barreiro\inst{67}
% ...
}
\institute{\small
APC, AstroParticule et Cosmologie, Universit\'{e} Paris Diderot, CNRS/IN2P3, CEA/lrfu, Observatoire de Paris, Sorbonne Paris Cit\'{e}, 10, rue Alice Domon et L\'{e}onie Duquet, 75205 Paris Cedex 13, France\\
\and
Aalto University Mets\"{a}hovi Radio Observatory, Mets\"{a}hovintie 114, FIN-02540 Kylm\"{a}l\"{a}, Finland\\
\and
African Institute for Mathematical Sciences, 6-8 Melrose Road, Muizenberg, Cape Town, South Africa\\
\and
Agenzia Spaziale Italiana Science Data Center, Via del Politecnico snc, 00133, Roma, Italy\\
\and
Agenzia Spaziale Italiana, Viale Liegi 26, Roma, Italy\\
\and
Astrophysics Group, Cavendish Laboratory, University of Cambridge, J J Thomson Avenue, Cambridge CB3 0HE, U.K.\\
\and
Astrophysics \& Cosmology Research Unit, School of Mathematics, Statistics \& Computer Science, University of KwaZulu-Natal, Westville Campus, Private Bag X54001, Durban 4000, South Africa\\
\and
Atacama Large Millimeter/submillimeter Array, ALMA Santiago Central Offices, Alonso de Cordova 3107, Vitacura, Casilla 763 0355, Santiago, Chile\\
\and
CITA, University of Toronto, 60 St. George St., Toronto, ON M5S 3H8, Canada\\
\and
CNRS, IRAP, 9 Av. colonel Roche, BP 44346, F-31028 Toulouse cedex 4, France\\
\and
California Institute of Technology, Pasadena, California, U.S.A.\\
\and
Centre for Theoretical Cosmology, DAMTP, University of Cambridge, Wilberforce Road, Cambridge CB3 0WA, U.K.\\
\and
Centro de Estudios de F\'{i}sica del Cosmos de Arag\'{o}n (CEFCA), Plaza San Juan, 1, planta 2, E-44001, Teruel, Spain\\
\and
Computational Cosmology Center, Lawrence Berkeley National Laboratory, Berkeley, California, U.S.A.\\
\and
Consejo Superior de Investigaciones Cient\'{\i}ficas (CSIC), Madrid, Spain\\
\and
DSM/Irfu/SPP, CEA-Saclay, F-91191 Gif-sur-Yvette Cedex, France\\
\and
DTU Space, National Space Institute, Technical University of Denmark, Elektrovej 327, DK-2800 Kgs. Lyngby, Denmark\\
\and
D\'{e}partement de Physique Th\'{e}orique, Universit\'{e} de Gen\`{e}ve, 24, Quai E. Ansermet,1211 Gen\`{e}ve 4, Switzerland\\
\and
Departamento de F\'{\i}sica Fundamental, Facultad de Ciencias, Universidad de Salamanca, 37008 Salamanca, Spain\\
\and
Department of Astronomy and Astrophysics, University of Toronto, 50 Saint George Street, Toronto, Ontario, Canada\\
\and
Department of Astrophysics/IMAPP, Radboud University Nijmegen, P.O. Box 9010, 6500 GL Nijmegen, The Netherlands\\
\and
Department of Electrical Engineering and Computer Sciences, University of California, Berkeley, California, U.S.A.\\
\and
Department of Physics \& Astronomy, University of British Columbia, 6224 Agricultural Road, Vancouver, British Columbia, Canada\\
\and
Department of Physics and Astronomy, Dana and David Dornsife College of Letter, Arts and Sciences, University of Southern California, Los Angeles, CA 90089, U.S.A.\\
\and
Department of Physics and Astronomy, University College London, London WC1E 6BT, U.K.\\
\and
Department of Physics, Florida State University, Keen Physics Building, 77 Chieftan Way, Tallahassee, Florida, U.S.A.\\
\and
Department of Physics, Gustaf H\"{a}llstr\"{o}min katu 2a, University of Helsinki, Helsinki, Finland\\
\and
Department of Physics, Princeton University, Princeton, New Jersey, U.S.A.\\
\and
Department of Physics, University of California, One Shields Avenue, Davis, California, U.S.A.\\
\and
Department of Physics, University of California, Santa Barbara, California, U.S.A.\\
\and
Department of Physics, University of Illinois at Urbana-Champaign, 1110 West Green Street, Urbana, Illinois, U.S.A.\\
\and
Dipartimento di Fisica e Astronomia G. Galilei, Universit\`{a} degli Studi di Padova, via Marzolo 8, 35131 Padova, Italy\\
\and
Dipartimento di Fisica e Scienze della Terra, Universit\`{a} di Ferrara, Via Saragat 1, 44122 Ferrara, Italy\\
\and
Dipartimento di Fisica, Universit\`{a} La Sapienza, P. le A. Moro 2, Roma, Italy\\
\and
Dipartimento di Fisica, Universit\`{a} degli Studi di Milano, Via Celoria, 16, Milano, Italy\\
\and
Dipartimento di Fisica, Universit\`{a} degli Studi di Trieste, via A. Valerio 2, Trieste, Italy\\
\and
Dipartimento di Fisica, Universit\`{a} di Roma Tor Vergata, Via della Ricerca Scientifica, 1, Roma, Italy\\
\and
Discovery Center, Niels Bohr Institute, Blegdamsvej 17, Copenhagen, Denmark\\
\and
Dpto. Astrof\'{i}sica, Universidad de La Laguna (ULL), E-38206 La Laguna, Tenerife, Spain\\
\and
European Southern Observatory, ESO Vitacura, Alonso de Cordova 3107, Vitacura, Casilla 19001, Santiago, Chile\\
\and
European Space Agency, ESAC, Planck Science Office, Camino bajo del Castillo, s/n, Urbanizaci\'{o}n Villafranca del Castillo, Villanueva de la Ca\~{n}ada, Madrid, Spain\\
\and
European Space Agency, ESTEC, Keplerlaan 1, 2201 AZ Noordwijk, The Netherlands\\
\and
Helsinki Institute of Physics, Gustaf H\"{a}llstr\"{o}min katu 2, University of Helsinki, Helsinki, Finland\\
\and
INAF - Osservatorio Astrofisico di Catania, Via S. Sofia 78, Catania, Italy\\
\and
INAF - Osservatorio Astronomico di Padova, Vicolo dell'Osservatorio 5, Padova, Italy\\
\and
INAF - Osservatorio Astronomico di Roma, via di Frascati 33, Monte Porzio Catone, Italy\\
\and
INAF - Osservatorio Astronomico di Trieste, Via G.B. Tiepolo 11, Trieste, Italy\\
\and
INAF Istituto di Radioastronomia, Via P. Gobetti 101, 40129 Bologna, Italy\\
\and
INAF/IASF Bologna, Via Gobetti 101, Bologna, Italy\\
\and
INAF/IASF Milano, Via E. Bassini 15, Milano, Italy\\
\and
INFN, Sezione di Bologna, Via Irnerio 46, I-40126, Bologna, Italy\\
\and
INFN, Sezione di Roma 1, Universit\`{a} di Roma Sapienza, Piazzale Aldo Moro 2, 00185, Roma, Italy\\
\and
IPAG: Institut de Plan\'{e}tologie et d'Astrophysique de Grenoble, Universit\'{e} Joseph Fourier, Grenoble 1 / CNRS-INSU, UMR 5274, Grenoble, F-38041, France\\
\and
IUCAA, Post Bag 4, Ganeshkhind, Pune University Campus, Pune 411 007, India\\
\and
Imperial College London, Astrophysics group, Blackett Laboratory, Prince Consort Road, London, SW7 2AZ, U.K.\\
\and
Infrared Processing and Analysis Center, California Institute of Technology, Pasadena, CA 91125, U.S.A.\\
\and
Institut N\'{e}el, CNRS, Universit\'{e} Joseph Fourier Grenoble I, 25 rue des Martyrs, Grenoble, France\\
\and
Institut Universitaire de France, 103, bd Saint-Michel, 75005, Paris, France\\
\and
Institut d'Astrophysique Spatiale, CNRS (UMR8617) Universit\'{e} Paris-Sud 11, B\^{a}timent 121, Orsay, France\\
\and
Institut d'Astrophysique de Paris, CNRS (UMR7095), 98 bis Boulevard Arago, F-75014, Paris, France\\
\and
Institute for Space Sciences, Bucharest-Magurale, Romania\\
\and
Institute of Astro and Particle Physics, Technikerstrasse 25/8, University of Innsbruck, A-6020, Innsbruck, Austria\\
\and
Institute of Astronomy and Astrophysics, Academia Sinica, Taipei, Taiwan\\
\and
Institute of Astronomy, University of Cambridge, Madingley Road, Cambridge CB3 0HA, U.K.\\
\and
Institute of Theoretical Astrophysics, University of Oslo, Blindern, Oslo, Norway\\
\and
Instituto de Astrof\'{\i}sica de Canarias, C/V\'{\i}a L\'{a}ctea s/n, La Laguna, Tenerife, Spain\\
\and
Instituto de F\'{\i}sica de Cantabria (CSIC-Universidad de Cantabria), Avda. de los Castros s/n, Santander, Spain\\
\and
Jet Propulsion Laboratory, California Institute of Technology, 4800 Oak Grove Drive, Pasadena, California, U.S.A.\\
\and
Jodrell Bank Centre for Astrophysics, Alan Turing Building, School of Physics and Astronomy, The University of Manchester, Oxford Road, Manchester, M13 9PL, U.K.\\
\and
Kavli Institute for Cosmology Cambridge, Madingley Road, Cambridge, CB3 0HA, U.K.\\
\and
LAL, Universit\'{e} Paris-Sud, CNRS/IN2P3, Orsay, France\\
\and
LERMA, CNRS, Observatoire de Paris, 61 Avenue de l'Observatoire, Paris, France\\
\and
Laboratoire AIM, IRFU/Service d'Astrophysique - CEA/DSM - CNRS - Universit\'{e} Paris Diderot, B\^{a}t. 709, CEA-Saclay, F-91191 Gif-sur-Yvette Cedex, France\\
\and
Laboratoire Traitement et Communication de l'Information, CNRS (UMR 5141) and T\'{e}l\'{e}com ParisTech, 46 rue Barrault F-75634 Paris Cedex 13, France\\
\and
Laboratoire de Physique Subatomique et de Cosmologie, Universit\'{e} Joseph Fourier Grenoble I, CNRS/IN2P3, Institut National Polytechnique de Grenoble, 53 rue des Martyrs, 38026 Grenoble cedex, France\\
\and
Laboratoire de Physique Th\'{e}orique, Universit\'{e} Paris-Sud 11 \& CNRS, B\^{a}timent 210, 91405 Orsay, France\\
\and
Lawrence Berkeley National Laboratory, Berkeley, California, U.S.A.\\
\and
Max-Planck-Institut f\"{u}r Astrophysik, Karl-Schwarzschild-Str. 1, 85741 Garching, Germany\\
\and
McGill Physics, Ernest Rutherford Physics Building, McGill University, 3600 rue University, Montr\'{e}al, QC, H3A 2T8, Canada\\
\and
National University of Ireland, Department of Experimental Physics, Maynooth, Co. Kildare, Ireland\\
\and
Niels Bohr Institute, Blegdamsvej 17, Copenhagen, Denmark\\
\and
Observational Cosmology, Mail Stop 367-17, California Institute of Technology, Pasadena, CA, 91125, U.S.A.\\
\and
Optical Science Laboratory, University College London, Gower Street, London, U.K.\\
\and
SB-ITP-LPPC, EPFL, CH-1015, Lausanne, Switzerland\\
\and
SISSA, Astrophysics Sector, via Bonomea 265, 34136, Trieste, Italy\\
\and
School of Physics and Astronomy, Cardiff University, Queens Buildings, The Parade, Cardiff, CF24 3AA, U.K.\\
\and
Space Sciences Laboratory, University of California, Berkeley, California, U.S.A.\\
\and
Special Astrophysical Observatory, Russian Academy of Sciences, Nizhnij Arkhyz, Zelenchukskiy region, Karachai-Cherkessian Republic, 369167, Russia\\
\and
Stanford University, Dept of Physics, Varian Physics Bldg, 382 Via Pueblo Mall, Stanford, California, U.S.A.\\
\and
Sub-Department of Astrophysics, University of Oxford, Keble Road, Oxford OX1 3RH, U.K.\\
\and
Theory Division, PH-TH, CERN, CH-1211, Geneva 23, Switzerland\\
\and
UPMC Univ Paris 06, UMR7095, 98 bis Boulevard Arago, F-75014, Paris, France\\
\and
Universit\'{e} de Toulouse, UPS-OMP, IRAP, F-31028 Toulouse cedex 4, France\\
% ...
}

\title{\textit{\Planck\ }2013 results. VI. High Frequency Instrument \\
 data processing}

\abstract{We describe the processing of the 531 billion raw data samples from
the High Frequency Instrument (hereafter HFI), which we performed
to produce six temperature maps from the first 473 days of \Planck-HFI
survey data. These maps provide an accurate rendition of the sky emission
at 100, 143, 217, 353, 545, and 857\GHz\ with an angular resolution
ranging from $9\parcm7$ to 4\parcm6. The detector noise per (effective)
beam solid angle is respectively, 10, 6 , 12, and 39\,$\mu\textrm{K}$
in the four lowest HFI frequency \rev{channels}\textcolor{red}{\normalsize{}
}(100--353\GHz) and 13 and 14\,kJy sr$^{-1}$ in the 545 and 857\,GHz
channels. Relative to the 143\,GHz channel, these two high frequency
channels are calibrated to within 5\,\% and the 353\,GHz channel
to the percent level. The 100 and 217\,GHz channels, which together
with the 143\,GHz channel determine the high-multipole part of the
CMB power spectrum ($50<\ell<2500$), are calibrated relative to 143\,GHz
to better than 0.2\,\%. }


\maketitle

\section{Introduction}

This paper, one of a set associated with the 2013 release of data
from the \Planck\footnote{\Planck\ (\url{http://www.esa.int/Planck}) is a project of the European Space Agency (ESA) with instruments provided by two scientific consortia funded by ESA member states (in particular the lead countries France and Italy), with contributions from NASA (USA) and telescope reflectors provided by a collaboration between ESA and a scientific consortium led and funded by Denmark.}
mission (\citealt{planck2013-p01}-\citealt{planck2013-p01a}), describes
the processing of data from the \Planck\ High Frequency Instrument
(HFI) to produce calibrated and characterized maps. HFI (\citealt{lamarre2010,planck2011-1.5})
observes in the 100, 143, 217, 353, 545, and 857\,GHz bands with
bolometers cooled to 0.1\,\hbox{K}. The HFI instrument comprises
50 signal bolometers, as well as two dark bolometers, 16 thermometers,
a resistor, and a capacitor used for monitoring and housekeeping.
The count of 50 bolometers includes 12 polarization sensitive bolometer
(PSB) pairs, four each at 100--$353\GHz$; the rest are unpolarized
spider-web bolometers (SWBs). We describe the steps taken by the HFI
data processing centre (hereafter DPC) to transform the packets sent
by the satellite into sky maps at HFI frequencies, with the help of
ancillary data, for example, from ground calibration. These are temperature
maps alone, as obtained from the beginning of the first light survey
on 13 August 2009, to the end of the nominal mission on 27 November
2010. 

\Planck\ defines a sky survey as the time over which the spin axis
rotates by 180\deg, a period close to six months in duration in which
about 95\,\% of the sky is covered at each frequency. During routine
operations, \Planck\ scans the sky by spinning in circles with an
angular radius \rev{of} roughly 85\deg. The spin axis follows a
cycloidal path on the sky by periodic step-wise displacements of 2\arcm,
resulting in typically 40 (35 to 70) circles of typical duration 46
minutes, constituting a \emph{stable pointing period} between repointings.
The scanning strategy is discussed in more detail in \citet{tauber2010a},
\citet{planck2011-1.1}, and \citet{planck2013-p01}. The 15.5~months
of nominal mission survey data then provide 2.5 sky surveys, and maps
are provided for the first two sky surveys separately, as well as
for the complete nominal mission. As a means to estimate aspects of
the noise distribution, we also deliver ``half-ring'' maps made
out of the first and second half of each stable pointing period. Maps
are produced for individual detectors, as well as for averages over
each band and for selected detector sets defined within each band
(see Table~\ref{tab:detsets}). 

The next section provides an overview of HFI data processing. Section~\ref{sec:TOI-PROCESSING}
is devoted to the processing of time-ordered information (hereafter
TOI) from individual detectors to produce cleaned timelines. These
timelines are used to estimate the temporal noise properties in Sect.~\ref{sec:Toi-Qualification}
and to determine the detector pointings and beams in Sects.~\ref{sec:DETECTOR-POINTINGS}
and \ref{sec:DETECTOR-BEAMS}. Section~\ref{sec:MAP-MAKING} discusses
the creation of maps and their photometric calibration, while Sect.~\ref{sec:VALIDATION}
presents tests applied to assess the consistency and accuracy of the
products. For completeness, component separation and further processing
are briefly described in Sect.~\ref{sec:Other-papers-validation}.
Section~\ref{sec:SUMMARY} concludes with a summary of the \rev{characteristics}
of the HFI data delivered, as currently processed. 

Some of the specific processing steps for HFI data are described more
fully elsewhere: \citet{planck2013-p03c} discusses the transfer function
and beams; \citet{planck2013-p03f} the calibration of HFI detectors;
\citet{planck2013-p03d} the determination of the spectral bands for
each detector and their combination; and \citet{planck2013-p03e}
the effect of so-called ``glitches'' such as cosmic-ray hits on
the detectors. The processing of data from the Low Frequency Instrument
(LFI) is discussed in \citet{planck2013-p02}. Technical details of
specific data products are discussed in \citet{planck2013-p28}. We
have applied to the delivered data products many consistency and validation
tests to assess their quality \rev{(see in particular} \citealt{planck2013-p03d,planck2013-p06,planck2013-p08,planck2013-p11}).
While the products meet a very high standard, as described here, we
did find limitations. Their mitigation, and related data products,
are left to future releases. In particular, HFI analysis revealed
that nonlinear effects in the on-board analogue-to-digital converters
(ADC) modified the recovered bolometer signal. In situ\emph{ }observations
over 2012--2013 are measuring this effect, and algorithms have been
developed to explicitly account for it in the data analysis. However,
the first-order effect of the ADC nonlinearity mimics a gain variation
in the bolometers, which the current release measures and removes
as part of the calibration procedures. This is discussed further in
Sects.~\ref{sec:Abs-calib} and \ref{sec:ADC-non-linearity}. 

The mapmaking procedure uses the full intensity and polarization information
from the HFI bolometers. The current analysis cannot guarantee that
the large-scale polarization signal is free from systematic effects.
However, the preliminary analysis shows that the small-scale maps
have the expected CMB content at high signal-to-noise, as discussed
in Sect.~\ref{sec:Polar} below, and in \citet{planck2013-p01} and
\citet{planck2013-p11}. Although we do not use these maps for cosmological
measurements, future work will use them for investigations of the
properties of polarized emission of the Galaxy.

\rev{Finally, since the March 2013 data release, we have found strong evidence that the $\ell\simeq 1800$ dip in some 217\,GHz detector cross-spectra is stronger in the first six-month survey than in subsequent surveys and that its amplitude may be reduced by additional data flagging targeting electromagnetic interference from the \HeJT\ (hereafter 4\,K) cooler drive and read-out electronics (see Sect.~\ref{sec:4Kline-removal}). This dip is therefore likely to be a (small) residual systematic effect in the data, which we show has little impact on cosmological parameter determination \citep{planck2013-p08, planck2013-p11}, but which contributes to the weak detection of a feature in the power spectrum reconstruction done in \cite{planck2013-p17}.}\foreignlanguage{english}{}
\begin{table*}[t]
\protect\caption{Detector sets (``ds'') used in this data release. \label{tab:detsets}}
\centering{}\begingroup
\newdimen\tblskip \tblskip=5pt
\nointerlineskip
\vskip -3mm
\footnotesize
\setbox\tablebox=\vbox{
   \newdimen\digitwidth 
   \setbox0=\hbox{\rm 0} 
   \digitwidth=\wd0 
   \catcode`*=\active 
   \def*{\kern\digitwidth}
   \newdimen\signwidth 
   \setbox0=\hbox{+} 
   \signwidth=\wd0 
   \catcode`!=\active 
   \def!{\kern\signwidth}
\halign{\hfil#\hfil\tabskip=2em&
\hfil#\hfil&
\hfil#\hfil&
\hfil#\hfil&
#\hfil&
#\hfil\/\tabskip=0pt\cr             % Template goes here.
%\hfil#\hfil&
%\hfil#\hfil\/\tabskip=0pt\cr             % Template goes here.
\noalign{\doubleline}
Set name& Frequency&  Type& Detectors in the set& Weights in the set& Products\cr
     &    [GHz]&      &     &     &     \cr
  % Table headings go here.
\noalign{\vskip 3pt\hrule\vskip 5pt}
                                    % Body of table goes here.
100-ds0& 100& MIX& All 8 detectors& 0.33, 0.44, 1.50, 0.76& N, B, F, H, S1, S2, HR1, HR2\cr
     &      &      &      & 2.08, 1.40, 0.97, \rev{0.52}& \hskip 1em Z-N, Z-S1, Z-S2\cr
\noalign{\vskip 3pt\hrule\vskip 3pt} 
100-ds1& 100& PSB& 1a+1b + 4a+4b& 0.58, 0.78, 1.71, 0.93& C\cr
100-ds2& 100& PSB& 2a+2b + 3a+3b& 1.05, 0.53, 1.45, 0.98& C\cr
\noalign{\vskip 3pt\hrule\vskip 3pt} 
143-ds0& 143& MIX& 11 detectors& 1.04, 0.44, 1.06, 0.93, 0.88, 0.94& N, B, F, H, S1, S2, HR1, HR2\cr
     &      &      &      & 0.80, 0.66, 1.32, 1.27, 1.66& \hskip 1em Z-N, Z-S1, Z-S2\cr
\noalign{\vskip 3pt\hrule\vskip 3pt} 
143-ds1& 143& PSB& 1a+1b + 3a+3b& 1.26, 0.53, 1.06, 1.15& C\cr
143-ds2& 143& PSB& 2a+2b + 4a+4b& 1.23, 1.08, 0.92, 0.76& C\cr
143-ds3& 143& SWB& 143-5&      & C\cr
143-ds4& 143& SWB& 143-6&      & C\cr
143-ds5& 143& SWB& 143-7&      & C\cr
\noalign{\vskip 3pt\hrule\vskip 3pt} 
217-ds0& 217& MIX& All 12 detectors& 1.54, 1.44, 1.62, 1.83, 0.60, 0.77& N, B, F, H, S1, S2, HR1, HR2\cr
     &      &      &      & 0.62, 0.67, 0.81, 0.85, 0.64, 0.59& \hskip 1em Z-N, Z-S1, Z-S2\cr
\noalign{\vskip 3pt\hrule\vskip 3pt} 
217-ds1& 217& PSB& 5a+5b + 7a+7b& 0.79, 1.02, 1.07, 1.12& C\cr
217-ds2& 217& PSB& 6a+6b + 8a+8b& 0.98, 1.06, 1.01, 0.94& C\cr
217-ds3& 217& SWB& 217-1&      & C\cr
217-ds4& 217& SWB& 217-2&      & C\cr
217-ds5& 217& SWB& 217-3&      & C\cr
217-ds6& 217& SWB& 217-4&      & C\cr
\noalign{\vskip 3pt\hrule\vskip 3pt} 
353-ds0& 353& MIX& All 12 detectors& 2.45, 2.38, 0.44, 0.65, 0.61, 0.57& N, B, F, H, S1, S2, HR1, HR2\cr
&      &      &      & 0.64, 0.64, 0.31, 0.34, 1.62, 1.35& \hskip 1em Z-N, Z-S1, Z-S2\cr
\noalign{\vskip 3pt\hrule\vskip 3pt} 
353-ds1& 353& PSB& 3a+3b + 5a+5b& 0.75, 1.09, 1.09, 1.07\cr
353-ds2& 353& PSB& 4a+4b + 6a+6b& 1.33, 1.25, 0.68, 0.73\cr
353-ds3& 353& SWB& 353-1&\cr
353-ds4& 353& SWB& 353-2&\cr
353-ds5& 353& SWB& 353-7&\cr
353-ds6& 353& SWB& 353-8&\cr
\noalign{\vskip 3pt\hrule\vskip 3pt} 
545-ds0& 545& SWB& 3 detectors (1, 2, 4)& 0.94, 1.10, 0.96& N, B, F, H, S1, S2, HR1, HR2\cr
     &      &      &      &       & \hskip 1em Z-N, Z-S1, Z-S2\cr
\noalign{\vskip 3pt\hrule\vskip 3pt} 
857-ds0& 857& SWB& All 4 detectors& 1.14, 1.11, 1.15, 0.60& N, B, F, H, S1, S2, HR1, HR2\cr
     &      &      &      &       & \hskip 1em Z-N, Z-S1, Z-S2\cr
\noalign{\vskip 5pt\hrule\vskip 3pt}}}
\endPlancktablewide                 % ends two-column \halign
Unpolarized SWBs are used alone, while PSBs (with individual bolometers denoted ``a''
and ``b'') are used by pair; \rev{``MIX'' denotes a combination of detector types}.
Here we consider only the temperature map extracted from the analysis
of four bolometers (two pairs of PSBs). The weights indicate the relative
weighting used in producing maps out of the TOI of several detectors.
The relative weights in a set are given in the numerical order of
each detector (e.g., 1a, 1b, 2a, 2b, 3a, 3b, \ldots{} for the 100-ds0
set). The last column details the specific products created for each
set. \rev{``B'': beam information; ``C'': detector-set spectra, corrected for the beam transfer function, for the detector sets used in the high-$\ell$ likelihood; ``F'': frequency band information; ``H'': pixel hit-count maps; ``HR1'', ``HR2'': maps made from the first or second half of each ring; ``N'': nominal mission maps; ``S1'', ''S2'': survey maps generated from the data collected during the first six months or the next six months; and ``Z'': ZLE/FSL-corrected maps.}
\endgroup
\end{table*}

\section{HFI data processing overview\label{sec:OVERVIEW}}

\begin{figure*}[t]
\centering{}\includegraphics{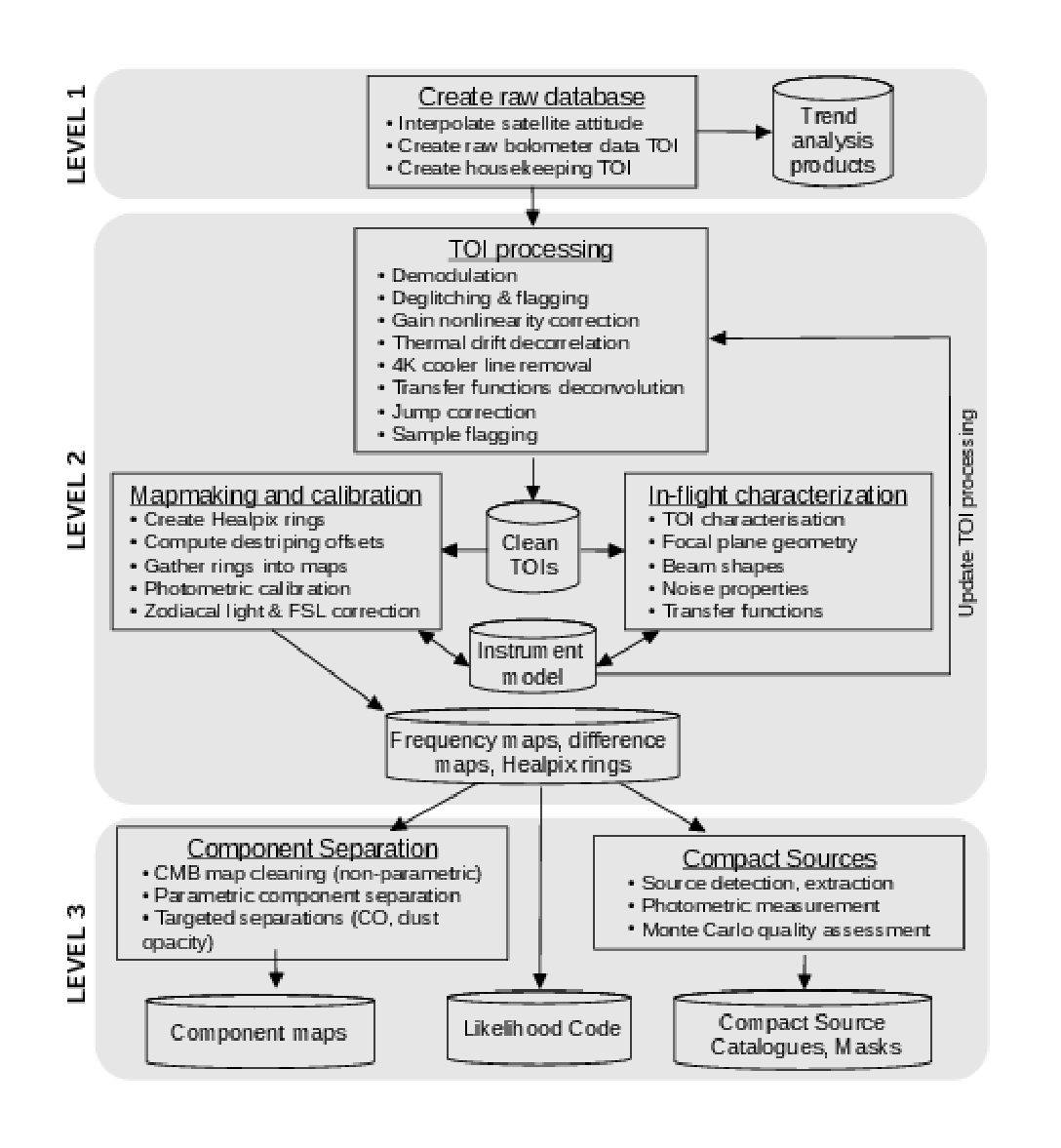}\protect\caption{\label{fig:Flowchart}Overview of the data flow and main functional
tasks of the HFI Data Processing Centre. Level 1 creates a database
of the raw satellite data as a function of time. Level 2 builds a
data model and produces maps sky maps at the six frequency of the
instrument. This flow diagram illustrates the crucial role of the
Instrument Model (\IMO), which is both an input and an output of
many tasks, and is updated iteratively during successive passes of
the data. Level 3 takes these instrument-specific products and derives
final astrophysical products. This paper is mostly concerned with
Level 2 processing and its validation. }
\end{figure*}

The processing of HFI data proceeds according to a series of levels,
shown schematically in Fig.~\ref{fig:Flowchart}. Level 1 (L1) creates
a database of the raw satellite data as a function of time (TOI objects).
The full set of TOI comprises the signals from each HFI bolometer,
ancillary information (e.g., pointing data), and associated housekeeping
data (e.g., temperature monitors). Level 2 (L2), the subject of this
paper, uses these data to build a model of the HFI instrument, the
Instrument Model (\IMO), produces cleaned, calibrated timelines for
each detector, and combines these into aggregate products such as
maps at each frequency. Level 3 (L3) takes these instrument-specific
results and derives various products: component-separation algorithms
transform the maps at each frequency into maps of separate astrophysical
components; source detection algorithms create catalogues of Galactic
and extragalactic objects; finally, a likelihood code assesses the
match between a cosmological and astrophysical model and the frequency
maps. 

Of course, these processing steps are not done completely sequentially:
HFI data are processed iteratively. In many ways, the \IMO\ is the
main internal data product from \textit{\Planck}, and the main task
of the HFI DPC is its iterative updating. Early versions of the \IMO\ were
derived from pre-launch data, and from the first-light survey of the
last two weeks of August 2009. Further revisions of the \IMO, and
of the pipelines themselves, were derived after the completion of
successive passes through the data.\textcolor{red}{{} }These new versions
included expanded information about the HFI instrument: for example,
the initial \IMO\ contained only coarse information about the shape
of the detector angular response (i.e., the full-width at half-maximum
of an approximating Gaussian); subsequent revisions included full
measured harmonic-space window functions. 

In somewhat more detail, L1 software fills the database and updates,
daily, the various TOI objects. Satellite attitude data, sampled at
8\Hz\ during science data acquisition and at 4\Hz\ otherwise, are
resampled by interpolation to the 180.37370\Hz\ (hereafter 180.4\Hz)
acquisition frequency of the detectors, corresponding to the integration
time for a single data sample; further information on L1 steps was
given in \citet{planck2011-1.7}. Raw timelines and housekeeping data
are then processed by L2 to compensate for the instrumental response
and to remove estimates of known artefacts. The various steps in TOI
processing are discussed in Sect.~\ref{sec:TOI-PROCESSING}. First,
the raw timeline voltages are demodulated, deglitched, and corrected
for the bolometer nonlinearity and for temperature fluctuations of
the environment using correlations with the signal TOI from the two
dark bolometers that serve as bolometer plate temperature monitors.
Narrow lines \rev{in the TOI frequency spectra} caused by the \HeJT\ (4\,K)
cooler are also removed before deconvolving the temporal response
of the instrument. Finally, various flags are set to mark unusable
samples.

Further use of the data requires knowledge of the pointing for individual
detectors, as discussed in Sect.~\ref{sec:DETECTOR-POINTINGS}. During
a single \emph{stable pointing period}, \Planck\   spins around an
axis pointing towards a fixed direction on the sky (up to an accounted-for
wobbling), repeatedly scanning approximately the same circle \citep{planck2013-p01}.
The satellite is re-pointed so that the spin axis follows the Sun,
and the observed circle sweeps through the sky at approximately one
degree per day. Assuming a focal plane geometry, i.e., a set of relations
between the satellite pointing and that of each of the detectors,
we build \emph{rings} of data derived by analysing the data acquired
by a detector during each stable pointing period (``ring'' refers
to the data obtained during a single stable pointing period). This
redundancy permits averaging of the data on rings to reduce instrument
noise. The resulting estimate of the sky signal can then be subtracted
from the timeline to estimate the temporal noise power spectral density,
a useful characterization of the detector data after TOI processing.
This noise may be described as a white noise component, dominating
at intermediate temporal frequencies, plus additional low- and high-frequency
noise. The effect on maps of the low-frequency part of the noise can
be partially mitigated by determining an offset for each ring. These
so-called ``destriping'' offsets are obtained by requiring that
the difference between intersecting rings be minimized. Once the offsets
are removed from each ring, the rings are co-added to produce sky
maps. 

As explained in Sect.~\ref{sec:MAP-MAKING}, a complication arises
from the fact that the detector data include both the contribution
from the\emph{ solar} dipole induced by the motion of the Solar System
through the CMB (sometimes referred to as the ``cosmological'' dipole),
and the \emph{orbital} dipole induced by the motion of the satellite
within the Solar System, which is not constant on the sky and must
therefore be removed from the rings before creating the sky map. The
solar dipole is used as a calibration source at lower HFI frequencies,
and bright planet fluxes at higher frequencies. Since we need this
calibration to remove the orbital dipole contribution to create the
maps themselves, the maps and their calibrations are obtained iteratively.
The dipoles are computed in the non-relativistic approximation. The
resulting calibration coefficients are also stored in the \IMO, which
can then be used, for instance, to express noise spectra in noise-equivalent
temperature (NET) units. The destriping offsets, once obtained through
a global solution, are also used to create local maps around planets.
As described in Sect.~\ref{sec:DETECTOR-POINTINGS} these are used
to improve our knowledge of the focal plane geometry stored in the
previous version of the \IMO\ and to improve measurements of the
``scanning'' beam (defined as the response to a point source of
the full optical and electronic system, \emph{after} the filtering
done during the TOI processing step, described in Sect.~\ref{sec:DETECTOR-BEAMS}).

The ring and mapmaking stages allow us to generate many different
maps, e.g., using different sets of detectors, the first or second
halves of the data in each ring, or data from different sky surveys.
Null tests using difference maps of the same sky area observed at
different times, in particular, have proved extremely useful in characterizing
the map residuals, described in Sect.~\ref{sec:VALIDATION}.

\begin{figure*}[t]
\centering{}\includegraphics[clip,width=1\textwidth]{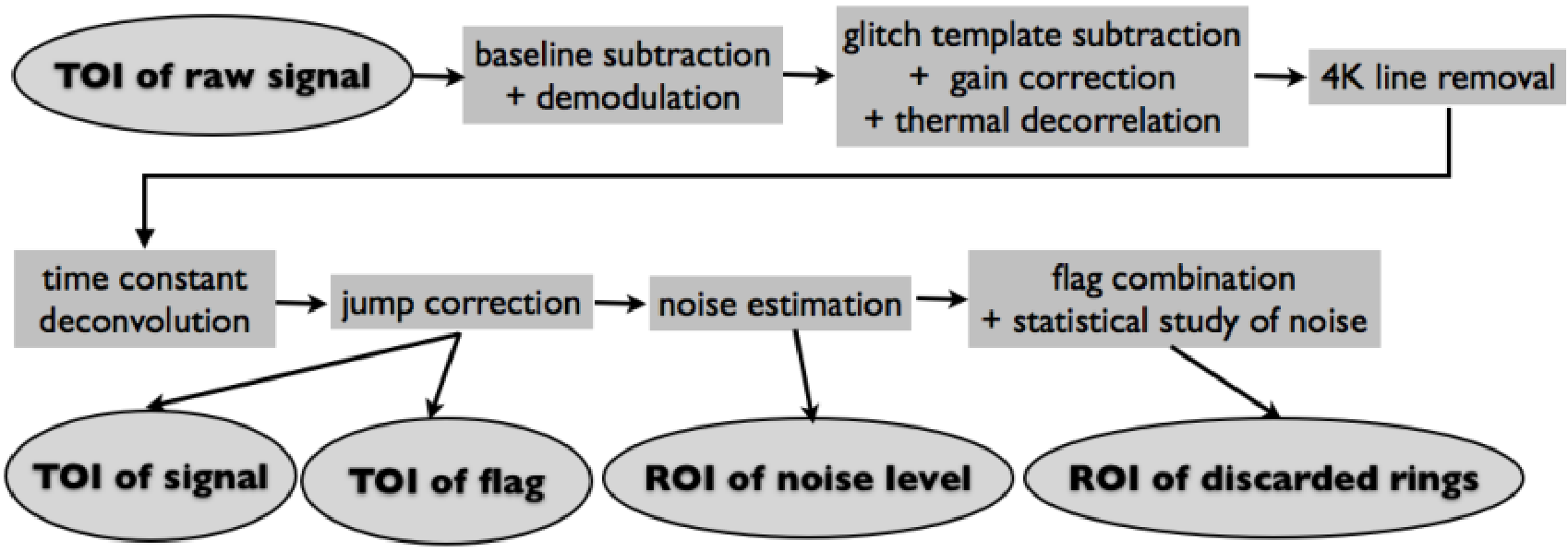}\protect\caption{Main TOI processing steps. Pipeline modules are represented as filled
squares. TOI: time ordered information. ROI: ring ordered information
(i.e., one piece of information per stable pointing period). Ring:
data obtained during a stable pointing period (typically 40 satellite
revolutions of one minute each).\label{fig:ToiProcSteps} }
\end{figure*}

\section{TOI processing\label{sec:TOI-PROCESSING} }

In the L1 stage of processing (previously described in Section 3 of
\citealt{planck2011-1.7}), the raw telemetry TOI are unfolded into
one time series for each bolometer. The signal is regularly sampled
at 180.4\,\Hz. We denote as \emph{TOI processing} the transformation
of the TOI coming from L1 into clean TOI objects, which can be used
for mapmaking after focal plane geometry reconstruction and photometric
calibration have been performed. The general philosophy of the TOI
processing is to modify the timelines as little as possible, and therefore
to flag regions contaminated by systematic effects (e.g., cosmic-ray
glitches). We deal with each bolometer signal separately. Aside from
allowing the possible flagging of known bright sources, the only pointing
information that is used in the main TOI processing is the phase (see
Sect.~\ref{sec:Deglitching}), so the TOI processing assumes perfect
redundancy of the data within a given pointing period. The output
of TOI processing is not only a set of clean TOI but also accompanying
qualifying flags and trend parameters used internally for detailed
statistics. Moreover, all data samples are processed, although only
clean samples will be projected on maps. For beam measurement (see
Sect.~\ref{sec:DETECTOR-BEAMS}), specific processing is performed
on pointing periods that are close to Mars, Jupiter or Saturn (see
Sect.~\ref{sec:Big-planet-TOI}).

A flow diagram is shown in Fig.~\ref{fig:ToiProcSteps}, illustrating
the TOI processing steps detailed in the following subsections. Section
4 of \citet{planck2011-1.7} presents the early version of the TOI
processing. The pipeline has changed sufficiently since then to warrant
a self-contained global description of the TOI processing. We refer
the reader to \citet{planck2013-p28} and the various companion papers
mentioned above for more details. The changes mostly reflect the improvement
in performance and in our understanding of the underlying effects.
The TOI are not delivered in the present data release, but their processing
is an essential, though hidden, ingredient in the delivered maps.
Some of the systematic effects arising in the map analysis can only
be understood by referring to the bolometer timeline behaviour and
processing. 

It was recently realised that some apparent gain variations, spotted
comparing identical pointing circles one-year apart, actually originate
in nonlinearities in the bolometer readout system ADCs. Note that
the ADC nonlinearity is not explicitly corrected for in the TOI processing,
but rather as an equivalent gain variation at the mapmaking level
(see Sect.~\ref{sec:Abs-calib} and Sect.~\ref{sec:ADC-non-linearity}).

\subsection{Input flags\label{sec:Input-flags}}

Strong signal gradients can adversely affect some stages of TOI processing.
We flag the data expected to have strong gradients using only the
pointing information, exclusively in the intermediate stages and not
for mapmaking. We also flag data where the pointing is known to be
unstable. Input flags come from the following.
\begin{enumerate}
\item A \emph{point-source} flag. This is based on the locations of sources
in the \Planck\ Early Release Compact Source Catalog (ERCSC; see
\citealt{planck2011-1.10}). A mask map is generated around%
\begin{comment}
\textcolor{red}{{} {[}give radius size ??{]}} 
\end{comment}
{} each of them from which flag TOI objects are created using the pointing
information.
\item A \emph{Galactic} flag. This is based on \emph{IRAS} maps with a threshold
\textcolor{red}{}%
\begin{comment}
\textcolor{red}{{[}Give threshold details ??{]} All details in the
XS}
\end{comment}
{} that depends on the frequency.
\item A \emph{BigPlanet} flag. Any sample that falls within a given distance
\textcolor{red}{}%
\begin{comment}
\textcolor{red}{{[}what distance, or call it just Planet Flag{]}}
\end{comment}
{} of Mars, Jupiter, or Saturn is flagged.
\item An \emph{Unstable pointing} flag (see Sect.~\ref{sec:DETECTOR-POINTINGS}).
This accounts for depointings between stable pointing periods and
other losses of pointing integrity. 
\end{enumerate}
Flagged data are either processed separately or ignored, depending
on the specific analysis. The flags are described in more detail in
\citet{planck2013-p28}.

\subsection{Demodulation\label{sec:Demodulation}}

The bolometers are AC square-wave modulated to put the acquisition
electronics $1/f$ noise at high temporal frequencies \citep{lamarre2010}.
\rev{The modulation frequency is $f_\textrm{mod}=f_\textrm{acq}/2=90.18685\,\mathrm{Hz}$}. A
demodulation step is done as follows. First a one-hour running average
of the modulated timeline is computed, known as the AC offset baseline.
This is carried out by excluding data that are masked due to glitches
(see below) or by the Galactic flag (see above). Once the AC offset
baseline is subtracted from the raw timeline, a simple $(+,-)$ demodulation
is applied. The overall sign of the signal is set to obtain a positive
signal on point sources and Galaxy crossings. This AC offset baseline
removal is needed in order to correct for the slow drift of the zero
level of the electronics. Any possible drift of the baseline on a
timescale smaller than one hour is dealt with at the filtering stage
(see below). The baseline varies very smoothly over the mission and
fluctuates by less than ten minimum resolution units from the middle
of the range of 65536 values allowed by the on-board ADC, discussed
in greater detail in Sect.~\ref{sec:ADC-non-linearity}.

\subsection{Deglitching and gap-filling\label{sec:Deglitching}}

The timelines are affected by obvious \emph{``}glitches'' (cosmic
ray hits and other large excursions) at a rate of about one per second.
This generates a huge Poisson noise if not dealt with. Such glitches
are detected as a large positive signal followed by a roughly exponential
tail. There are three basic classes of glitches affecting bolometers.
The statistical and physical understanding of the different populations
is given by \citet{planck2011-1.7}, and revised and expanded by \citet{planck2013-p03e}.
Here we present the general algorithm. 

The type of TOI used in the deglitching is slightly different from
the one used in the main processing described in the following sections.
Before deglitching, the timeline is demodulated and digitally filtered
with a three-point (0.25, 0.5, 0.25) moving kernel. Linear interpolation
is performed on those parts of the timeline when the bolometer pointing
nears Mars, Jupiter or Saturn, as determined by the \emph{BigPlanet}
flag (see Sect.~\ref{sec:Input-flags}). This step is done to treat
the pointing periods containing a planet crossing whose large gradients
would otherwise be \rev{confused} with glitches. 

The algorithm also requires an estimate of the sky signal in order
to assess the magnitude of excursions about the mean. It relies on
the fact that the signal component of the timeline is periodic within
a stable pointing period (up to the slight wobbling of the satellite
spin axis). We construct a phase-binned ring (PBR), a useful estimate
of the sky signal obtained by averaging the unflagged TOI samples
in bins of constant satellite rotation phase. The phase of a sample
is given by the pointing reconstruction pipeline and varies continuously
from 0 to $2\pi$ for each scan circle (about one minute). The bin
size is about the width of a single sample, i.e., 1\parcm7, and the
definition of the zero of the phase is irrelevant.

The algorithm treats each pointing period separately and one bolometer
at a time. The localization of glitches is performed with a sigma-clipping
method applied to the sky-subtracted TOI. A template fitting method
is used to identify the type of each glitch. After masking and subtracting
the series of fitted glitches from the original timeline, the PBR
is then recomputed as the average of unflagged samples. Several iterations
(generally six) are performed until the variation of $\chi^{2}$ becomes
negligible. The spike part of each glitch is flagged and the exponential
tail, below the $3.3\,\sigma$ level, is subtracted using the last
iteration of the glitch template fitting process.

\begin{figure}[!tp]
\centering{}\includegraphics[bb=80bp 80bp 700bp 550bp,clip,angle=180,width=0.95\columnwidth]{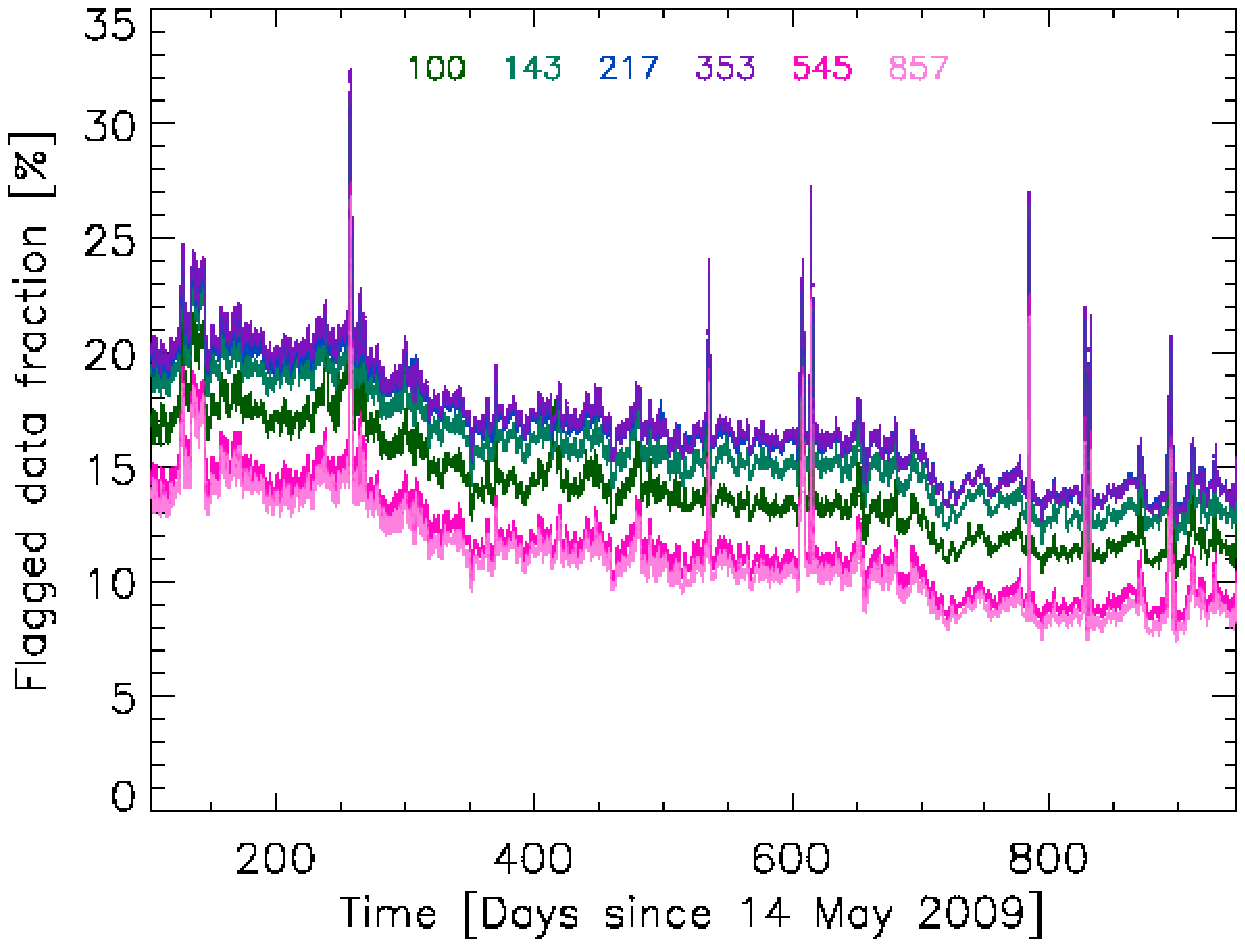}\protect\caption{Evolution of the fraction of the flagged data per bolometer averaged
over a channel for the six HFI channels. A running average over 31
pointing periods (approximately a day) is shown. \rev{The general decrease of the fraction of flagged data comes from the increase in the solar activity and the correlated decrease of the cosmic ray flux.}
\rev{The sharp peaks }are due to solar flares.\label{fig:Evolution-of-theFlagData}
}
\end{figure}

Most of the samples flagged by this process are due to cosmic ray
hits. Figure~\ref{fig:Evolution-of-theFlagData} shows the mean evolution
per channel of the fraction of flagged samples over the full mission.
More glitch statistics are given in \citet{planck2013-p03e} and \citet{planck2013-p28}.

All gaps due to flagged samples within a ring of data are replaced
with an estimate of the signal given by the PBR. For samples around
planets, we fill the timeline with the values read from a frequency
map (made by excluding planets) from a previous iteration of the data
processing at the corresponding pointing coordinates. \textcolor{red}{}%
\begin{comment}
\textcolor{red}{{[}CRL: why do we only need to fill with signal, and
not signal+noise? FXD:the change of the noise properties in valid
pixels is negligible{]}}
\end{comment}

\subsection{Bolometer nonlinearity correction\label{sec:Non-linearity-correction}}

If the environment of a bolometer changes, its response will change
accordingly. The extreme thermal stability of \Planck\ is shown by
the measured total power level, which is seen as almost constant (see
the next paragraph). Nevertheless, the timelines must be corrected
to account for the slightly varying power absorbed by the bolometer
coming from the sky load and $100\,\mathrm{mK}$ bolometer plate temperature
fluctuations. For that purpose, the voltage-to-power conversion step
is assumed to be a simple second-order polynomial applied to the timeline:
\begin{equation}
P=\frac{v}{S_{0}}\left(1+\frac{v}{v_{0}}\right)\ ,\label{eq:NLin}
\end{equation}
where $v$ is the out-of-balance \textcolor{red}{}%
\begin{comment}
\textcolor{red}{{[}CRL, AJ: define FXD: done{]}}
\end{comment}
{} demodulated voltage (i.e.\emph{, }the voltage difference between
the square-wave compensating input voltage and the total bolometer
voltage), and $S_{0}$ is the responsivity (typically of order $10^{9}\,\mathrm{V\, W}^{-1}$)
for the fiducial $v=0$ voltage. It is measured on ground-based I-V
curves, but its exact value is unimportant as it is calibrated out.
The parameter $v_{0}$ is determined for each bolometer from measurements
made during the calibration and performance verification (CPV) phase.
In general the $v_{0}$ values (5 to 200\,mV) are equal to or larger
than expected using a static bolometer model. The main reason is that
a TOI sample is an average of 40 on-board measurements (half a modulation
period), which includes an electronic transient \citep{catalano2010}.
The non-linearity correction, as measured by $v/v_{0}$ in Eq.~\ref{eq:NLin},
is static and accurate for a slowly varying signal, in particular
for dipole calibration. The term $v/v_{0}$ is at most $10^{-3}$,
the effect being largest in the $100\GHz$ band. 

During the mission, the drift of the zero level (as measured by $v/S_{0}$)
is at most roughly $7\ \mathrm{\mathrm{fW}}$ for bolometers up to
the $353\ \mathrm{GHz}$ channels, and $50\ \mathrm{fW}$ for the
$545$ and $857\ \mathrm{GHz}$ channels, such that the nonlinearity
correction implies a drift in the gain of typically $10^{-3}$ at
most, for all detectors. The correction has a relative precision better
than $10\,\%$, so that the gain uncertainty is at the $10^{-4}$
level. The effective gain and uncertainty is, however, further affected
by the offset in the ADC electronics, discussed in Sect.~\ref{sec:ADC-non-linearity}\textbf{.
}The dynamical bolometer nonlinearity on very strong sources (Jupiter
and the Galactic centre) is not corrected in the nominal TOI.

In addition, there are limits associated with the saturation of the
ADC. As discussed in \citet{planck2013-p28}, Jupiter and the Galactic
centre are the only signals on the sky strong enough to trigger this
effect.

\subsection{Thermal drift decorrelation\label{sec:Thermal-drift-decorrelation}}

The removal of the common mode due to temperature fluctuations of
the 100\,mK cooler stage is based on measured coupling coefficients
between the bolometers and the bolometer plate temperature. The coefficients
were measured during the CPV phase (typically $50\,\mathrm{pW\, K^{-1}}$
for bolometers up to the $353\,\mathrm{GHz}$ channel, and $300\ \mathrm{pW\, K^{-1}}$
for the $545$ and $857\ \mathrm{GHz}$ channels). The two dark bolometers
are used as a proxy for the bolometer plate temperature fluctuations,
as HFI 100\,mK thermometers have too many cosmic ray hits to be used.
The two dark bolometer timelines are deglitched (in the same way as
the other bolometers) then smoothed with a one minute flat kernel.
From each bolometer timeline, a linear combination of the two smoothed
dark timelines is subtracted. During this stage, the timelines of
all signal bolometers are flagged during periods where any of the
dark bolometers is flagged for at least 30 seconds. This is done in
order to suppress the impact of large glitches happening on the bolometer
plate. We found empirically that this method automatically excludes
the worst solar flares and in particular the rising common mode induced
by massive cosmic ray events. Note that the total range of the temperature
fluctuations of the bolometer plate is less than $80\,\mu\mathrm{K}$,
except for some strong solar flares. As shown in Fig.~6 of \citet{planck2011-1.5},
the temperature fluctuations of the 1.6\,K and 4\,K stages have
a negligible influence on the TOI noise properties.

\subsection{4\,K cooler line removal\label{sec:4Kline-removal}}

\revv{Electromagnetic interference (by conduction) from the drive electronics of the 4K cooler can affect the HFI data. The 4K cooler main frequency (40\, Hz) is an harmonic of the signal sample frequency; it leads to very narrow lines in the power spectral density (PSD) of the signal. In system tests before launch, the design of the grounding system of this part was found to be out of specification, but could not be corrected without unacceptable cost and delay.}
For a given stable 4\,K cooler line, only one frequency is affected in
the PSD, i.e., a narrow line is produced, only broadened by the pointing
period integration time. Nine individual frequencies are detectable,
each of which can be traced to some 4\,K cooler harmonic unfolded
by the AC modulation. In the signal domain, they are at 10, 20, 30,
40, 50, 60, 70, 80, and 17\,Hz, which are these fractions of the
modulation frequency: $1/9$,~$2/9$,~$3/9$,~$4/9$,~$5/9$,~$6/9$,~$7/9$,~$8/9$,~$5/27$.
\rev{The modulation frequency itself is filtered out} (see Sect.~\ref{sec:Filtering-and-deconvolution}).
Some thermal instability in the service module makes the line properties
change on time scales larger than ten minutes. 

We now describe the method used to correct the 4\,K cooler line contamination
in the timelines. In order to prepare the data from which to compute
the Fourier coefficients, we build an intermediate set of \rev{of TOI, denoted TOI4K, containing the original TOI except in glitch-flagged areas which are noise-filled, and for Solar System objects (SSO) where the signal is estimated from SSO-free maps.}

\emph{}%
\begin{comment}
\emph{signal-removed} TOI4K objects. We ``unroll'' the PBR by using
as a signal estimate the PBR value from the sky location observed
at each time. We then subtract this unrolled signal from the original
TOI objects. Point sources are linearly interpolated in these intermediate
TOI4K. 
\end{comment}
Then, we measure the cosine and sine Fourier coefficients on the TOI4K
for each of the nine lines, once per pointing period. We thus neglect
variations of these components within the duration of a pointing period. 

We can then subtract from the original TOI objects a timeline made
of the nine reconstructed Fourier components. This notch filter scheme
produces ripples around strong point sources \citep[see][]{planck2011-1.7}.
The origin of this problem lies in the position of the 4\,K cooler
line frequencies with respect to the harmonics of the spin frequency.
Most of the time (typically \rev{90 \%} of the pointing periods),
a given 4\,K cooler line does not overlap with any spin frequency
harmonics and thus does not affect the signal at all. The spin frequency
is very stable within a pointing period and varies from one pointing
period to another by a factor of at most $10^{-4}$. Hence, in so-called
resonant pointing periods, a 4\,K cooler line overlaps one harmonic
of the spin frequency. \rev{Operationally, overlap is declared when the 4\,K cooler line frequency is within twice the reciprocal of the pointing period duration of one of the spin frequency harmonics. In the resonant rings, } the
Fourier coefficients are perturbed by the signal and no longer represent
the systematic effect \rev{alone, hence resulting in ripples around point sources}.
For those pointing periods only, we interpolate the Fourier coefficients
from adjacent non-resonant pointing periods and subtract the systematic
timeline from the original TOI objects in the same way as the other
pointing periods. 

While tests with simulations (described in \citealt{planck2013-p28})
demonstrate that any residual 4\,K line contamination is reduced
\rev{to} less than 3\,\% of other sources of noise, these lines
also affect our ability to characterize and remove the ADC nonlinearity
(discussed in Sect.~\ref{sec:ADC-non-linearity}), and are thus still
a subject of active analysis. 

The efficiency of our removal procedure may be judged by comparing
map power spectra with this step switched on or off. It is found that
the affected multipoles are at $\ell\simeq60\,(f/1\,\mathrm{Hz)}$,
i.e., 600, 1020, 1200, 1800, \ldots. When the module is switched
on, the line residuals amount to no more than $2.5\,\%$ of the noise
level at those multipoles and much less elsewhere (see \citealt{planck2013-p28}).

\subsection{Fourier transform processing\label{sec:Filtering-and-deconvolution}}

The bolometer response suffers from time-constant effects, which must
be corrected. These effects, modelled as a complex transfer function
whose parameters are determined on planet data (see Sect.~\ref{sec:Big-planet-TOI}),
are deconvolved at the end of TOI processing. An additional filter
is applied in order to avoid a large increase in noise (without much
signal) produced by the deconvolution in the last 20\,Hz near the
modulation frequency. \rev{This filter has a perfect zero at the modulation frequency.}
We refer the reader to a dedicated accompanying paper for a complete
description \citep{planck2013-p03c}. Filtering and deconvolution
are performed during the same Fourier and inverse Fourier transform
stage. Chunks of $2^{19}$ samples are used at a time, with a standard
fast-Fourier transform (FFT) and inverse. The deconvolved and filtered
data are then computed and saved for the middle $2^{18}$ samples.
The process is continued by shifting the input samples by $2^{18}$
until the end of the timeline. This ensures continuity in the recovered
final timelines. 

For the 545 and 857\,GHz channels, the filtering is digital and consists
of a simple local smoothing kernel: 0.25, 0.5, 0.25. For these channels,
the filtering is applied first, then the deconvolution is done with
the FFT module.

\subsection{Jump corrections\label{sec:Jump-corrections}}

We now build another set of intermediate \emph{signal-removed} TOI
from the deconvolved data. These consist mostly of noise, except in
regions of strong gradients, for example near the Galactic plane.
Some pointing periods are seen to be affected by a sudden jump (either
positive or negative) in the \emph{signal-removed} TOI. We therefore
correct for that jump by subtracting a piecewise-constant template
from the timeline, while preserving the mean level. \rev{Since the exact jump location cannot be determined very precisely,}
the TOI are flagged around the recovered position with 100 samples
on each side.

We find on average 17 jumps per day, over all bolometers, with jumps
affecting a single bolometer at a time. Hence, a fraction  less than
$10^{-5}$ of data is lost in the flagging process. The jump rate
fluctuates during the mission, with a peak-to-peak variation of nine
jumps per day. So far, there is no real explanation for these events,
although we suspect violent cosmic ray hits on the warm electronics. 

We have checked the jump correction process on simulated jumps of
various intensities added to pointing periods without jumps. All jumps
above half the local standard deviation of the \emph{signal-removed}
TOI are found.

\subsection{Flagging samples\label{sec:Flagging-the-data}}

We now build a total flag out of several flags in order to qualify
the TOI for mapmaking. A sample with any of the following flags is
considered \emph{invalid data.}
\begin{enumerate}
\item The \emph{unstable pointing} flag described in Sect.~\ref{sec:Input-flags}
\citep[see also ][]{planck2013-p01}, typically accounting for three
minutes per pointing period (roughly 7\,\% of data).
\item The \emph{missing or compression error} data flag, discarding a fraction
of $10^{-9}$ of the whole mission \citep[see][]{planck2011-1.5}. 
\item The \emph{bolometer plate temperature fluctuation} flag (see Sect.~\ref{sec:Thermal-drift-decorrelation}).
One to two percent of the data are flagged this way depending on time.
\item The \emph{glitch} flag. Typically between 8 and 20\,\% of the data
are flagged depending on the time and the bolometer \citep{planck2013-p03e}.
For PSBs, both detectors are flagged even if, in a few cases, only
one of them exhibits the glitch.
\item The \emph{jump} flag. Two hundred samples per jump are flagged, a
fraction less than $10^{-5}$ (see Sect.~\ref{sec:Jump-corrections}).
\textcolor{red}{}%
\begin{comment}
\textcolor{red}{{[}Fraction of data?{]}FXD: DONE}
\end{comment}

\item The Solar System object (\emph{SSO}) flag. A zone of exclusion is
defined around Mars, Jupiter, Saturn, Uranus, Neptune, and around
the 24 asteroids detected at 857\,GHz (see the complete discussion
of asteroids in Appendix A of \citealt{planck2013-pip88}). The coordinates
of the objects are obtained from the JPL Horizons database%
\footnote{\url{http://ssd.jpl.nasa.gov/}%
} \citep{horizons1997}, which uses the actual position of the \Planck\ 
satellite in its orbit around the L2 point. As the Solar System objects
are moving in celestial coordinates, the data (although valid) must
be discarded and this masking must be done in the time domain. One
survey map thus has up to 35 holes, which are filled by information
from other surveys. Overall, the final maps have almost no holes:
less than $10^{-5}$ of pixels are missing.%
\begin{comment}
 \textcolor{red}{{[}Make sure Olivier talks about PStotalFlag. FXD:
No he does not. Too bad{]}}
\end{comment}

\end{enumerate}

\subsection{TOI Qualification\label{sec:Toi-Qualification}}

On top of flagged samples in the TOI, we show now that some flagging
also has to be done at the pointing period level (i.e.,\emph{ }considering
entire rings of data). 

We will later make maps by projecting the PBR into sky coordinates.
We therefore base our criteria for the acceptance or rejection of
a given pointing period for mapmaking by using the statistics of only
the \emph{signal-removed} TOI as an estimate of the noise. Hence,
the qualification is minimally biased by \rev{the variation of} the
signal itself. 

This noise estimate\emph{ }timeline is first analysed in the Fourier
domain. In \citet{planck2011-1.7} we described how the detector white
noise level (equivalently, the NET) is determined from noise periodograms
in the frequency region beyond the low frequency excess noise component
and before the increase introduced by the time-constant deconvolution%
\begin{comment}
, namely in the 1.0--2.5\,Hz region
\end{comment}
. \citet{planck2011-1.7} \rev{describes other properties of the noise for each detector, including the knee frequency and estimated spectral index of the $1/f^\alpha$ noise component.}

This white noise level differs from that deduced directly from the
maps.%
\begin{comment}
 FXD: looks incorrect (using the ``half-ring'' maps --- see columns
2 and 3 of Table~\ref{tab:Total-noise}) AHJ: changed from ``lower''
to ``different''
\end{comment}
{} We have therefore adopted a method for determining the ``total''
noise: this is derived directly from the standard deviation of the
\emph{signal-removed} TOI after flagging the Galactic plane and point
sources for each pointing period. The standard deviation is then corrected
upward by a term that depends on the duration of the pointing period.
This term has the form $\sqrt{fd/(fd-1)}$, where $d$ is the pointing
period duration in minutes and $f$ is the fraction of valid data
within a pointing period (typically 0.8). It accounts for the fact
that a fraction of the noise remains in the PBR, and that fraction
is smaller in longer pointing periods, where the signal is better
estimated. The final value of this total noise level for each detector
(third column of Table~\ref{tab:Total-noise}) is then taken to be
the peak of the distribution of the root-mean-square (rms) noise of
the valid pointing periods. The typical bolometer NETs, measured on
the deconvolved TOI between 0.6 and 2.5\,Hz, are also given in Table~\ref{tab:Total-noise}
(second column). The total noise can be viewed as proportional to
the square root of the integral of the \rev{square of the} noise-equivalent
temperature across the total frequency bandpass (including filtering
effects) from 0 to 91\,Hz. It is normalized to one second of integration
time, so that pure white noise would have numerically identical NET
and total noise. Deconvolution and filtering effects can produce a
NET larger or smaller than the total noise.

\begin{table}[htb]
\protect\caption{Noise characteristics.  \label{tab:Total-noise}}
\centering{}\begingroup
\newdimen\tblskip \tblskip=5pt
\nointerlineskip
\vskip -3mm
\footnotesize
\setbox\tablebox=\vbox{
   \newdimen\digitwidth 
   \setbox0=\hbox{\rm 0} 
   \digitwidth=\wd0 
   \catcode`*=\active 
   \def*{\kern\digitwidth}
   \newdimen\signwidth 
   \setbox0=\hbox{+} 
   \signwidth=\wd0 
   \catcode`!=\active 
   \def!{\kern\signwidth}
\halign{\hfil#\hfil\tabskip=2em&
\hfil#\hfil&
\hfil#\hfil&
\hfil#\hfil&
\hfil#\hfil\/\tabskip=0pt\cr             % Template goes here.
\noalign{\doubleline}
Band& NET& Total Noise& Goal& Units\cr   % Table headings go here.
\noalign{\vskip 3pt\hrule\vskip 5pt}
                                    % Body of table goes here.
100P& *71& 132& 100& $\mu\mathrm{K_{CMB}\, s}^{1/2}$\cr
143P& *58& *65& *82& $\mu\mathrm{K_{CMB}\, s}^{1/2}$\cr
143S& *45& *49& *62& $\mu\mathrm{K_{CMB}\, s}^{1/2}$\cr
217P& *88& 101& 132& $\mu\mathrm{K_{CMB}\, s}^{1/2}$\cr
217S& *74& *66& *91& $\mu\mathrm{K_{CMB}\, s}^{1/2}$\cr
353P& 353& 397& 404& $\mu\mathrm{K_{CMB}\, s}^{1/2}$\cr
353S& 234& 205& 277& $\mu\mathrm{K_{CMB}\, s}^{1/2}$\cr
\noalign{\vskip 3pt\hrule\vskip 3pt}
545S& 0.087& 0.052& 0.116& $\mathrm{MJy\, sr^{-1}\, s^{1/2}}$\cr
857S& 0.085& 0.056& 0.204& $\mathrm{MJy\, sr^{-1}\, s^{1/2}}$\cr
\noalign{\vskip 5pt\hrule\vskip 3pt}}}
\endPlancktable                    % ends one-column \halign
%\endPlancktablewide                 % ends two-column \halign
%\tablenote a Footnote a.\par
%\tablenote b Footnote b.\par
The quoted value is the noise that is obtained
on the map after one second of integration, i.e., about 180 hits for
one bolometer. This gives the noise characteristics of bolometers,
inverse quadratically averaged over the similar ones within a channel
(``P'' is for polarization sensitive bolometers and ``S'' for
spider-web unpolarized bolometers). The second column gives the white
noise level measured on the power spectrum in the range of 0.6 to
$2.5\,\mathrm{Hz}$. The third column gives the total noise, i.e.,
the rms noise at the map level for one second of integration time
of a single detector. The next column recalls the goal stated before
launch by \citet{lamarre2010}. Note that the mean integration time
per detector per $(1.7\arcm)^{2}$ pixel is 0.56\,s for the 100--353\,GHz
channels and 0.63\,s for 545 and 857\,GHz, for the nominal mission
(see Fig.~\ref{fig:Imaps-IntegTime}).
\endgroup
\end{table}

\begin{figure*}[t]
\begin{centering}
\includegraphics[width=1\columnwidth]{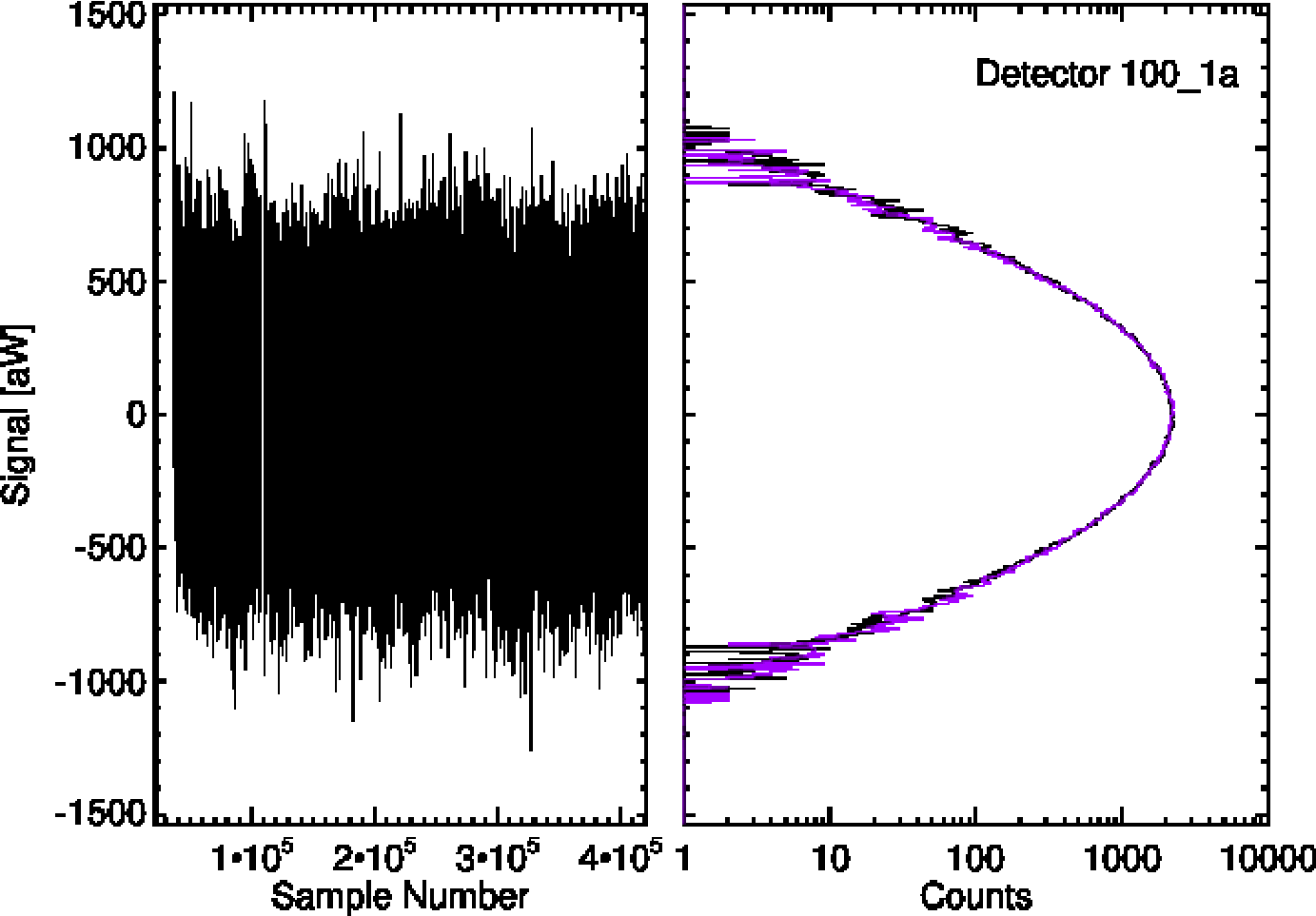}\includegraphics[width=1\columnwidth]{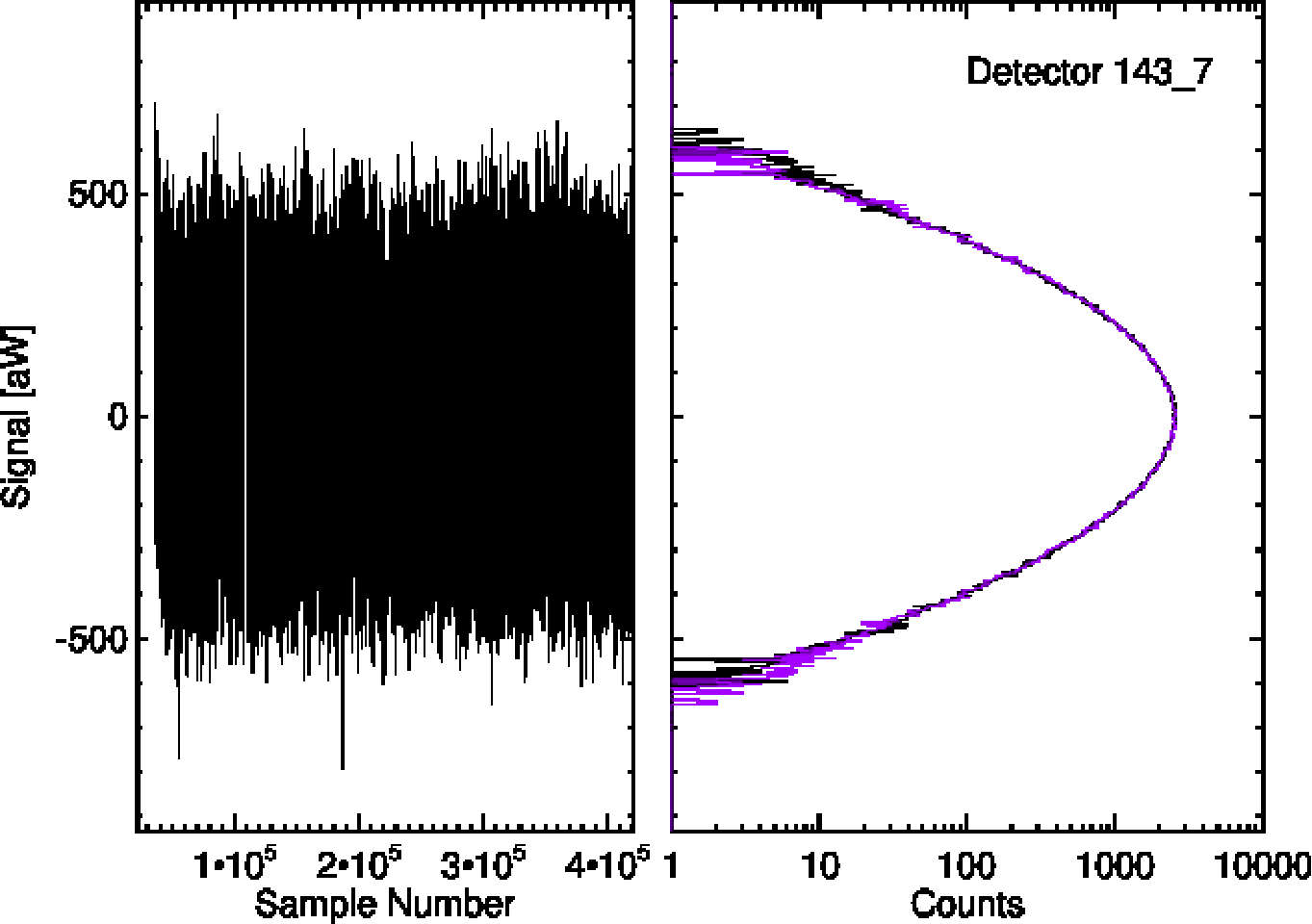}
\par\end{centering}

\centering{}\bigskip{}
\includegraphics[width=1\columnwidth]{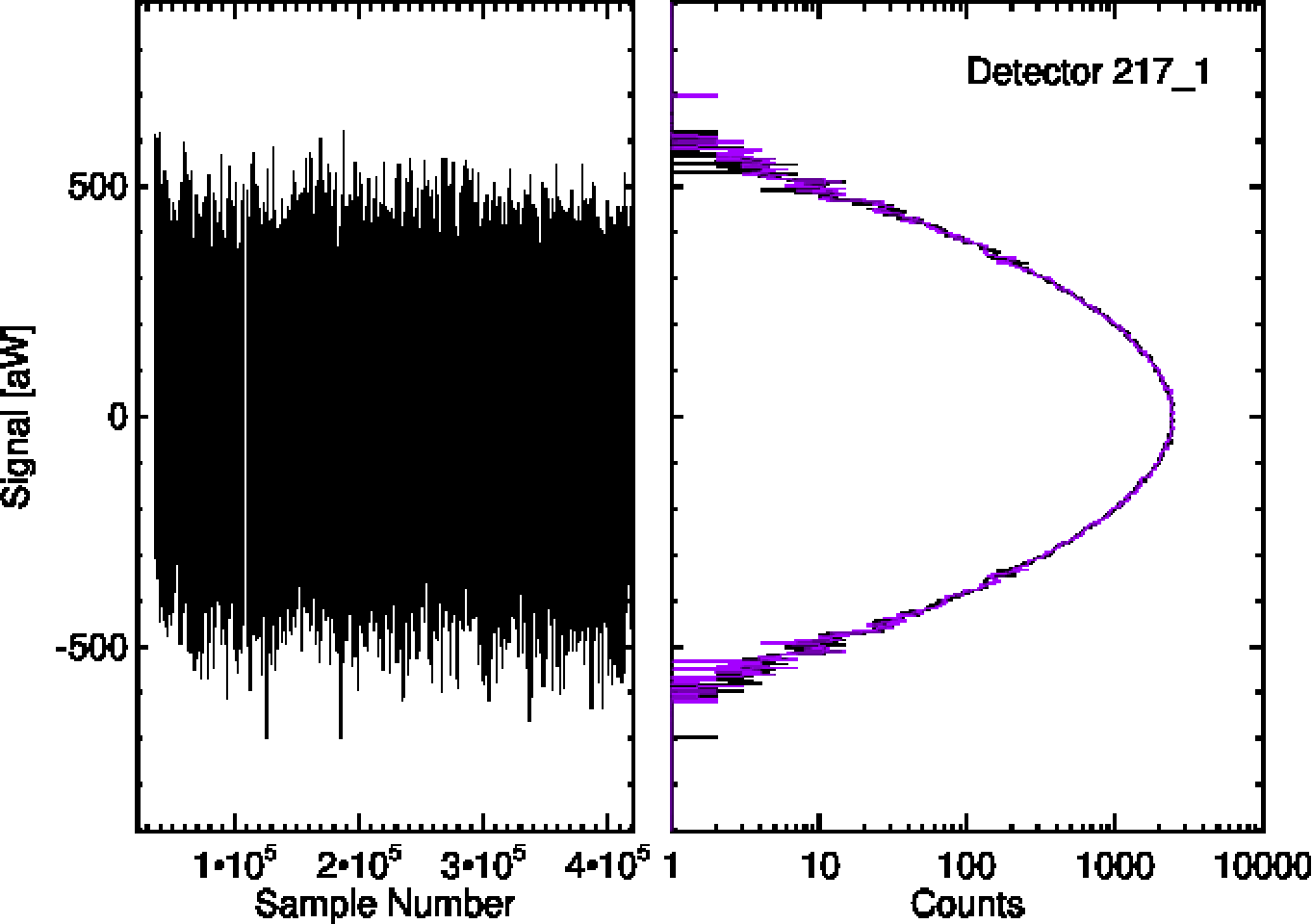}\includegraphics[width=1\columnwidth]{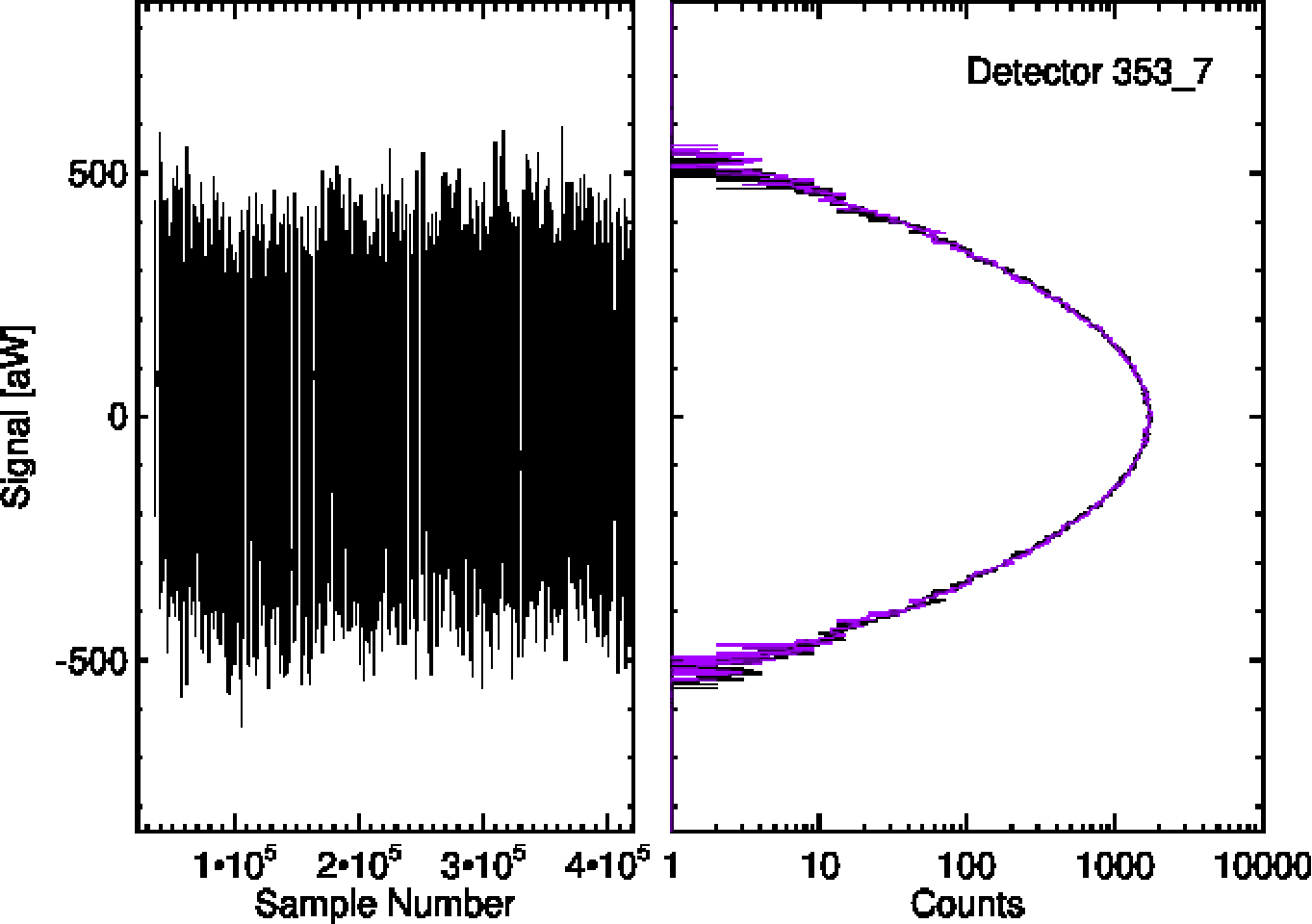}\protect\caption{The \emph{signal-removed} TOI sample values are displayed for one
pointing period and four detectors (left panel in each of the four
sets). Flagged samples are not shown. Samples falling near the Galactic
plane or point sources are not shown (corresponding to one-minute
periodic blanks in the plots). The right panel of each graph shows
the histogram (in black) of the samples. The coloured curve is \rev{the vertically reflected version of the histogram}
around the mean. \label{fig:rmsigAndHisto}  }
\end{figure*}

The noise estimate is also analysed in the time domain. Figure~\ref{fig:rmsigAndHisto}
shows one pointing period of the \emph{signal-removed} TOI for four
detectors, along with the histogram of the samples.

\begin{figure*}[t]
\begin{centering}
\includegraphics[bb=0bp 0bp 828bp 453bp,clip,width=1\textwidth]{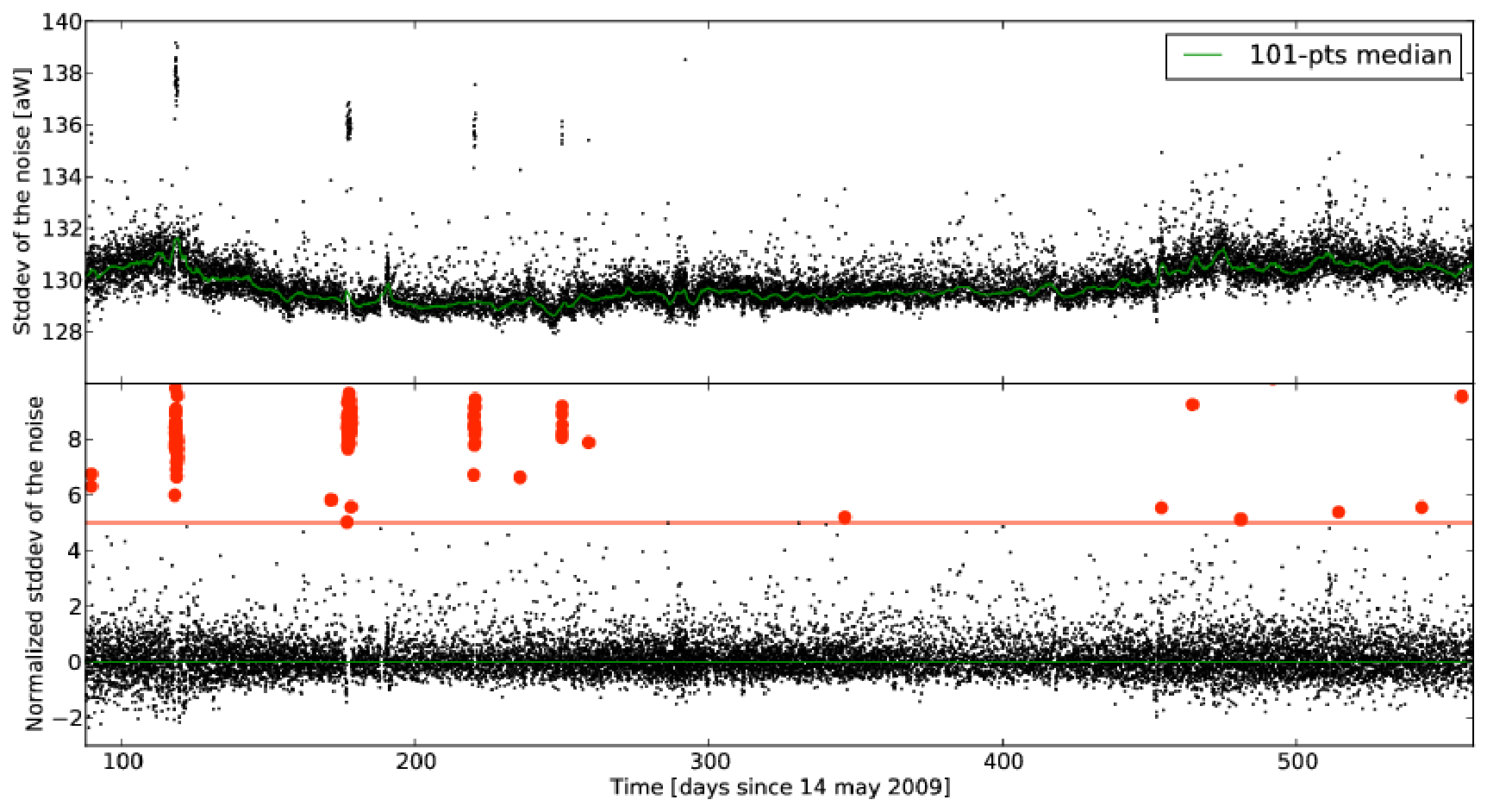}
\par\end{centering}

\protect\caption{\emph{Top}: Total noise estimated for each stable pointing period
for a 143\,GHz PSB. A smoothed version is shown in green. \emph{Bottom}:
\rev{A version of the upper plot shifted to zero mean and normalized to unit variance shows how outliers, the red dots,}
are picked up (above 5~$\sigma$ of the moving median in green).
\label{fig:Total-noise} }
\end{figure*}

Figure~\ref{fig:Total-noise} shows an example of the evolution of
this total noise (uncalibrated at this stage) during the first four
surveys (hence beyond the nominal mission). The bias linked to the
pointing period duration has been corrected for. Notice that the total
noise is mostly constant, with peak-to-peak variations at the percent
level, suspected to be induced by ADC non-linearities (see Sect.~\ref{sec:Abs-calib}). 

A pointing period that deviates significantly from the others is flagged
in the qualification process and not used in further processing. Three
criteria were used to quantify the deviation: 
\begin{itemize}
\item the absolute value of the difference between the mean and the median
value of the \emph{signal-removed} TOI; 
\item the total noise value; and 
\item the Kolmogorov-Smirnov test deviation value --- when a given pointing
period is stamped as anomalous for more than half the bolometers,
it is discarded for all of them. 
\end{itemize}
The common anomalous pointing periods are linked to some spacecraft
events or strong solar flares, while the individual anomalous pointing
periods correspond to incidents in the overall level of the timeline
(like strong drifts). 

There are two detectors (one each at 143 and 545\,GHz, not counting
toward the total of 50 HFI detectors) that are not used at all because
they show a permanent non-Gaussian noise structure, which we call
random telegraphic signal (RTS). The RTS bolometer \emph{signal-removed}
TOI show an accumulation of values at two to five discrete levels.
The measured \emph{signal-removed} TOI jump at random intervals of
several seconds between the different levels. The jumps are uncorrelated.
A third detector (at 857\,GHz) shows the RTS phenomenon only for
some well-delimited periods \rev{amounting to 7.4\,\% of the nominal mission. Other detectors show a significant but smaller RTS in episodes of short duration (2.1\,\% and 1.2\,\% of the data from two} 217\GHz\ bolometers). A
dedicated module searches for these episodes, which are then added
to the list of discarded pointing periods. Ten pointing periods of
exceptionally long duration are discarded for all bolometers, because
noise stationarity is not satisfied in those cases. The discarded
pointing periods represent less than 0.8\,\% of the total integration
time for the nominal mission.

\begin{comment}
sur mission totale, - 25 rings > 90' - 0.09\% en nombre de ring -
1.4\% du temps total

sur mission nominale - 11 rings > 90' - 0.075\% en nombre de ring
- 0.6\% du temps total

--> source dominante de discarded ...

Other non-stationarities have been observed: In some detectors, the
total noise level is observed to change by typically 2~\%. Some times,
there are only two levels of noise attained by the detector for each
ring. As the power on the detector does not change in a correlated
way, we think that an excess noise may come from the electronic part
of the readout (either in the JFET preamplifier or above).
\end{comment}
Many examples of the characterization and qualification of the TOI
processing can be found in \citet{planck2013-p28}, where simulations
are used to limit the possible effects of undetected RTS periods to
a fraction of the overall noise level. In addition, \citet{planck2013-p28}
presents a number of simulations and other tests which have been done
to set limits on possible contamination from other possible systematic
effects.

\subsection{Big planet TOI\label{sec:Big-planet-TOI}}

As the TOI of the standard pipeline are interpolated around Mars,
Jupiter, and Saturn, they cannot be used for the beam reconstruction
(see Sect.~\ref{sec:DETECTOR-BEAMS}). Therefore a dedicated pipeline
is run on appropriate pointing period ranges. The difference from
the standard pipeline is mainly a special deglitching step performed
on the planet samples to cope with such a strong signal and its variations
during a given pointing period. Also, the jump correction is not applied
to avoid being triggered by the planet transit.

\section{Detector pointing\label{sec:DETECTOR-POINTINGS}}

\begin{figure}[th]
\centering{}\includegraphics[clip,width=0.95\columnwidth]{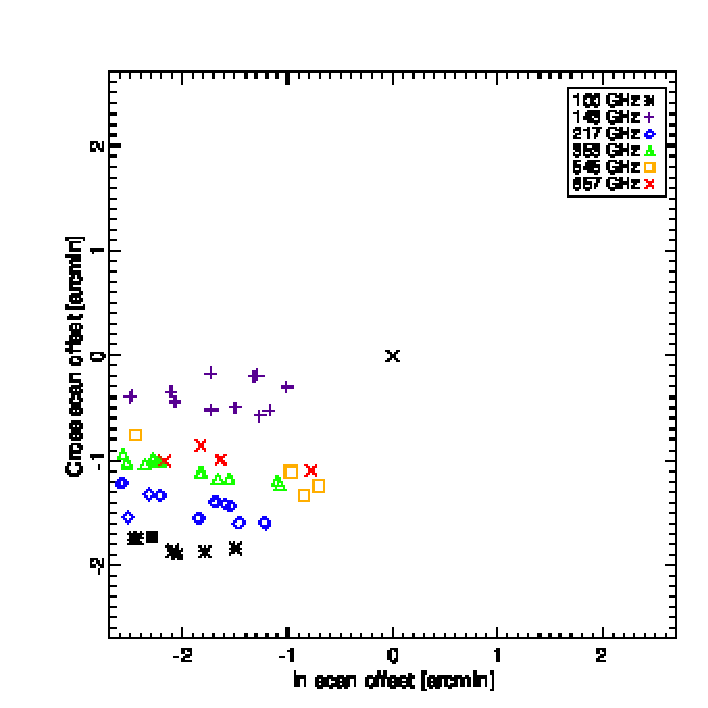}\protect\caption{Positions of the individual HFI detectors with respect to a pre-launch
model, as measured on the first pass of Mars. The horizontal axis
is the direction of the scan (``in-scan''); vertical is perpendicular
to this (``cross-scan''). \label{fig:Mars1Pointing} }
\end{figure}

Here we summarize the pointing solution we use, determined with the
overall goal of limiting detector pointing errors to less than 0\parcm5
(rms). The satellite pointing comes from the star tracker camera subsystem,
which gives the location of a fiducial boresight direction as a function
of time, sampled at\textcolor{red}{{} }8\,Hz. In practice, science
data analysis requires the pointing of each detector, sampled at the
same rate as the detector signal. This, in turn, requires a number
of separate steps:
\begin{enumerate}
\item resampling to the acquisition frequency (180.4\,Hz);
\item rotation from the fiducial boresight to the detector line of sight;
and
\item correction for detector aberration\\ \citep{planck2013-pipaberration}.

\end{enumerate}
The second step must be calibrated in situ: the detailed geometry
must be measured by comparison with the positions of known objects
(planets and other bright sources whose extent is small compared to
the width of the beam). The primary source of this geometrical calibration
is the planet Mars, which is bright (but not so bright as to drive
the detectors into a nonlinear response) and nearly pointlike (with
a mean disk radius of 4\parcs1 during the nominal mission). Figure~\ref{fig:Mars1Pointing}
shows the pointing of each detector relative to the pre-launch optical
model. 

Note that the in-scan pointing is degenerate with any phase shift
induced by the combined detector time response and optical beam and
any attempt to remove them by deconvolution \citep{planck2013-p03c}.
For that reason, this calibration of the pointing cannot be considered
as a measurement of the focal plane geometry per se. 

Because the transfer of the pointing from the star tracker to the
detectors depends on the satellite rotation axis and hence the moment
of inertia tensor, we do not expect the detector pointing relative
to the star tracker to be constant in time. We therefore reconstruct
the pointing solution relative to the star tracker as a function of
time, and in the following discuss the accuracy of the reconstruction
in different frequency regimes.

High-frequency fluctuations in the spacecraft and focal plane orientation
are inhibited by inertia, but are still present in the raw star tracker
pointing. Having no useful signal content, this regime is treated
by applying a low pass filter to the reconstructed pointing. Similarly,
high-frequency errors can be assessed by studying the noise-dominated
high frequency pointing power spectrum.

\begin{figure*}[th]
\centering{}\includegraphics[width=0.45\textwidth]{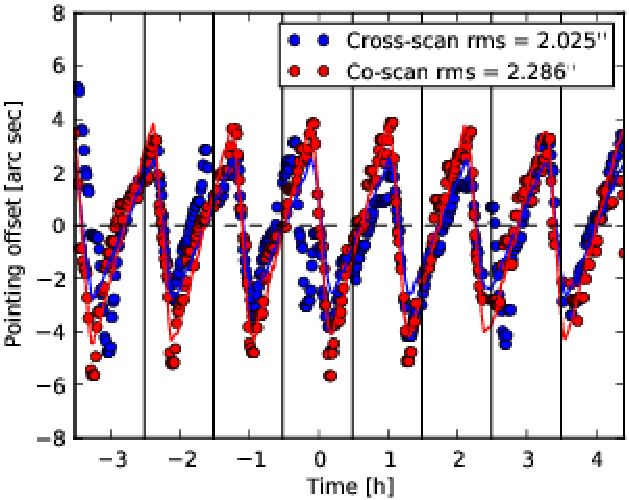}\qquad{}\includegraphics[width=0.45\textwidth]{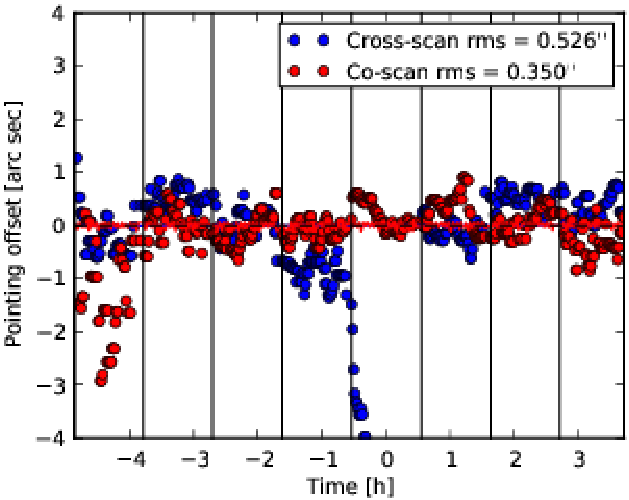}\protect\caption{Measured intermediate frequency pointing errors at one minute intervals.
\rev{The plotted time span covers the period that Jupiter is within one FWHM of the 143-1a bolometer beam center.}
The offsets were smoothed with a five-point median filter. Vertical
lines mark the pointing period boundaries. The overplotted solid line
is the scaled and translated REBA temperature. REBA thermal control
was adjusted between Jupiter transits two and three. \emph{Left}:
Pointing fluctuations during the second Jupiter transit. \emph{Right}:
Fluctuations during the third Jupiter transit, after the REBA thermal
control was adjusted. Note the different vertical scales.\label{fig:pointing-intermediateErr}\textcolor{red}{{} }}
\end{figure*}

Intermediate-frequency pointing errors, on one minute time scales,
are characterized with Jupiter transits. We make use of the optimal
combination of high signal-to-noise ratio and relatively wide beam
at the \Planck\ $143\GHz$ channel and first perform a global fit
of the planet position in the transit data. We then consider a small
pointing offset to each 60\,s scanning circle. This successfully
recovers an expected interference signal from the radiometer electronics
box assembly (REBA; see \citealt{planck2011-1.3} and \citealt{planck2013-p01}).
Thermal control on the REBA was adjusted on day 540 after launch and
subsequently Jupiter transit three does not exhibit the interference.
We find that the intermediate-frequency pointing error is less than
3\arcs\  (rms) before the REBA adjustment and less than 1\arcs\ beyond
day 540 (see Fig.~\ref{fig:pointing-intermediateErr}).

\begin{figure}[th]
\centering{}\includegraphics[clip,width=1\columnwidth]{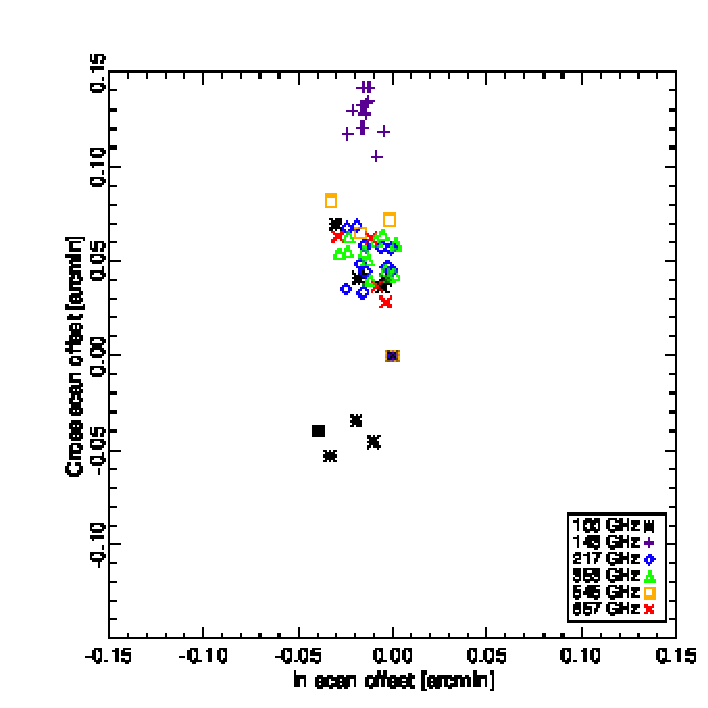}\protect\caption{Pointing differences between the first and second crossings of Mars,
as in Figure~\ref{fig:Mars1Pointing} (but note the different axis
scales).\label{fig:Mars2vsMars1FPG}}
\end{figure}

The pointing is easiest to monitor over long time scales, using observations
of bright planets. In Fig.~\ref{fig:Mars2vsMars1FPG} we show the
difference between the first two observations of Mars; differences
along the scan direction have a mean of $0\parcs85$ and a standard
deviation of $0\parcs59$, while cross-scan differences in the direction
perpendicular to this are $3\parcs7\pm2\parcs8$ and are systematic
with frequency. Indeed, analysis of further planet-crossings and of
aggregate high-frequency point-source crossings shows arcminute-scale
evolution over the course of the mission. Fitting for an overall offset
for each planetary transit allows us to measure long time scale (several
days) pointing fluctuations. We complement this sparse sampling of
the overall offset by time domain position fits of the brightest compact
radio sources as well as monitoring of many low-flux-density high-frequency
sources in aggregate. In the process we learned that each radio source
position was biased by a few arcseconds from its catalogued position.
We assume the bias results from convolving the source emission with
our effective beams and correct for it by assuming that the low frequency
pointing correction derived from planet transits alone is already
accurate enough for debiasing the positions. We solve for the apparent
position of each compact source in the \Planck\  data by estimating
a single constant offset from its catalogued position.

The planet positions alone indicate a change in the average level
of the low frequency fluctuations between Jupiter transit two and
Neptune transit three (days 418 and 540 from launch). The radio sources,
particularly QSO J0403$-$3605, further narrow the transition before
day 456. It is likely that the offset is due to thermoelastic deformation
caused by the sorption cooler system switchover on day 460 \citep{planck2011-1.3,planck2013-p01}.
If entirely untreated, the low frequency pointing error dominates
the error budget with around $15\arcs$ (rms) error. Two independent
focal plane solutions bring the low-frequency rms error well below
$10\arcs$ and applying a time-dependent correction reduces the low
frequency error to a few arcseconds. 

\begin{figure*}[th]
\centering{}\includegraphics[width=0.5\textwidth]{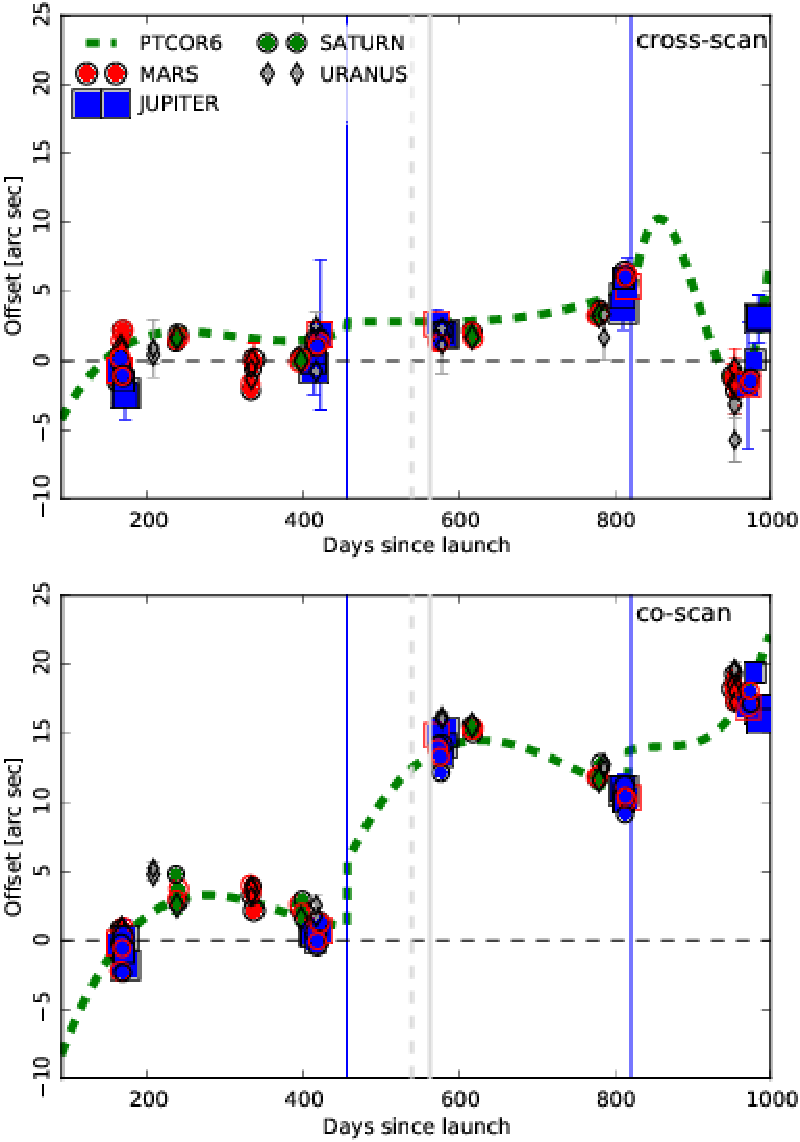}\includegraphics[width=0.5\textwidth]{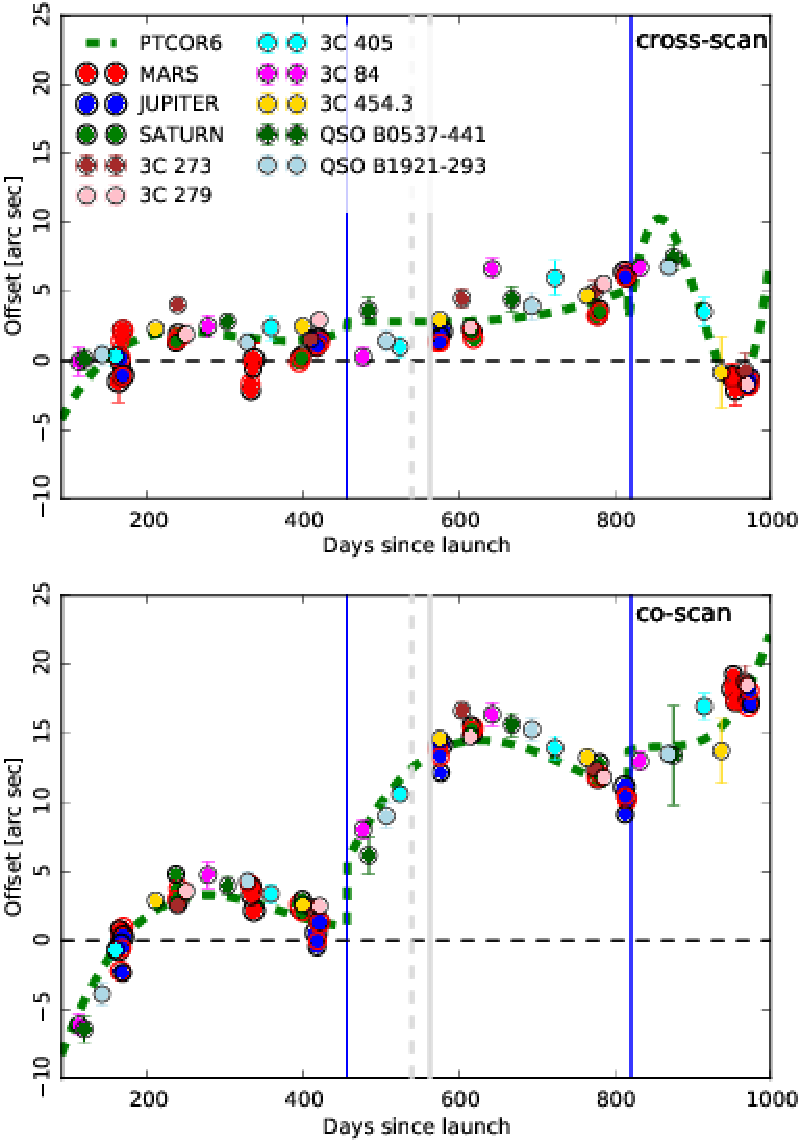}\protect\caption{Compilation of planet and bright point-source positions versus time
since the start of the mission. We mark the REBA thermal control adjustment
on day 540 with a dashed grey line and the end of the nominal mission
(day 563) with a solid grey line. Note the two discontinuities added
to the pointing correction model at days 455 and 818. The dashed green
line denotes the spline fit to the planet positions used to correct
the long-term pointing variations. \emph{Left}: Compilation of measured
planet position offsets across the \Planck\  frequencies. Usable
planets change by frequency: the blue spectral energy distribution
of the planets renders all but Jupiter too dim for required positioning
in the LFI frequencies and Jupiter and Saturn positions above $217\GHz$
are compromised due to the nonlinear bolometer response. \emph{Right}:
Debiased bright source position offsets in \Planck\ 100--$217\GHz$
data compared to a trend line fitted to the planet position offsets.
Each source was fitted for an apparent position rather than using
the catalogued position to measure pointing offset. Two successive
observations of the same source have opposite scanning directions,
so having the wrong apparent position produces opposite pointing errors
that we have corrected. \label{fig:pointing-solution}}
\end{figure*}

\begin{figure*}[th]
\centering{}\includegraphics[clip,width=1\textwidth]{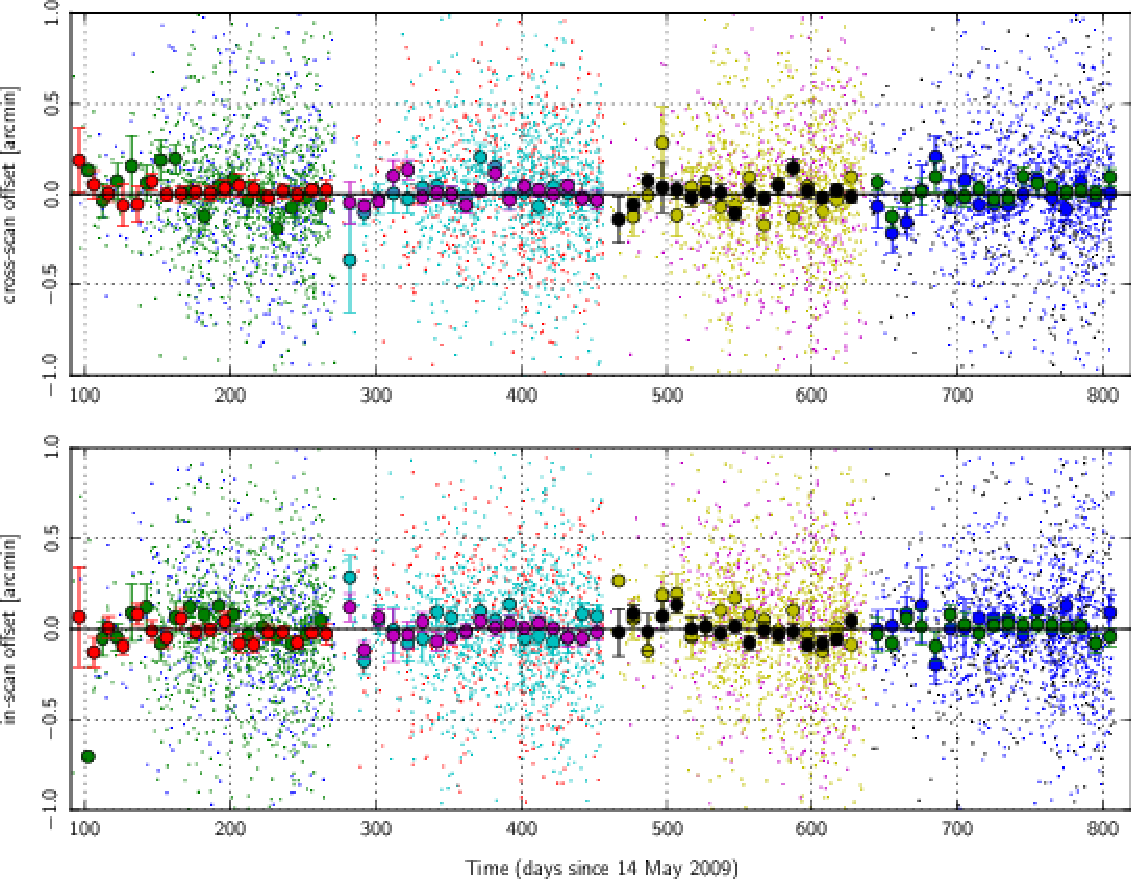}\protect\caption{Pointing evolution from aggregate high-frequency point sources, measured
by comparing point sources seen at 545 and 857\,GHz to known \emph{IRAS}
positions. Light-coloured points show individual source deviations,
points with error bars give ten-day errors. Different colours correspond
to the two frequencies and four individual \Planck\ sky surveys.\foreignlanguage{english}{\label{fig:srcPointingEvolution}
}}
\end{figure*}

Figure~\ref{fig:pointing-solution} shows the uncorrected planet
positions with the debiased bright point-source positions along with
the resulting \Planck\  pointing solution, which smoothly interpolates
with a spline fit between planet observations. Figure~\ref{fig:Mars2vsMars1FPG}
shows the offset of Mars for all detectors between the first and second
surveys after correction by the pointing solution and Fig.~\ref{fig:srcPointingEvolution}
the effect of the corrected pointing on the high-frequency point sources
at 545 and 857\,GHz. 

After these corrections, the pointing error is about 2\parcs5 (rms)
over the mission, less than $1\arcs$ during planet observations,
and increasing between planet crossings. The individual HFI detector
locations (along with their respective beam patterns) are shown in
Fig.~\ref{fig:scanningbeam_focalplane}.


\section{Detector beams\label{sec:DETECTOR-BEAMS}}

Here we provide a brief overview of the measurement of the \Planck\ HFI
beams and the resulting harmonic-space window functions that describe
the net optical and electronic response to the sky signal\rev{, as well as the impact of processing}
(for a full discussion see \textcolor{red}{\citealp{planck2013-p03c}}).

At any given time, the response to a point source is given by the
combination of the optical response of the \Planck\ telescope (the
\emph{optical beam}) and the electronic transfer function. The latter
is partially removed as discussed in Sect.~\ref{sec:TOI-PROCESSING}
and \citet{planck2013-p03c}, but any residual is taken into account
as part of the beam. This response pattern is referred to as the \emph{scanning
beam\rev{, which includes any further effect due to the temporal data processing}}.
However, the mapmaking procedure discussed in Sect.~\ref{sec:MAP-MAKING}
below \rev{implies} that any map pixel is the sum of many different
\rev{samples in} the timeline, \rev{each of which contributes in a different location within the pixel and with a}
different scan direction. Thus, the \emph{effective beam} \citep{mitra2010}
takes into account the details of the scan pattern (and even this
is a somewhat simplified picture, as long-term noise correlations
accounted for by \rev{destriping in the mapmaking procedure described in Sect.~}\ref{sec:MAP-MAKING}
mean that even samples taken when the telescope was looking elsewhere
contribute to a given pixel). Finally, the multiplicative effect on
the angular power spectrum is encoded in the \emph{effective beam
window function} \citep{hivon2002}, which includes the appropriate
weights for analysing aggregate maps across detector sets or frequencies.

\subsection{Scanning beams\label{sub:Scanning-beams}}

For the single-mode HFI channels, the scanning beam is well-described
by a two-dimensional elliptical Gaussian with small perturbations
at a level of a few percent of the peak \citep{huffenberger2010}.
We measure the beams on observations of Mars, and model them with
B-spline fits to the timeline supplemented with a smoothing criterion
as discussed in \citet{planck2013-p03c}.

\begin{figure*}[th]
\centering{}\includegraphics[width=1\textwidth]{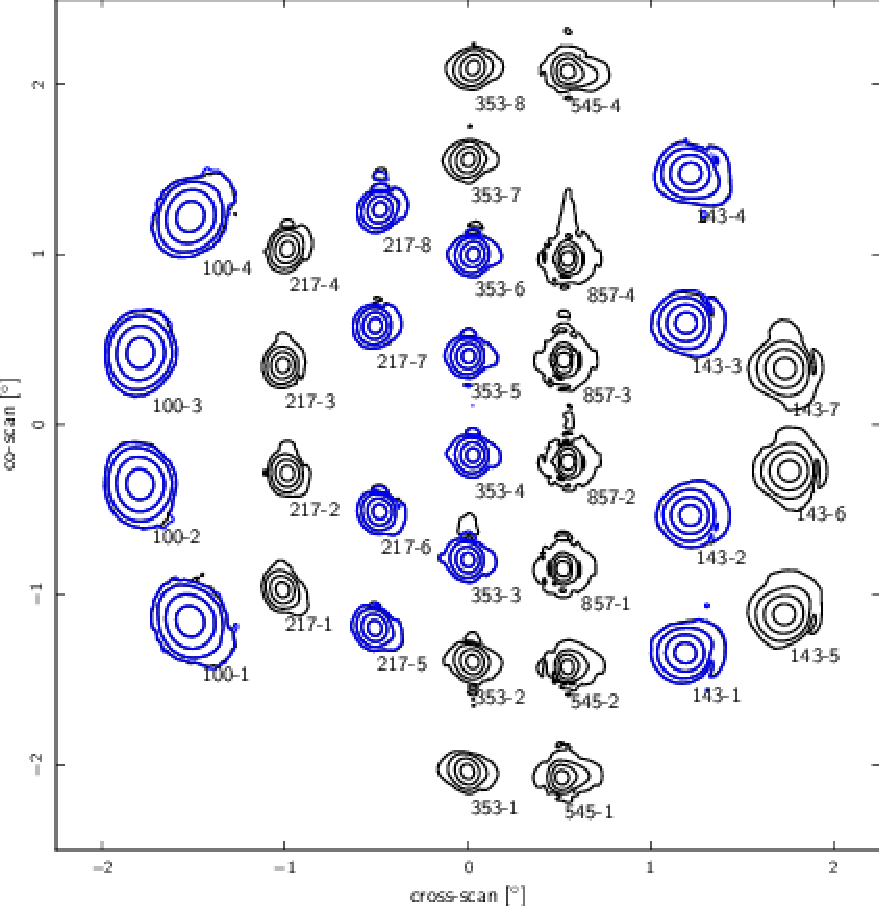}\protect\caption{The scanning beam patterns reconstructed from planet observations
at their respective positions in the HFI focal plane. The contours
are plotted at $-3$, $-10$, $-20$, and $-30\,$dB from the peak
of each beam. Polarization-sensitive bolometers are indicated as a
pair of contours, one black and one blue (in most cases the underlying
black contour is invisible, indicating essentially identical beam
shapes). \label{fig:scanningbeam_focalplane} }
\end{figure*}

We measure the beam on a square patch $40\arcm$ on a side. This main
beam pattern must be augmented by several effects that are visible
within roughly 1\deg\ of the beam centre due to variations of the
mirror surface. At all frequencies we see evidence of un-modelled
effectively random dimpling, well-fit by a sum of \citet{ruze1966}
models. At 353\,GHz and higher we see evidence of the hexagonal backing
structure of the mirrors, included directly in the B-spline fits.
The scanning beam patterns for each detector are shown in Fig.~\ref{fig:scanningbeam_focalplane}.

Other steps in the data processing pipeline also affect the scanning
beam pattern. We see residual effects from the deconvolution as a
shoulder in the beam localized to the ``trailing'' side of the scan.
The residual pointing uncertainty results in a slight broadening of
the beam which is modelled in the Monte Carlo simulations we use to
estimate the beam errors. Further simulations indicate that undetected
or inaccurately removed glitches (Sect.~\ref{sec:Deglitching}) result
in a negligible additional bias, as do the uncorrected ADC non-linearities.
Because the response of the detector depends on the spectral shape
of the source, there is a colour correction (see Sect.\ \ref{sec:Correct-color})
when converting from the approximately known planet spectrum (roughly
$\nu^2$) to the nominal $\nu^{-1}$ shape. 

All of these give a change in the beam solid angle of roughly 0.5\,\%
at 100\,GHz and less than 0.2\,\% for 143\,GHz and above. However,
there is an additional contribution to the scanning beam from near
sidelobe response beyond the square patch on which the beam is measured
(e.g., from the deconvolution of the time response). Simulations show
that the near sidelobes contribute less than 0.2\,\% of the beam
power. \rev{\cite{planck2013-p01a} shows that accounting for this contribution would further improve the good agreement between HFI and LFI at adjacent frequencies.} 

\subsection{Effective beams\label{sub:Effective-beams}}

The scanning beams are used to calculate the effective beam response
at a given pixel. A symmetric beam in a uniformly-sampled pixel (and
ignoring effects induced by long-time-scale noise correlations accounted
for in the mapmaking procedure) would give an effective beam that
is the convolution of the scanning beam pattern with a top-hat pixel
window function. Departures from these idealities, due to scanning
beam asymmetry and the scan strategy, mean that the effective beam
will be asymmetric and will depend upon sky location: ignoring long-term
noise correlations, the effective beam is given by averaging the scanning
beam pattern at the observed locations and orientations of the real
scan pattern in each pixel. In practice, this must be approximated
by using a coarse pixelization of the scanning beam or restricting
the calculation to those components with low spherical-harmonic multipoles
$m$. We have developed a set of methods in real and harmonic space
to account for these effects, described more completely in \citet{planck2013-p03c}
and references therein.

\begin{figure}[th]
\begin{centering}
\includegraphics[clip,width=1\columnwidth]{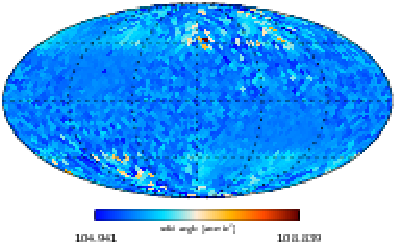}
\par\end{centering}

\centering{}\includegraphics[clip,width=1\columnwidth]{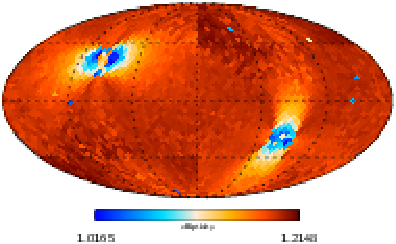}\protect\caption{Effective beam solid angle (upper panel) and the best-fit Gaussian
ellipticity (lower panel) of the 100\,GHz effective beam across the
sky in Galactic coordinates. \label{fig:eff_beam}}
\end{figure}

\begin{figure}[th]
\begin{centering}
\includegraphics[bb=0bp 0bp 233bp 264bp,clip,width=0.95\columnwidth]{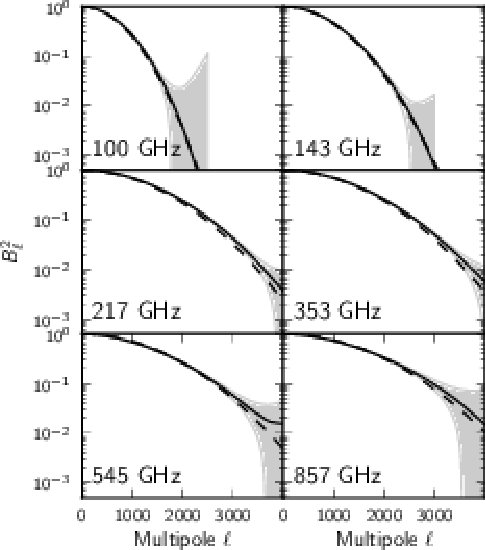}
\par\end{centering}

\protect\caption{Effective beam window functions (solid lines) for each HFI frequency.
The shaded region shows the $\pm 1 \sigma$ error envelope. Dashed
lines show the effective beam window function for Gaussian beams with \textsc{FWHM} 9\parcm65, 7\parcm25, 4\parcm99, 4\parcm82, 4\parcm68, and 4\parcm32
for 100, 143, 217, 353, 545, and 857\,GHz, respectively. We compute
the 100\,GHz window function to $\ell=2500$, 143\,GHz to $\ell=3000$,
and the rest to $\ell=4000$. \label{fig:eff_beam_window}}
\end{figure}

We propagate the scanning beam pattern from each detector, along with
the largest-variance eigenmodes of the Monte Carlo error covariance
matrix, using the \Planck\ scanning strategy. The error eigenmodes
are scaled by a factor of 2.7 to account for the unmodelled near sidelobe
bias in the scanning beam. The eventual calculation of power spectra
requires the cross-correlation of various detector pairs and more
generally arbitrary weighted combinations of detector pairs used to
construct the final CMB power spectrum. Our algorithm results in both
mean beam patterns (Fig.~\ref{fig:eff_beam}) and sky-averaged window
functions, along with error eigenmodes on the latter (Fig.\ \ref{fig:eff_beam_window}),
for the required detector combinations. The far sidelobe response
of the instrument (response to signal more than $5\deg$ from the
beam centroid, discussed in Sect.~\ref{sec:Correct-FSL=000026Zodi})
negligibly biases the calibration and the effective beam window function
\citep{planck2013-p03c,planck2013-pip88}. The full window functions
are available with the \Planck\ data release, and the key numerical
results are shown in Table~\ref{tab:summary}. 

\section{Mapmaking and photometric calibration\label{sec:MAP-MAKING} }

This section gives an overview of the mapmaking and photometric calibration
procedure. More details are described in \citet{planck2013-p03f}.

\subsection{Map projection \& calibration techniques }

The HFI mapmaking scheme is very similar to that described in \citet{planck2011-1.7}.
The inputs for each detector come from the TOI processing (see Sect.~\ref{sec:TOI-PROCESSING})
with the associated \emph{invalid data} flags. Mapmaking is done in
three steps. We first take advantage of the redundancies during a
stable pointing period (ring of data) by averaging each detector's
measurements in \texttt{HEALPix}%
\footnote{\url{http://healpix.sourceforge.net}%
} \citep{gorski2005} pixels, building a new data structure, the \texttt{HEALPix}
pixel ring (HPR). We prefer to use this structure for mapmaking, rather
than the phased binned rings previously described, because the latter
introduce an additional smoothing in the reconstructed map power spectra.
We set the \texttt{HEALPix} resolution for the HPR at the same level
as for the final maps, $N_{\mathrm{side}}=2048$, for the same reason.
In the second step we use these HPRs to perform the photometric calibration
of each detector's data. Once this is done, the mapmaking per se is
performed. As for the previous release, the mapmaking procedure algorithm
we used is an implementation of a destriping algorithm, which is made
possible by the characteristics of the HFI detector noise.

In the destriping framework, detector noise is modelled as a set of
constants, called offsets, representing the low-frequency drift of
the signal baseline over given time intervals, and a white noise component
uncorrelated with these constants. The general mapmaking scheme may
thus be written in this approximation as a solution to the equation

\begin{equation}
\vec{{d}}=G\times\tens{A}\cdot\vec{{T}}+\tens{\Gamma}\cdot\vec{{o}}+\vec{{n}},\label{eq:MM2}
\end{equation}
where $\vec{d}$ \rev{is the vector of TOI data}, $\vec{T}$ represents
the pixelized sky {[}which may be a vector $(I,Q,U)$ at each pixel
if polarization is accounted for{]}, $G$ is the detector gain, $\tens{A}$
is the pointing matrix indexed by sample number and pixel, equal to
zero when the pixel is unobserved at that sample, one when the pixel
is observed by an unpolarized detector, or the vector $(1,\cos2\alpha,\sin2\alpha)$
for a polarized detector at an angle $\alpha$ with respect to the
axis defining the Stokes parameters, and $\tens{\Gamma}$ is the matrix
folding the ring onto the sky pixels. From the above equation, the
offsets $\vec{o}$ are derived through maximum likelihood, imposing
an additional constraint, in our case that the sum of the offsets
has to be equal to zero (arbitrarily, since we do not measure the
absolute temperature on the sky, only differences). The performance
of this implementation has been evaluated using simulations in \citet{tristram2011}.

We produce temperature and polarization maps for a number of data
sets: 
\begin{itemize}
\item individual detectors;
\item independent detector sets at the same frequency (see Table~\ref{tab:detsets});
and
\item the combination of all detectors at each frequency. 
\end{itemize}
To enable systematic checks, we build maps for the whole time interval
spanned by the HFI data, as well as more restricted intervals corresponding
to individual surveys. We also build maps in each of the above cases
using data from the first and second half of the rings independently.
In each case (a full ring or its two halves) we first determine the
offsets on the total time range and apply them to produce maps on
any more restricted time range. Together with each temperature map,
we build maps containing the hit count (number of TOI samples in each
pixel) and the propagation of the TOI sample variance, as measured
by the total noise NET (see Sect.~\ref{sec:Toi-Qualification}).
In total, more than 6500 maps are produced.

\subsection{Absolute photometric calibration\label{sec:Abs-calib}}

We use two techniques to perform the photometric calibration of the
HFI detectors, one for lower frequency channels (100--353\,GHz, typically
given in temperature units, $\mathrm{K_{\mathrm{CMB}}}$), another
for higher frequency channels (545 and 857\,GHz, given in flux density
units, $\mathrm{MJy\, sr^{-1}}$). In both cases, significant changes
have occurred with respect to the early data processing. The photometric
calibration processes and their performance are described in detail
in \citet{planck2013-p03f}. Our philosophy is to calibrate each detector
individually, without relative calibration between them. 

Previously (for the early results from \Planck), the lower frequencies
were calibrated using the solar dipole, and the higher frequencies
using FIRAS dust spectra measurements, assuming constant gain. When
more data were accumulated, comparisons between measurements taken
one year apart in identical detector configuration unambiguously showed
that the detectors' gains exhibit variations of 1 to 2\,\%, over
a wide range of time intervals (one day to one year).

\begin{figure}[th]
\centering{}\includegraphics[width=1\columnwidth]{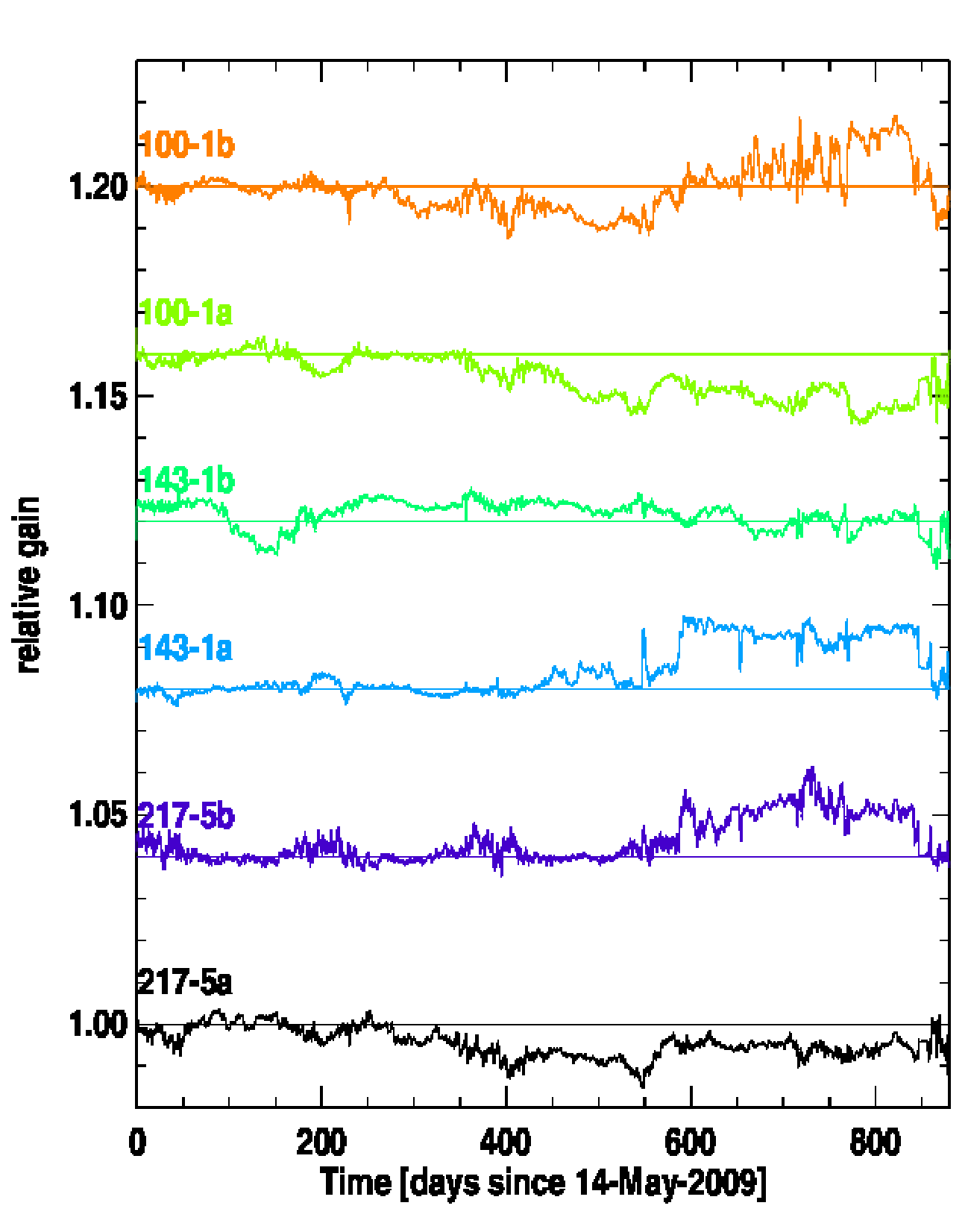}\protect\caption{\label{fig:bogopix-gains} Samples of relative gain variations reconstructed
for six CMB channel bolometers. Data for each bolometer have been
translated vertically for clarity, with a 3\,\% spacing. \rev{In each case, thin lines show the average level over days 150 to 280, during which the solar dipole calibration is determined}.
The observed variations range in general over $\pm 1.5\,\%$. }
\end{figure}

We \rev{also found} that these apparent gain variations are \rev{primarily}
consequences of the nonlinearities of the ADC used in the read-out
electronic chain (see Sect.~\ref{sec:ADC-non-linearity}). Correcting
for such effects requires a precise knowledge of the ADC transfer
function. To acquire these we characterized the read-out response
using warm data after the end of the 100\,mK cooling in January 2012.
This data-taking was not completed in time for the 2013 HFI data release.
At the frequencies where the solar dipole has a high enough amplitude
(100 to 217\,GHz) we evaluated the apparent gain variations of the
detectors by solving the nonlinear equation $data=gain\times sky+noise$
for both gain and sky signal. Examples of gain-variation measurements
are shown in Fig.~\ref{fig:bogopix-gains}. Note the wide range of
behaviour, from slow drifts (e.g., 100-1a) to apparent jumps (e.g.,
143-1a).

These measurements have been used to correct the bolometer data in
the present data release. As these gain variations are just the first-order
consequence of the ADC nonlinearities, these corrections will not
remove all ADC nonlinearity effects. We estimate \citep{planck2013-p03f}
that the remaining apparent gain variations (or other residual time-variable
systematics) after correction are lower than 0.3\,\%, as measured
by comparison with the solar dipole.

The presence of residual nonlinear systematic biases in our data precludes
the use of potentially more precise techniques, such as those discussed
in \citet{tristram2011} using the orbital CMB dipole anisotropy.
In \citet{planck2013-p03f} we show that using a calibration derived
from the orbital dipole would lead to larger detector-to-detector
relative calibration dispersion, which induces large-angular-scale
patterns in polarization through dipole leakage. In addition, for
detectors for which the Solar and orbital dipole difference is large
(typically around 0.5\,\% but in few cases as large as 1\,\%), a
residual dipole is apparent after subtracting the \emph{WMAP} prediction.
Thus, the calibration for 100 to 353\,GHz is based on the solar dipole
as measured by \emph{WMAP} (we use a non-relativistic calculation;
the roughly 0.1\,\% relativistic correction is smaller than other
residuals such as the uncorrected ADC non-linearities).

For the two highest frequency channels, the calibration scheme that
was used for the early data release was based on FIRAS data. As described
in \citet{planck2013-p03f}, several studies since the first release
indicated that the calibration scale of these two channels was somewhat
\rev{inaccurate}. These include measurements of the dust spectral
energy distribution and of the amplitude of the CMB anisotropy itself
(still detectable at 545\,GHz). Re-analysis of the FIRAS data \citep{liang2012}
also seems to indicate that the FIRAS spectra might contain additional
systematic errors. Since we also observed systematic differences between
HFI and FIRAS data, we decided to revise our calibration strategy.
We therefore now supplement our calibration using the reconstructed
fluxes of planets, compared with models of their emission \citep{moreno2010},
in order to renormalize the relative FIRAS calibration at 545 and
857\,GHz. We used observations of Uranus, Neptune, and Mars (Jupiter
and Saturn, although observed in \Planck\ data, are too bright and
hence affected by detector nonlinearities). Compared to previous data
releases, this new calibration scheme amounts to a decrease in flux
densities of 15\,\% and 7\,\% at 545 and 857\,GHz, respectively.

Various techniques have been used to evaluate the relative (intra-
or inter-frequency) and absolute calibration accuracy \citep{planck2013-p03f,planck2013-p08}.
Results are summarized in Table~\ref{tab:summary}. The main limitations
on the photometric calibration accuracy that we have identified result
from the residual nonlinearities for low frequency channels, and from
the accuracy of the models of planetary brightness at 545 and 857\,GHz.

For this data release, the zero levels of the maps, which \Planck\
cannot determine internally, have not been set. Still, we estimate
in \citet{planck2013-p03f} a Galactic and an extragalactic zero level
(Tables 4 and 5). For the Galactic case, we compute the map brightness
corresponding to zero gas column density as traced by the 21\,cm
emission from neutral hydrogen. For the monopole term of the CIB,
we use an empirical model, which can be used to enable total emission
analysis. These estimates can be used to set the appropriate zero
levels of HFI maps and are also summarized in Table~\ref{tab:summary}.

\subsection{Overview of HFI map properties\label{sub:Maps-overview}}

\begin{figure*}[!t]
\begin{centering}
\includegraphics[bb=280bp 130bp 780bp 400bp,clip,width=0.75\textwidth]{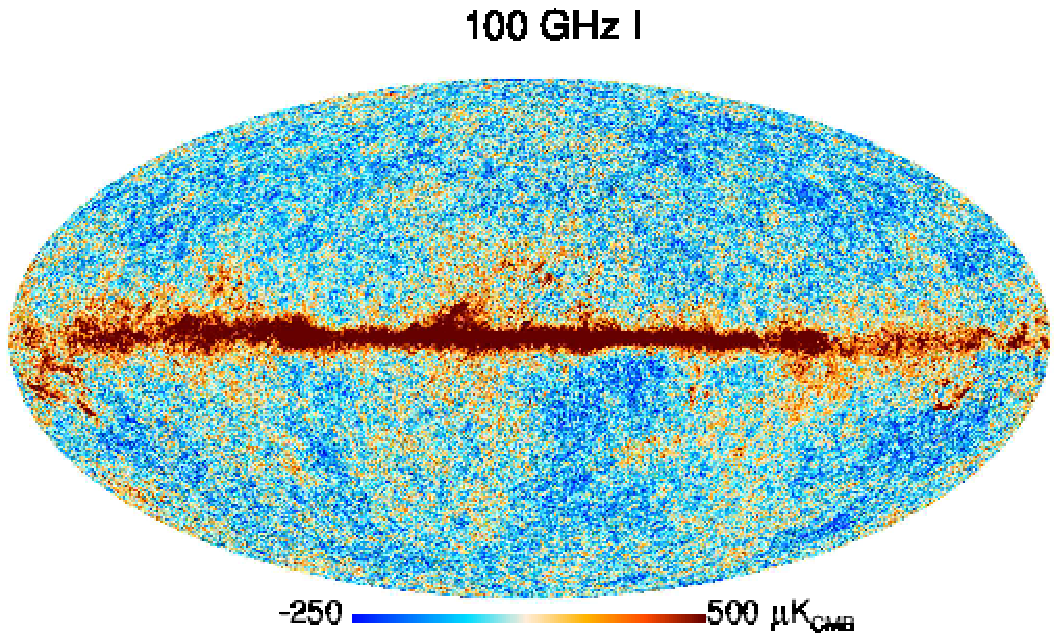}
\par\end{centering}

\begin{centering}
\includegraphics[bb=280bp 130bp 780bp 400bp,clip,width=0.75\textwidth]{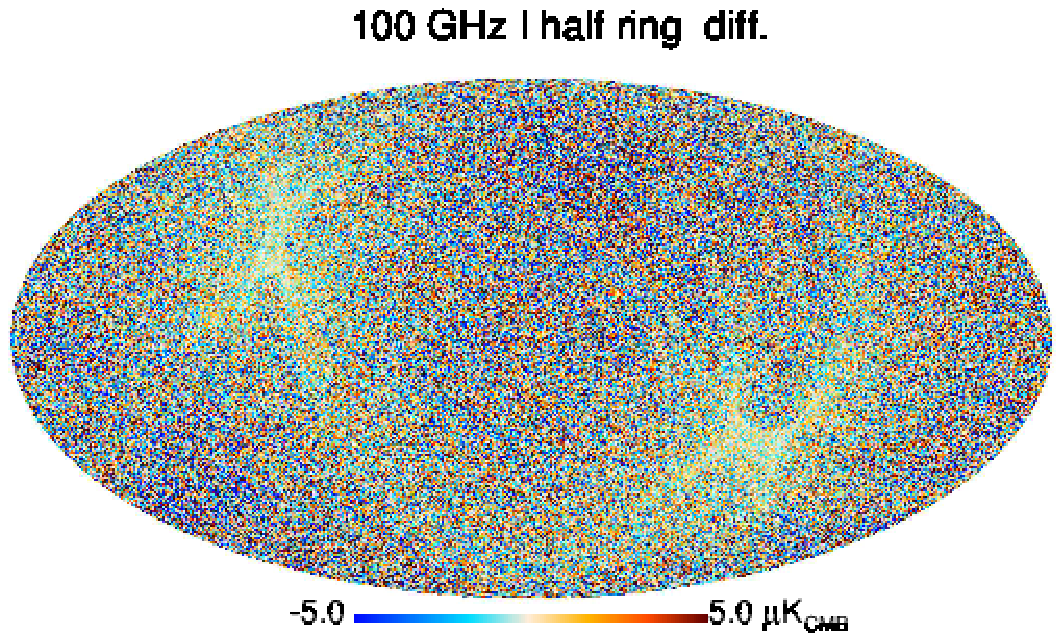}
\par\end{centering}

\begin{centering}
\includegraphics[bb=280bp 130bp 780bp 400bp,clip,width=0.75\textwidth]{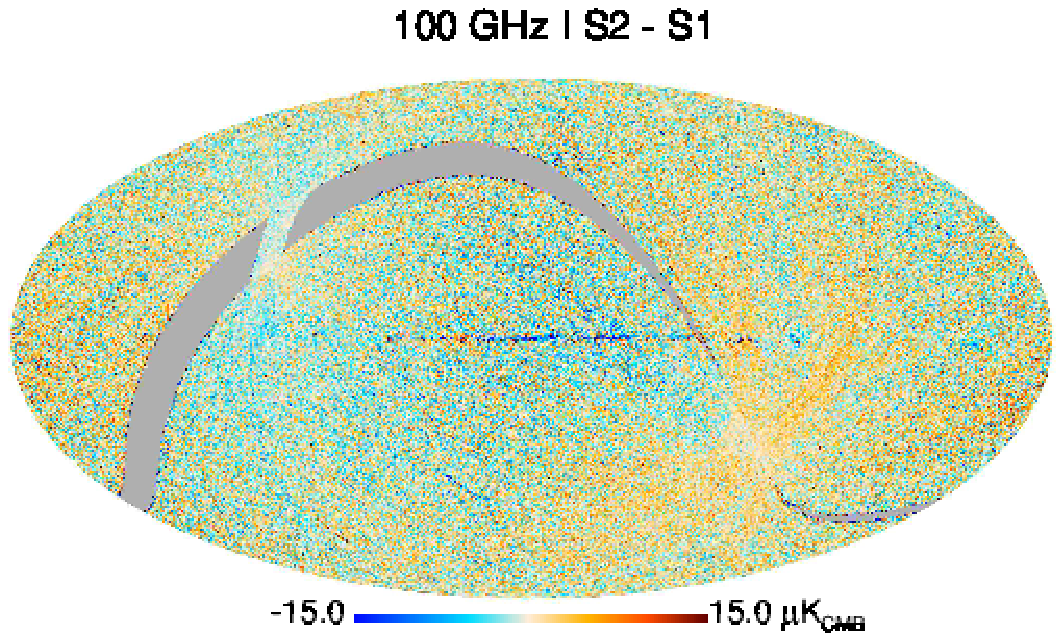}
\par\end{centering}

\centering{}\protect\caption{\label{fig:Imaps-100} HFI maps at 100\,GHz. The top panel gives
the intensity in $\mu\mathrm{K_{CMB}}$. The middle panel shows the
difference between maps made from the first and the second halves
of each stable pointing period (i.e., half-ring maps). The bottom
panel shows the difference between Survey 1 and Survey 2. }
\end{figure*}

\begin{figure*}[!t]
\begin{centering}
\includegraphics[bb=280bp 130bp 780bp 400bp,clip,width=0.75\textwidth]{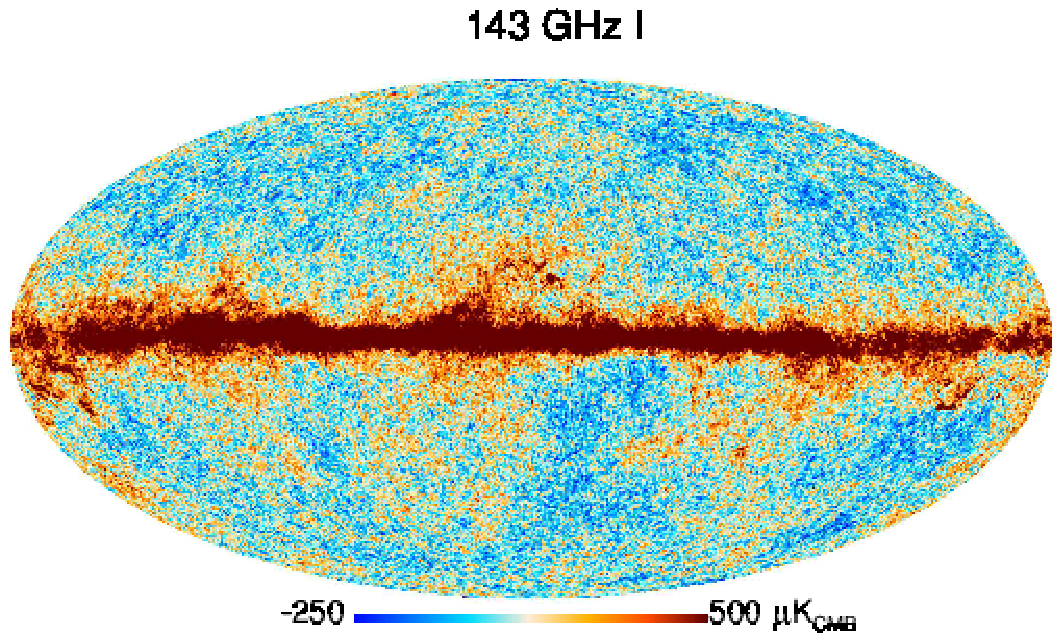}
\par\end{centering}

\begin{centering}
\includegraphics[bb=280bp 130bp 780bp 400bp,clip,width=0.75\textwidth]{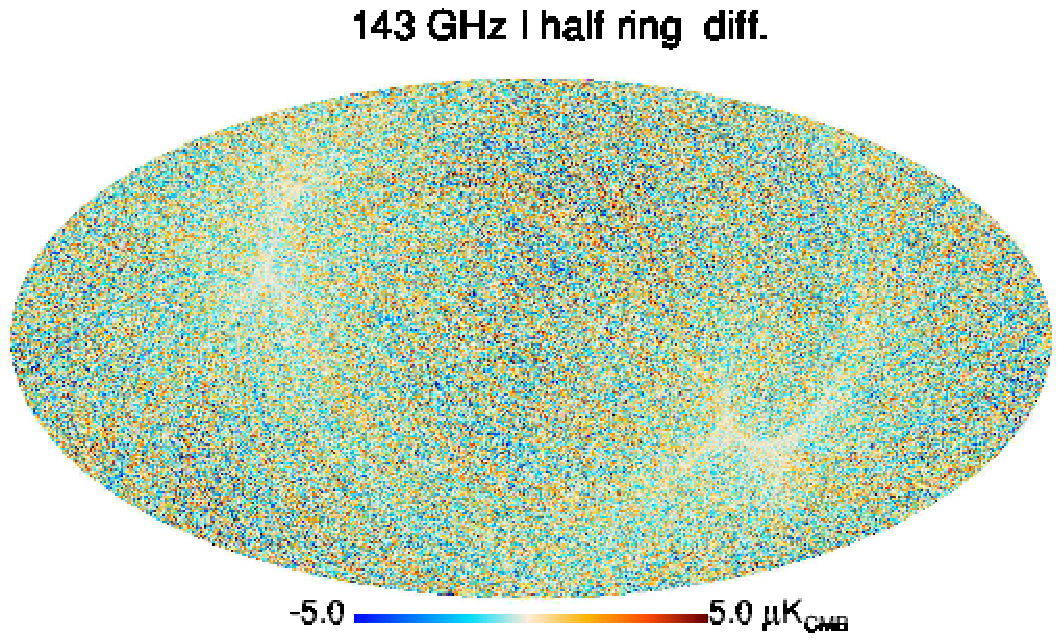}
\par\end{centering}

\begin{centering}
\includegraphics[bb=280bp 130bp 780bp 400bp,clip,width=0.75\textwidth]{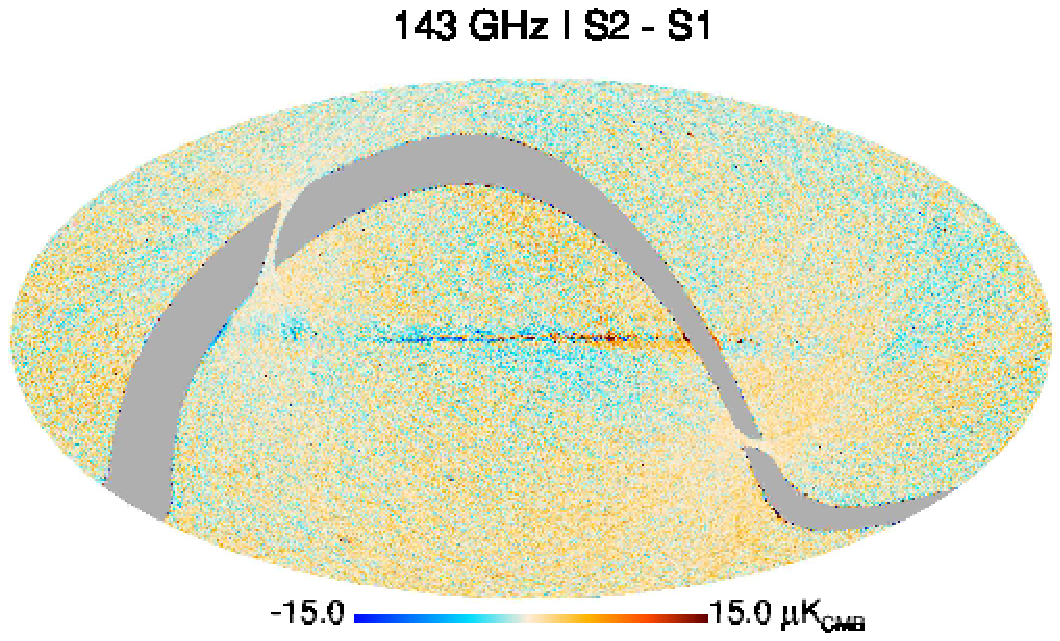}
\par\end{centering}

\centering{}\protect\caption{\label{fig:Imaps-143} HFI maps at 143\,GHz. The top panel gives
the intensity in $\mu\mathrm{K_{CMB}}$. The middle panel shows the
difference between maps made from the first and the second halves
of each stable pointing period (i.e., half-ring maps). The bottom
panel shows the difference between Survey 1 and Survey 2.}
\end{figure*}

\begin{figure*}[!t]
\begin{centering}
\includegraphics[bb=280bp 130bp 780bp 400bp,clip,width=0.75\textwidth]{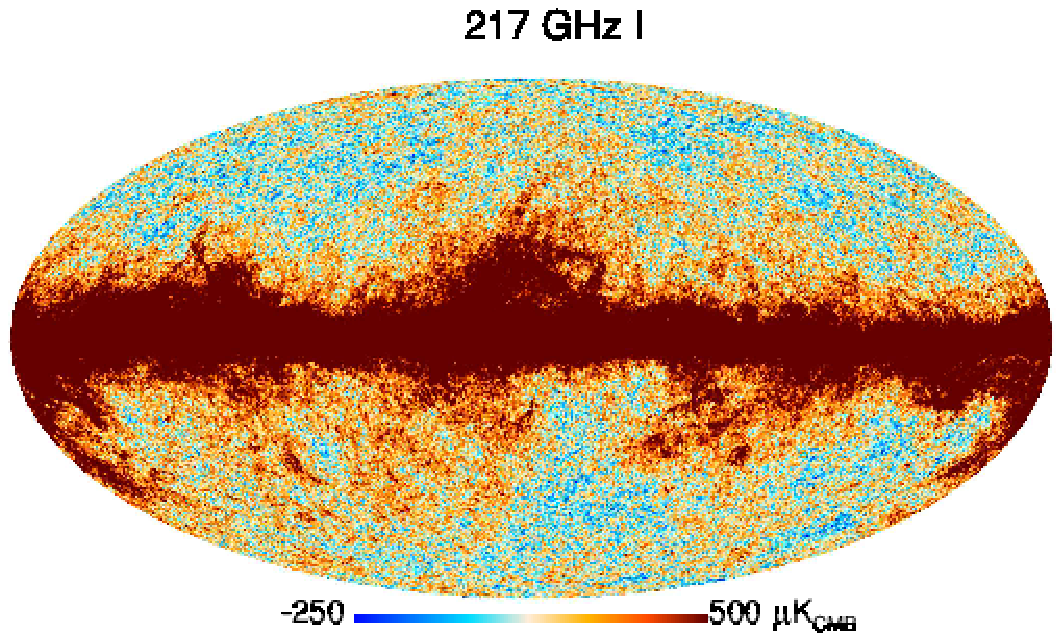}
\par\end{centering}

\begin{centering}
\includegraphics[bb=280bp 130bp 780bp 400bp,clip,width=0.75\textwidth]{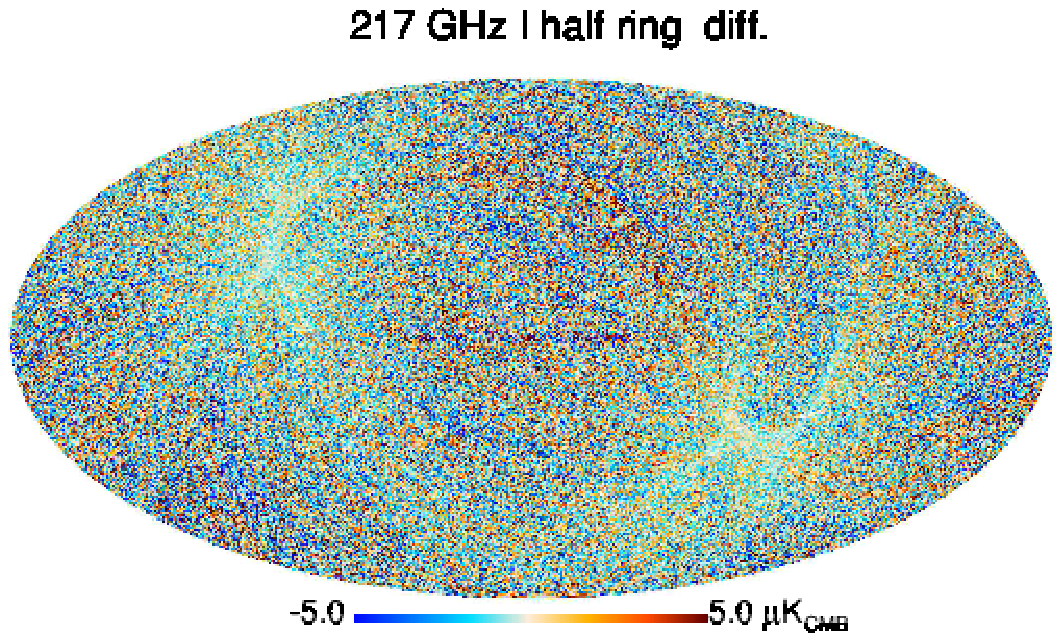}
\par\end{centering}

\begin{centering}
\includegraphics[bb=280bp 130bp 780bp 400bp,clip,width=0.75\textwidth]{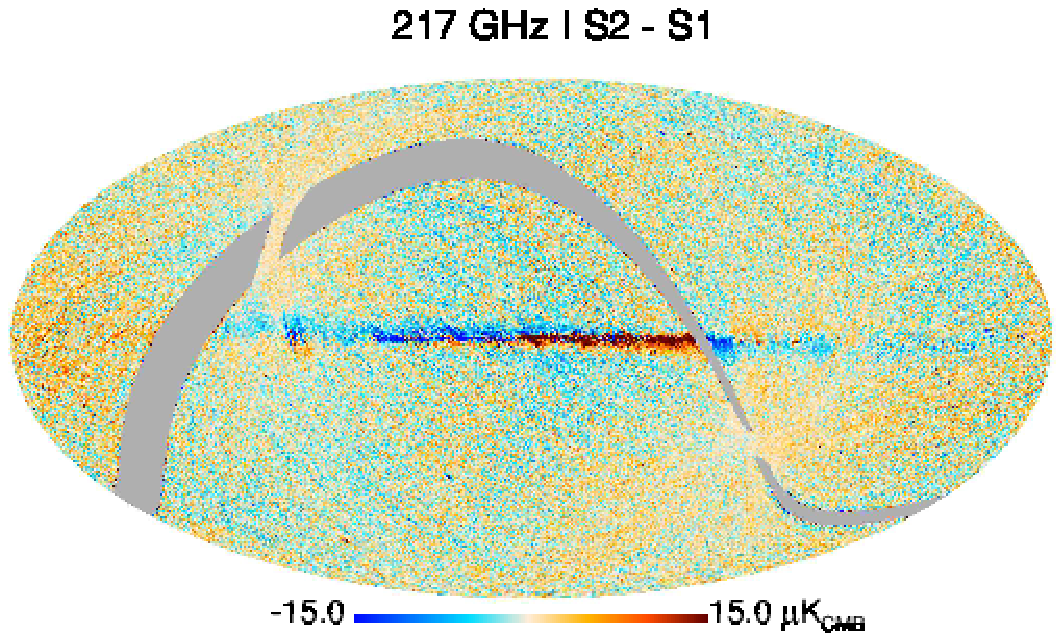}
\par\end{centering}

\centering{}\protect\caption{\label{fig:Imaps-217} HFI maps at 217\,GHz. The top panel gives
the intensity in $\mu\mathrm{K_{CMB}}$. The middle panel shows the
difference between maps made from the first and the second halves
of each stable pointing period (i.e., half-ring maps). The bottom
panel shows the difference between Survey 1 and Survey 2.}
\end{figure*}

\begin{figure*}[!t]
\begin{centering}
\includegraphics[bb=280bp 130bp 780bp 400bp,clip,width=0.75\textwidth]{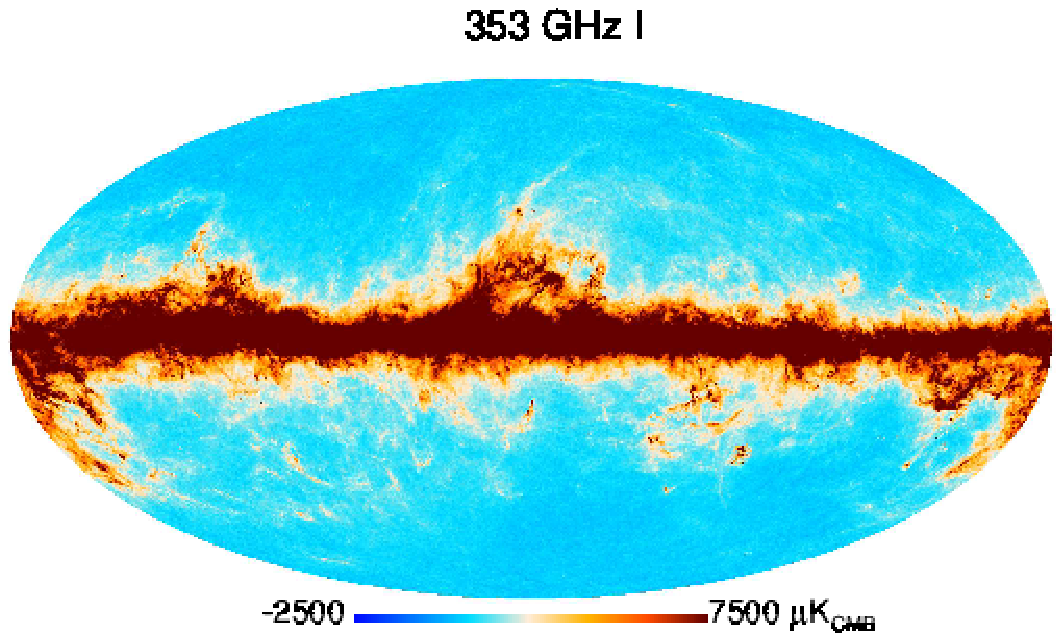}
\par\end{centering}

\begin{centering}
\includegraphics[bb=280bp 130bp 780bp 400bp,clip,width=0.75\textwidth]{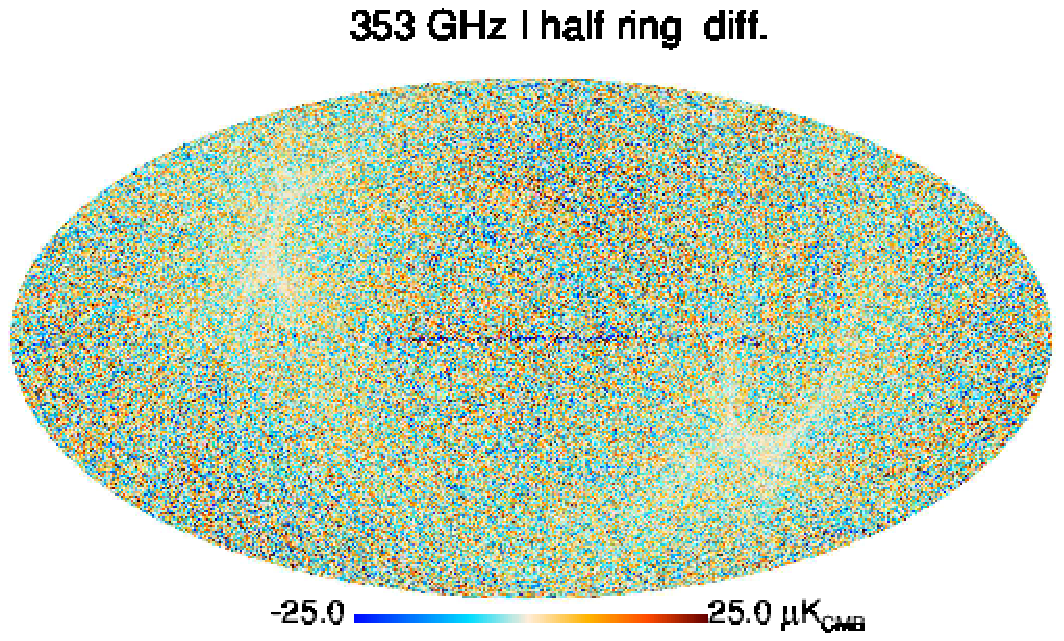}
\par\end{centering}

\begin{centering}
\includegraphics[bb=280bp 130bp 780bp 400bp,clip,width=0.75\textwidth]{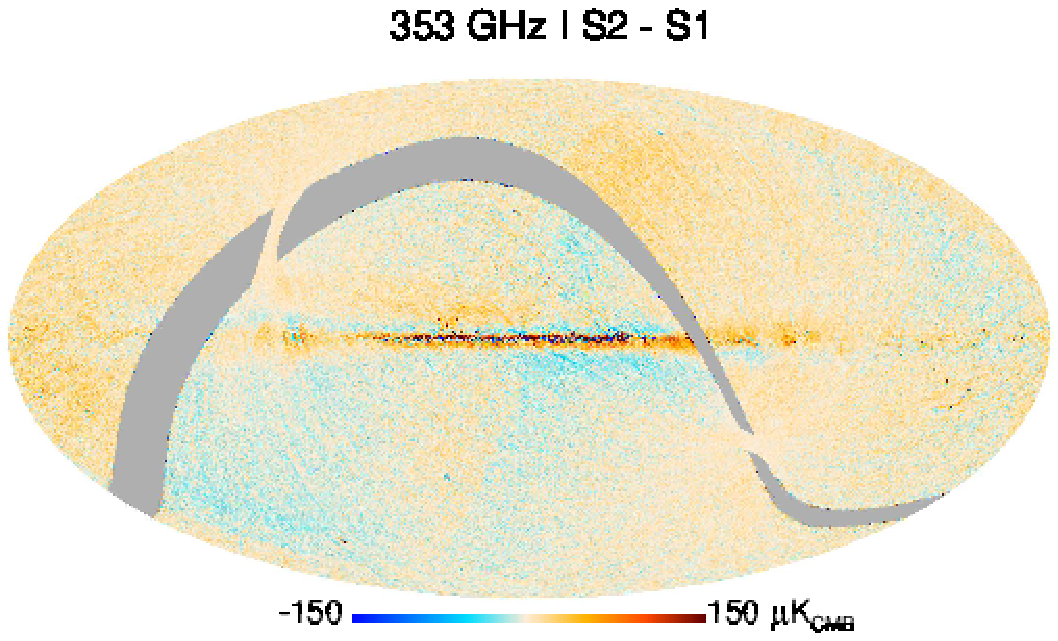}
\par\end{centering}

\centering{}\protect\caption{\label{fig:Imaps-353} HFI maps at 353\,GHz. The top panel gives
the intensity in $\mu\mathrm{K_{CMB}}$. The middle panel shows the
difference between maps made from the first and the second halves
of each stable pointing period (i.e., half-ring maps). The bottom
panel shows the difference between Survey 1 and Survey 2.}
\end{figure*}

\begin{figure*}[!t]
\begin{centering}
\includegraphics[bb=280bp 130bp 780bp 400bp,clip,width=0.75\textwidth]{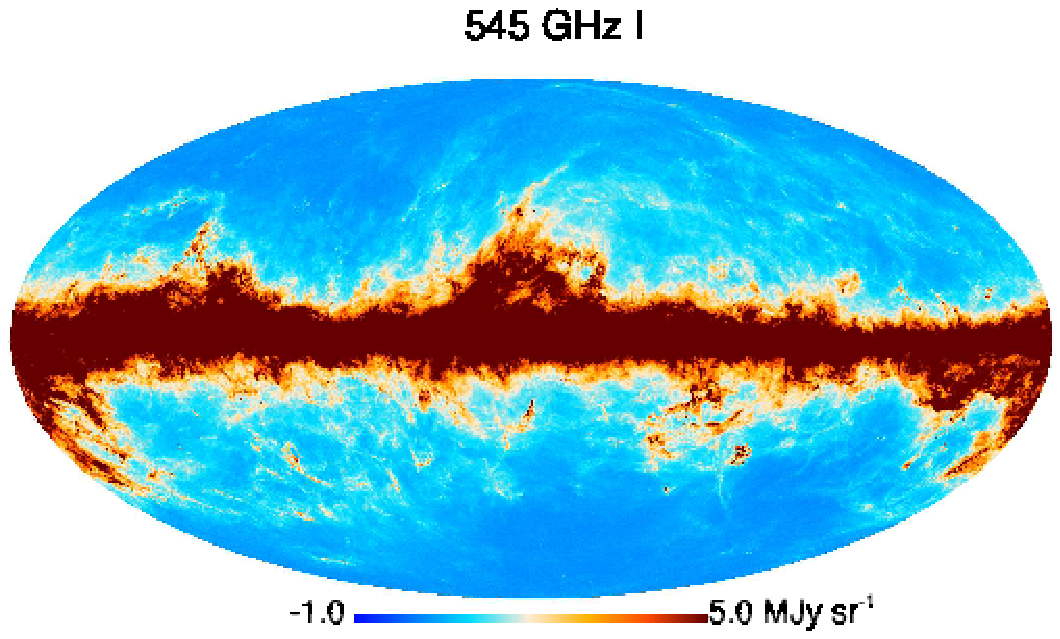}
\par\end{centering}

\begin{centering}
\includegraphics[bb=280bp 130bp 780bp 400bp,clip,width=0.75\textwidth]{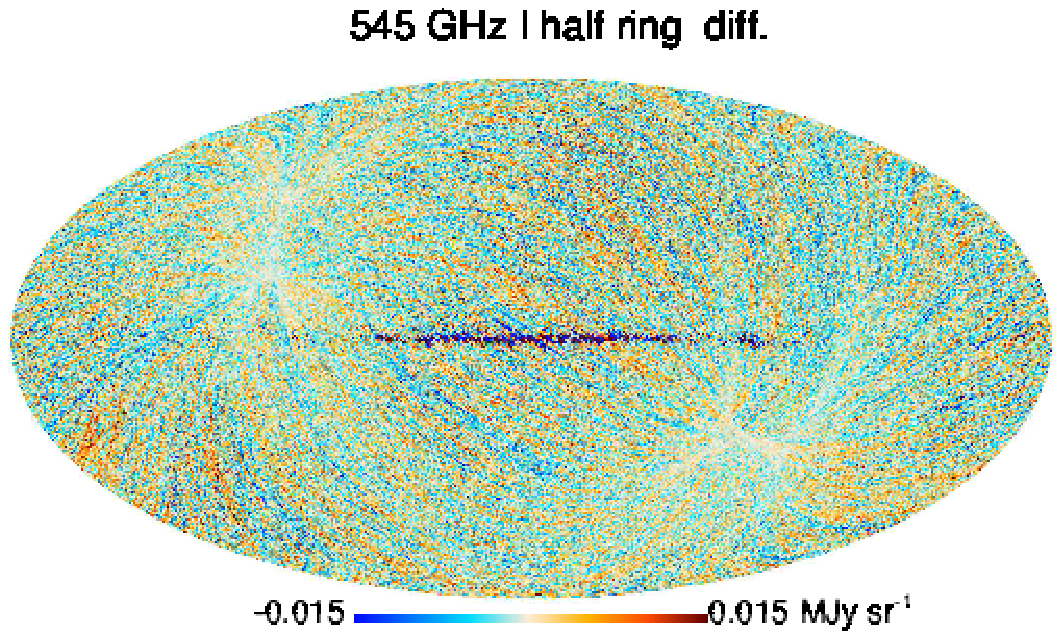}
\par\end{centering}

\centering{}\includegraphics[bb=280bp 130bp 780bp 400bp,clip,width=0.75\textwidth]{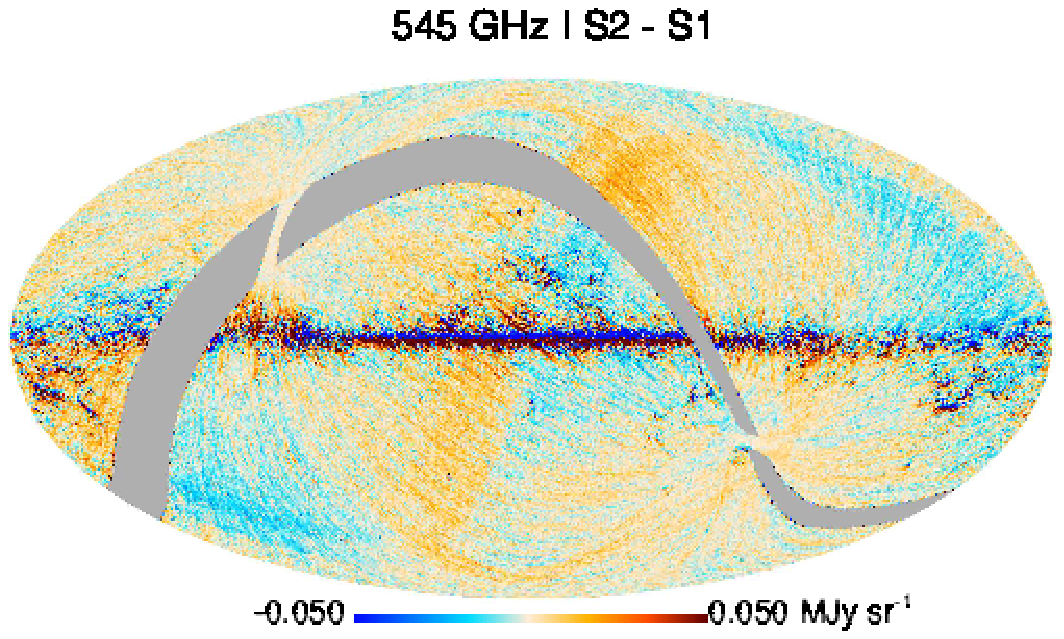}\protect\caption{\label{fig:Imaps-545} HFI maps at 545\,GHz. The top panel gives
the intensity in MJy\,sr$^{-1}$. The middle panel shows the difference
between maps made from the first and the second halves of each stable
pointing period (i.e., half-ring maps). The bottom panel shows the
difference between Survey 1 and Survey 2.}
\end{figure*}

\begin{figure*}[!t]
\begin{centering}
\includegraphics[bb=280bp 130bp 780bp 400bp,clip,width=0.75\textwidth]{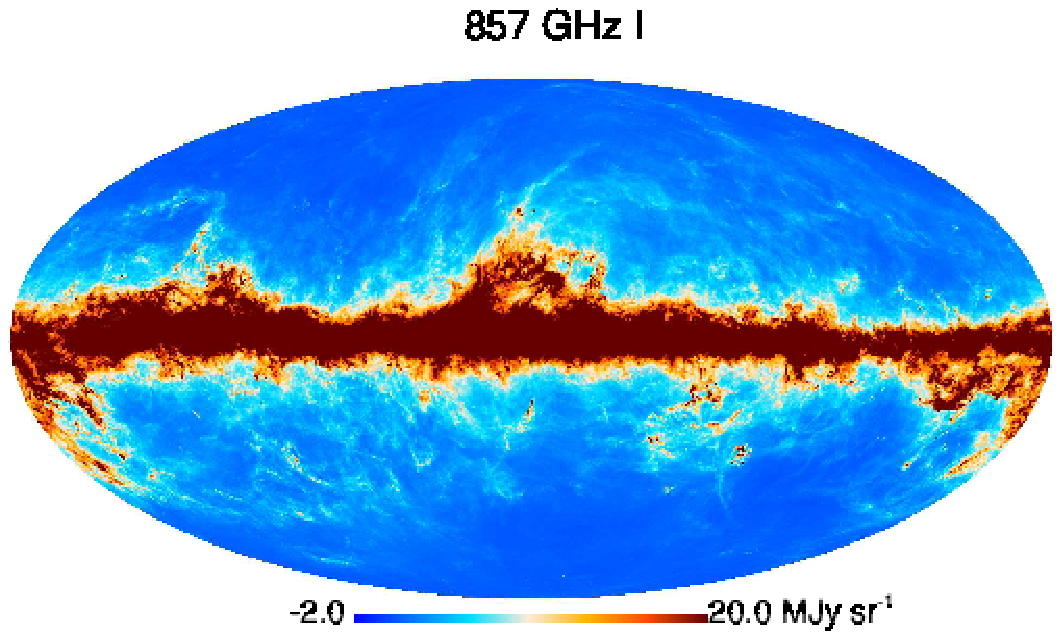}
\par\end{centering}

\begin{centering}
\includegraphics[bb=280bp 130bp 780bp 400bp,clip,width=0.75\textwidth]{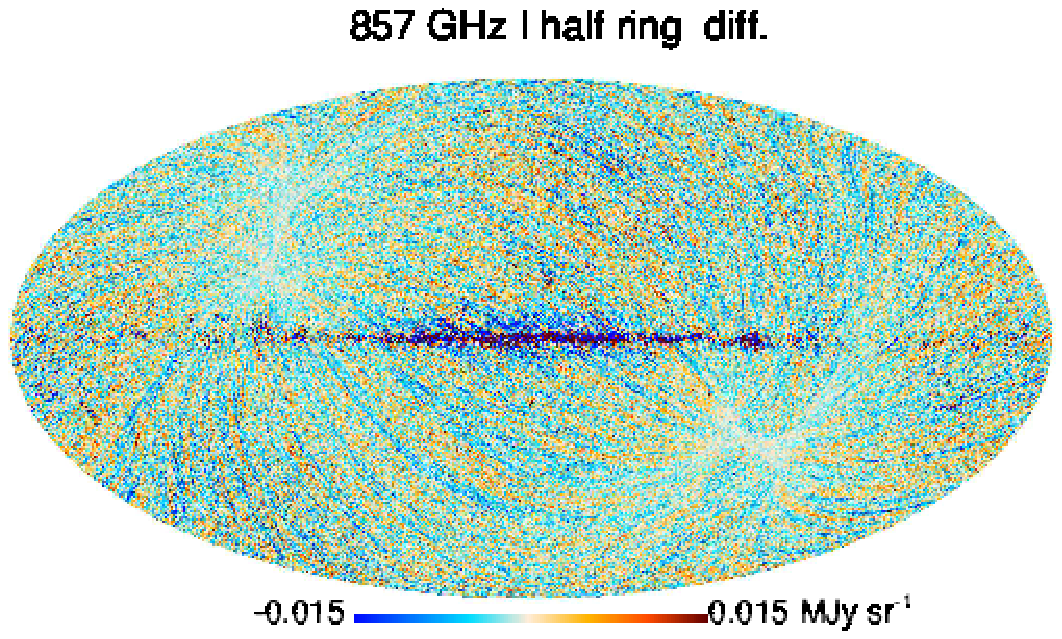}
\par\end{centering}

\centering{}\includegraphics[bb=280bp 130bp 780bp 400bp,clip,width=0.75\textwidth]{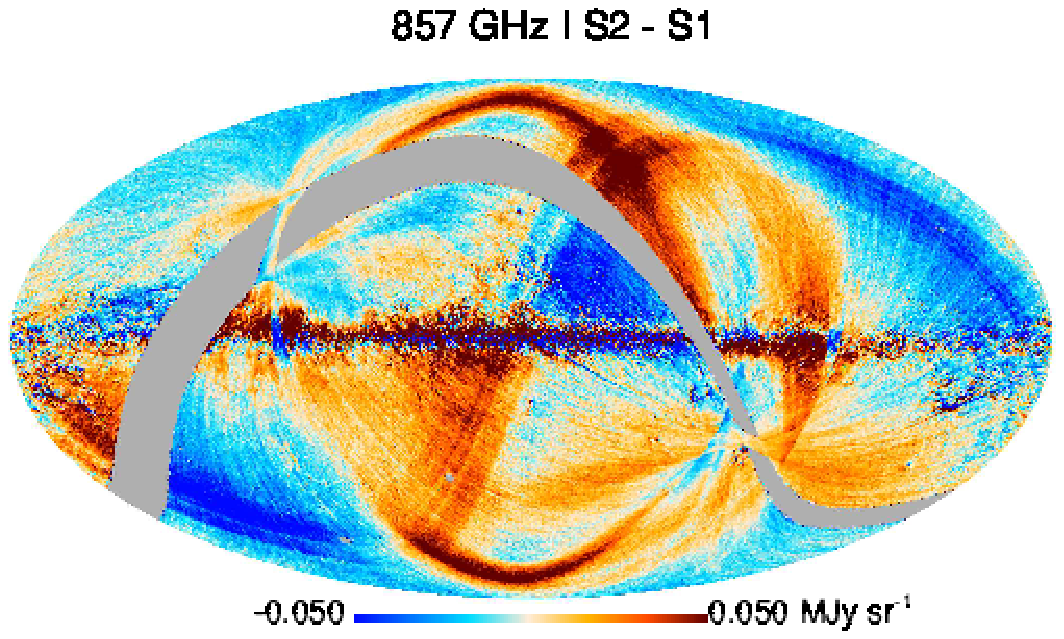}\protect\caption{\label{fig:Imaps-857} HFI maps at 857\,GHz. The top panel gives
the intensity in MJy\,sr$^{-1}$. The middle panel shows the difference
between maps made from the first and the second halves of each stable
pointing period (i.e., half-ring maps). The bottom panel shows the
difference between Survey 1 and Survey 2.}
\end{figure*}

Figures~\ref{fig:Imaps-100} to \ref{fig:Imaps-857} show the sky
maps constructed from HFI data. These correspond to maps uncorrected
for zodiacal light emission and far sidelobe pick-up (discussed in
the next section). The top row of the figure series shows the intensity
maps reconstructed from the nominal mission, while the second row
shows the difference between maps made from the first and the second
halves of each stable pointing period (i.e., half-ring maps), which
provides visual information on the level and distribution of the residuals,
since these maps illustrate the difference between maps constructed
from rings taken about 20 minutes apart; therefore they illustrate
the variation of the detector signal along a sky circle on such a
timescale, which certainly dominates the rms. Longer timescale variations
can be judged from the third row of plots showing the difference in
the sky between the Survey 1 and Survey 2. This time one sees variations
on a six-month time scale, which maximizes systematic effects by comparing
measurements taken with a different scan orientation of the satelitte.
Note that for 100--217\,GHz, the colour scale of the second row is
enlarged by a factor of 75, by 200 at 353 and 545\,GHz, and by a
factor of approximately 730 at 857\,GHz.


\begin{figure}[th]
\begin{centering}
\includegraphics[bb=80bp 60bp 650bp 460bp,clip,width=1\columnwidth]{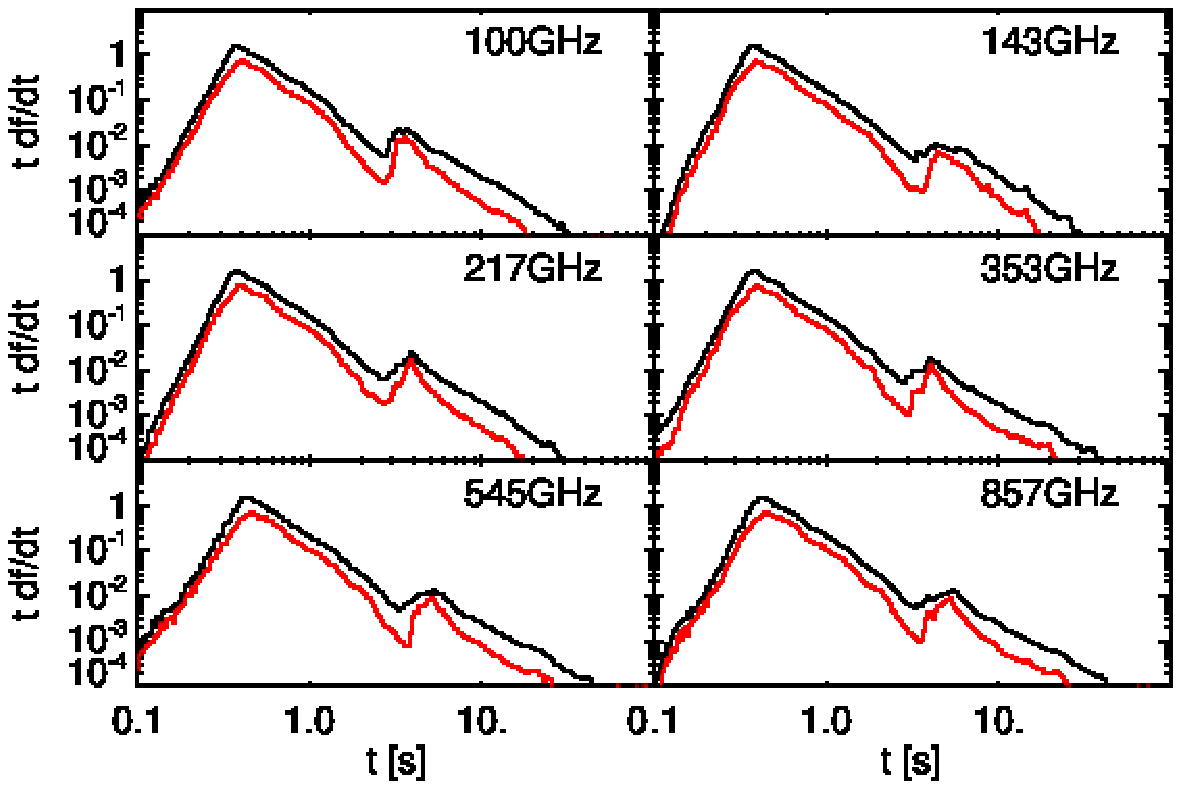}
\par\end{centering}

\centering{}\protect\caption{\label{fig:Imaps-IntegTime} HFI channel integration time distribution
function. For each HFI channel, the histogram of the integration time
$t$ per pixel of area $(1.7\arcm){}^{2}$ is shown. The black curve
is for the nominal all-sky ($f_{\mathrm{sky}}=1$) mission. The red
curve represents the masked sky, using the CL39 ($f_{\mathrm{sky}}=0.48$
\rev{for the Galactic part}) mask defined in \citet{planck2013-p08}.
The normalization is such that $\int df=f_{\mathrm{sky}}$. The median
integration time is about 0.42\,s whereas the mean is 0.56\,s. Notice
that about 0.5\,\% of the pixels are observed for more than 5\,s;
these mostly lie around the Ecliptic poles. }
\end{figure}

\begin{figure}[th]
\begin{centering}
\includegraphics[bb=30bp 625bp 340bp 842bp,width=1\columnwidth]{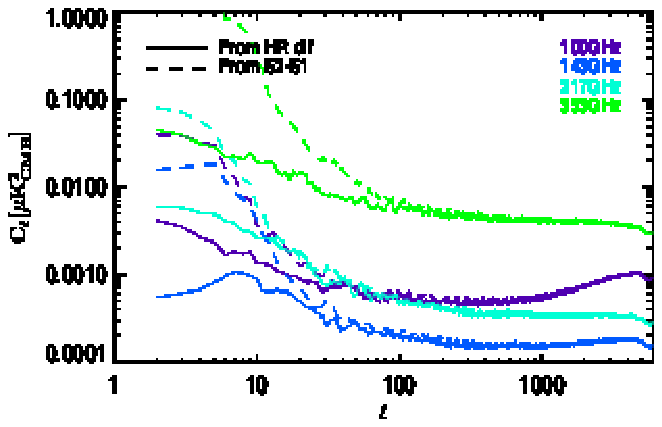}
\par\end{centering}

\begin{centering}
\includegraphics[bb=30bp 625bp 340bp 842bp,width=1\columnwidth]{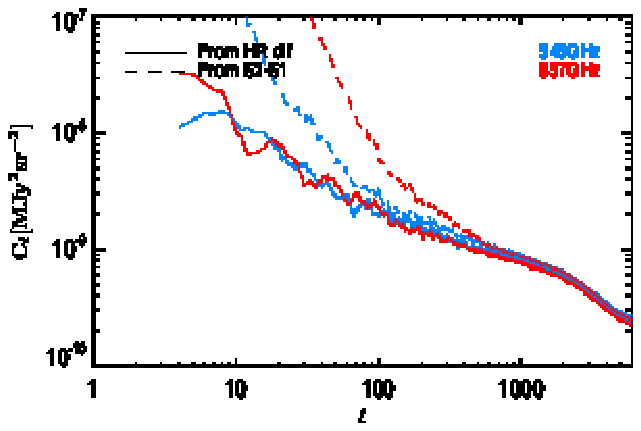}
\par\end{centering}

\centering{}\protect\caption{\label{fig:maps-half-ring-noise} Power spectra of the intensity maps
reconstructed from the difference between the first and second half
of each ring (denoted ``HR dif''), or from the difference between
Survey 1 and Survey 2. }
\end{figure}

The distribution of integration time per channel is shown in Fig.~\ref{fig:Imaps-IntegTime}.
The map noise properties may be evaluated from variance maps computed
from the integration time per pixel, but a better approach uses the
half-ring map differences (middle row of Figs.~\ref{fig:Imaps-100}
to \ref{fig:Imaps-857}), which gives a better rendition of the noise
at map level due to the non-white structure of the timeline noise
after time-response deconvolution. In Fig.~\ref{fig:maps-half-ring-noise}
we show the angular power spectra of these differences, for each frequency.
With respect to the \Planck\  early data release, these noise spectra
are notably flatter at high $\ell$, thanks to the use of a better
low pass filter in the TOI processing \rev{(although the 100\GHz\ channels still show a rise at high $\ell$ due to imperfect deconvolution)}.
The figure also shows the spectra of the differences between maps
made from the first and second sky surveys (bottom row of Figs.~\ref{fig:Imaps-100}
to \ref{fig:Imaps-857}), giving an indication of the contribution
to the noise from longer timescales. The two sets of spectra converge
at high $\ell$.

The average of the half-ring power spectra from $\ell=100$ to $6000$\textcolor{red}{{}
}are reported in Table~\ref{tab:summary}. One should note, however,
that the half-ring maps give an estimate of the noise that is biased
low (by a couple of percent), due to small correlations induced by
the way the timelines have been degliched. This is discussed in detail
with the help of simulations in Sect.~\ref{sec:Noise-Estimation-Bias}.
Estimates of the noise level from the timelines and from the maps
were discussed in Sect.~\ref{sec:Toi-Qualification} and given in
Table~\ref{tab:Total-noise}. They are consistent with alternative
descriptions of the noise that were computed in the HFI calibration
and mapmaking paper \citep{planck2013-p03f}. The summary in Table\ \ref{tab:summary}
(lines \emph{c1}--\emph{c2}) gives sensitivity levels derived from
these analyses.

\subsection{Far-side lobe and zodiacal light correction\label{sec:Correct-FSL=000026Zodi}}

\begin{figure*}[p]
\begin{centering}
\includegraphics[clip,width=0.33\textwidth]{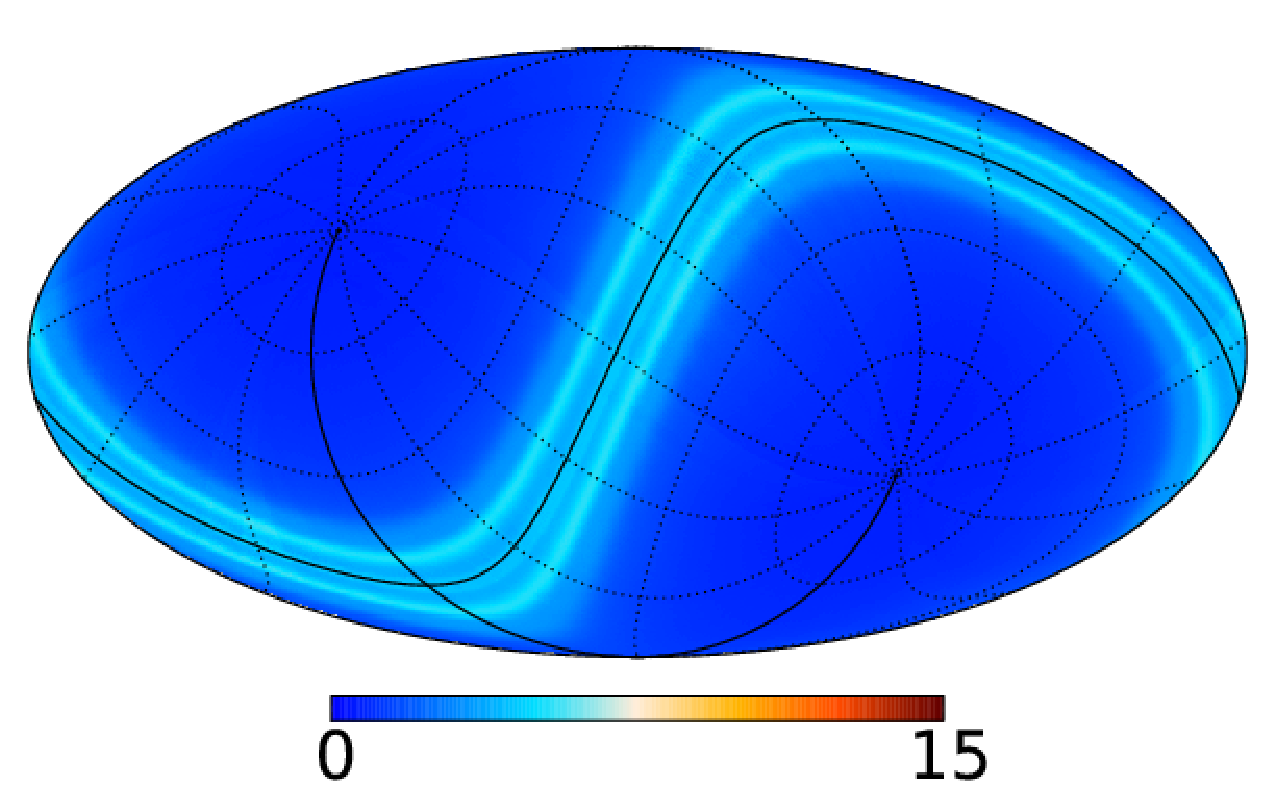}\includegraphics[clip,width=0.33\textwidth]{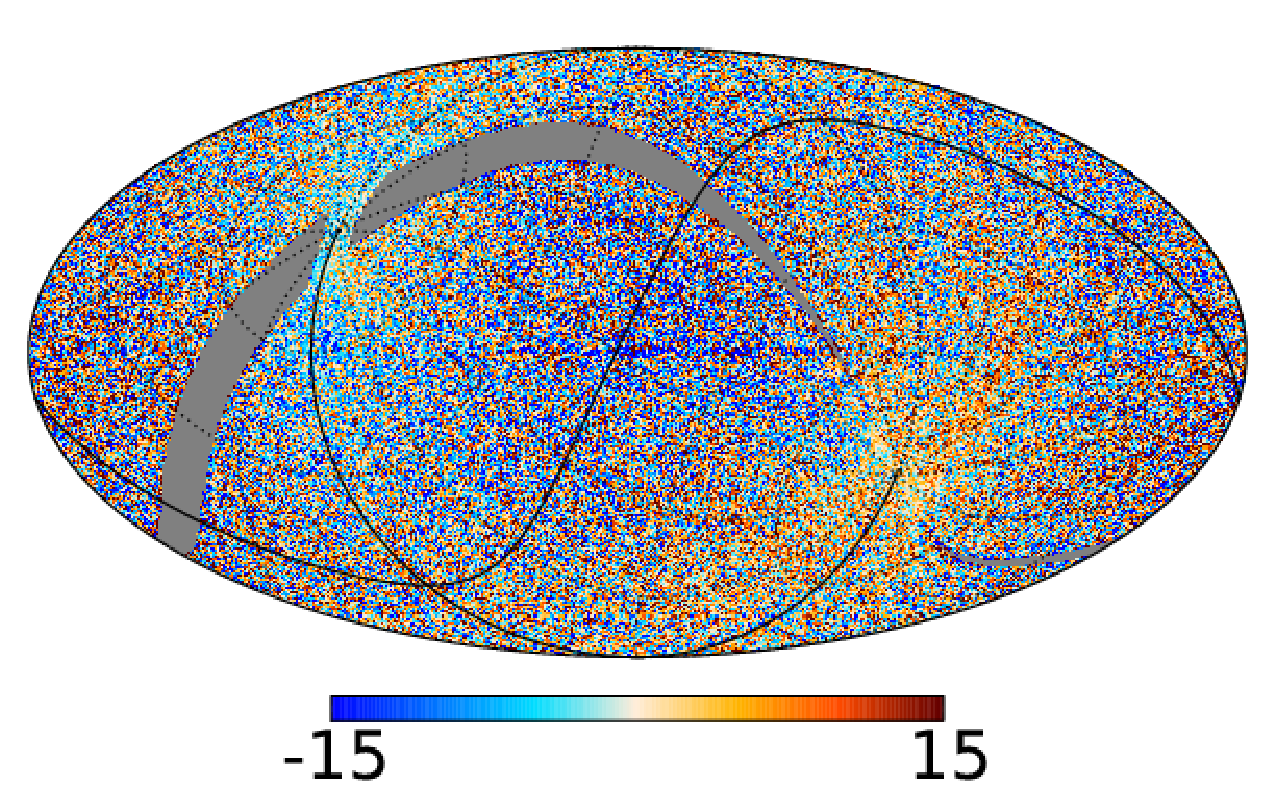}
\par\end{centering}

\begin{centering}
\includegraphics[clip,width=0.33\textwidth]{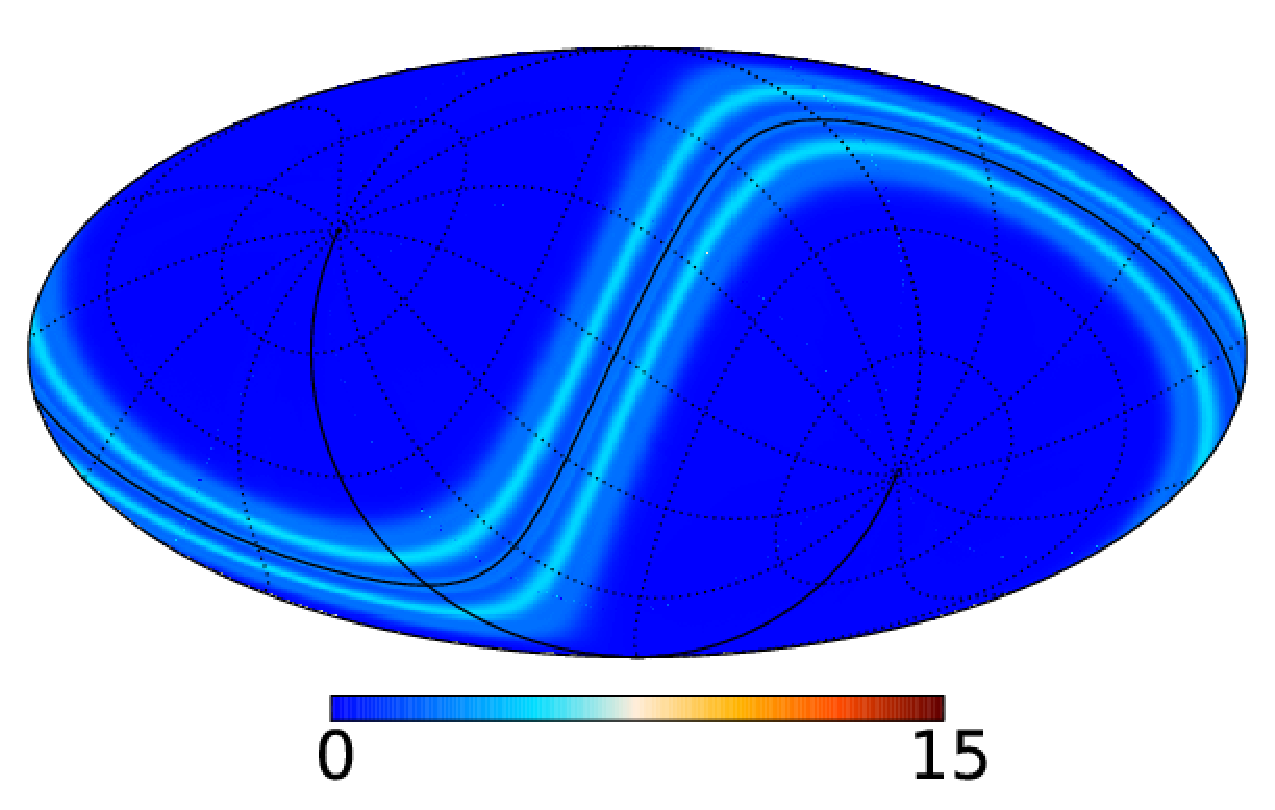}\includegraphics[clip,width=0.33\textwidth]{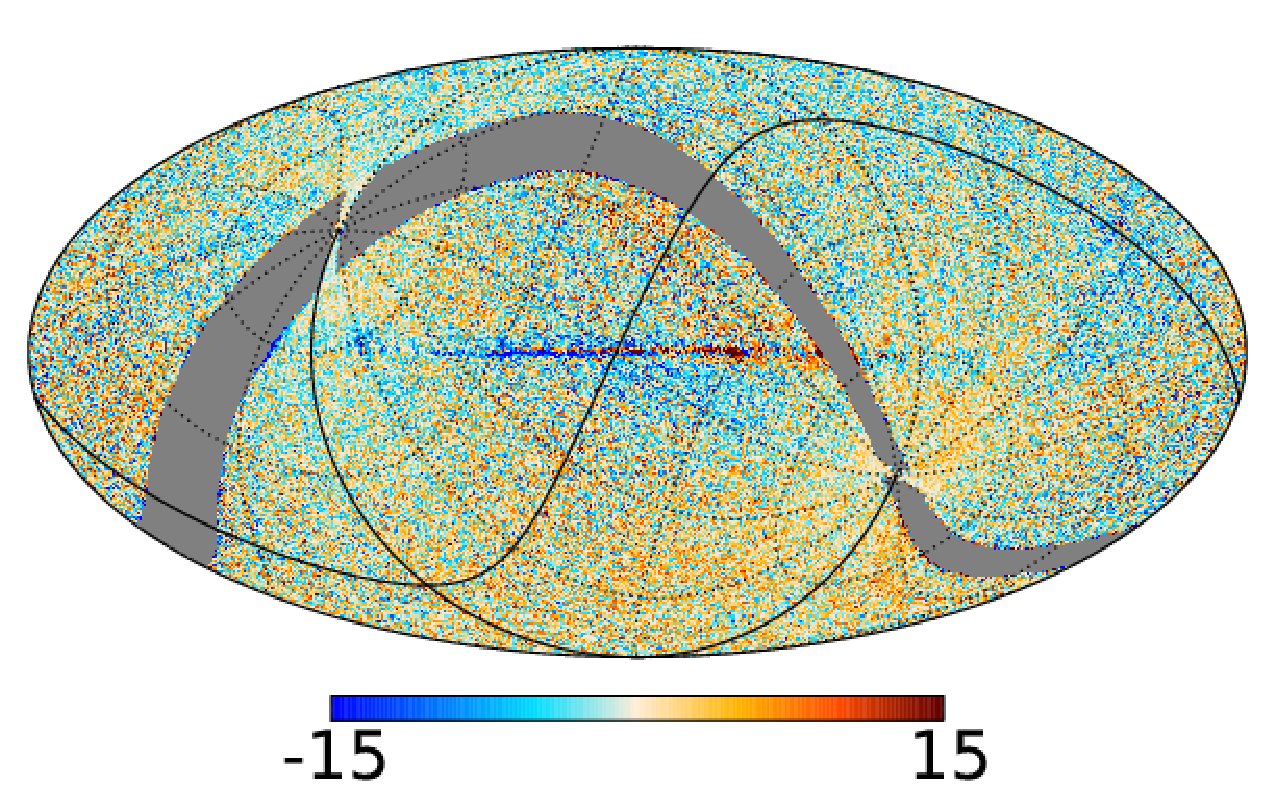}
\par\end{centering}

\begin{centering}
\includegraphics[clip,width=0.33\textwidth]{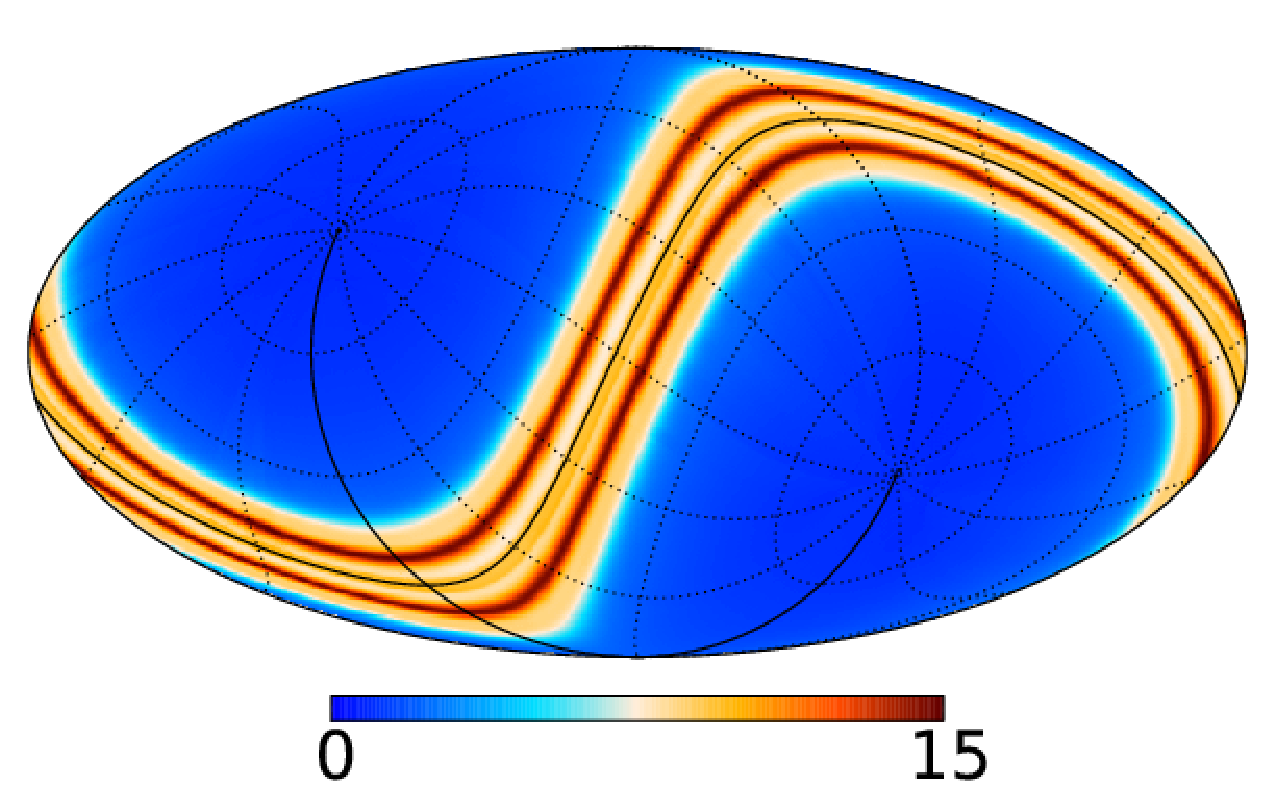}\includegraphics[clip,width=0.33\textwidth]{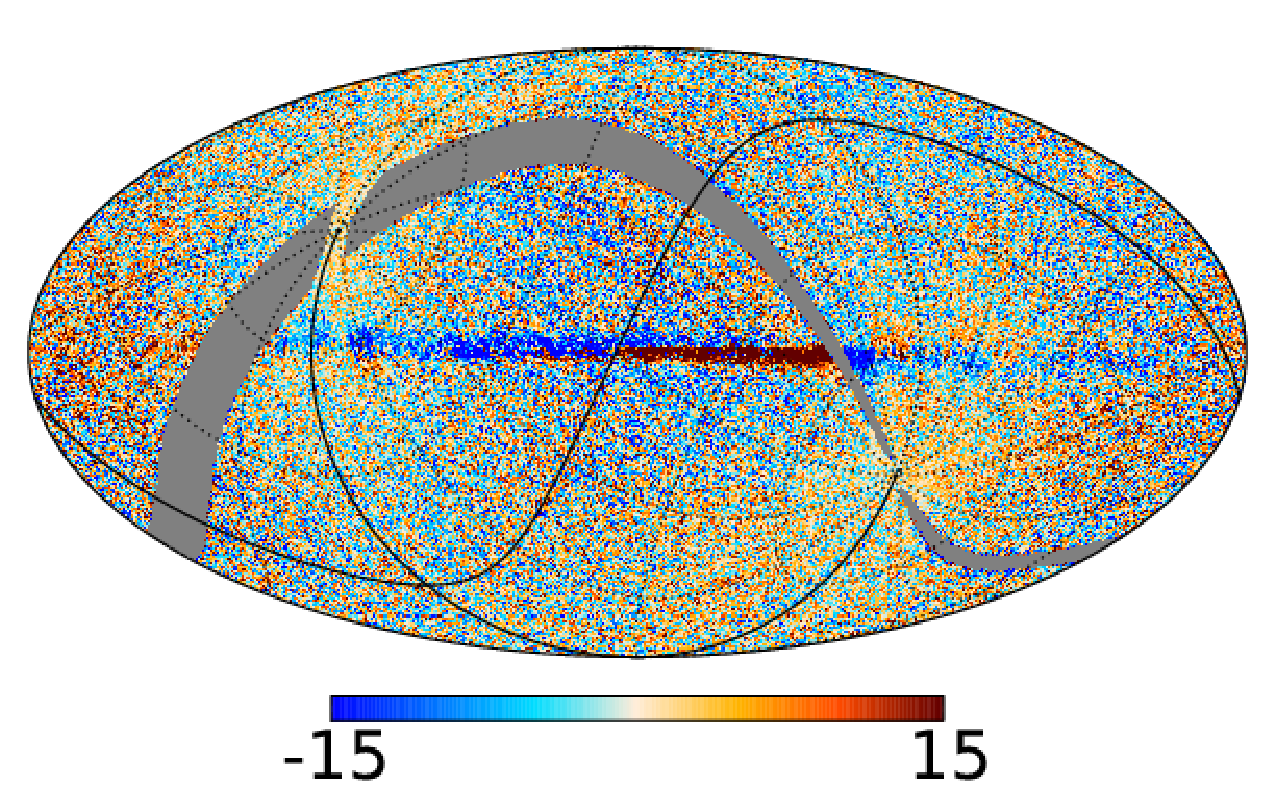}
\par\end{centering}

\begin{centering}
\includegraphics[clip,width=0.33\textwidth]{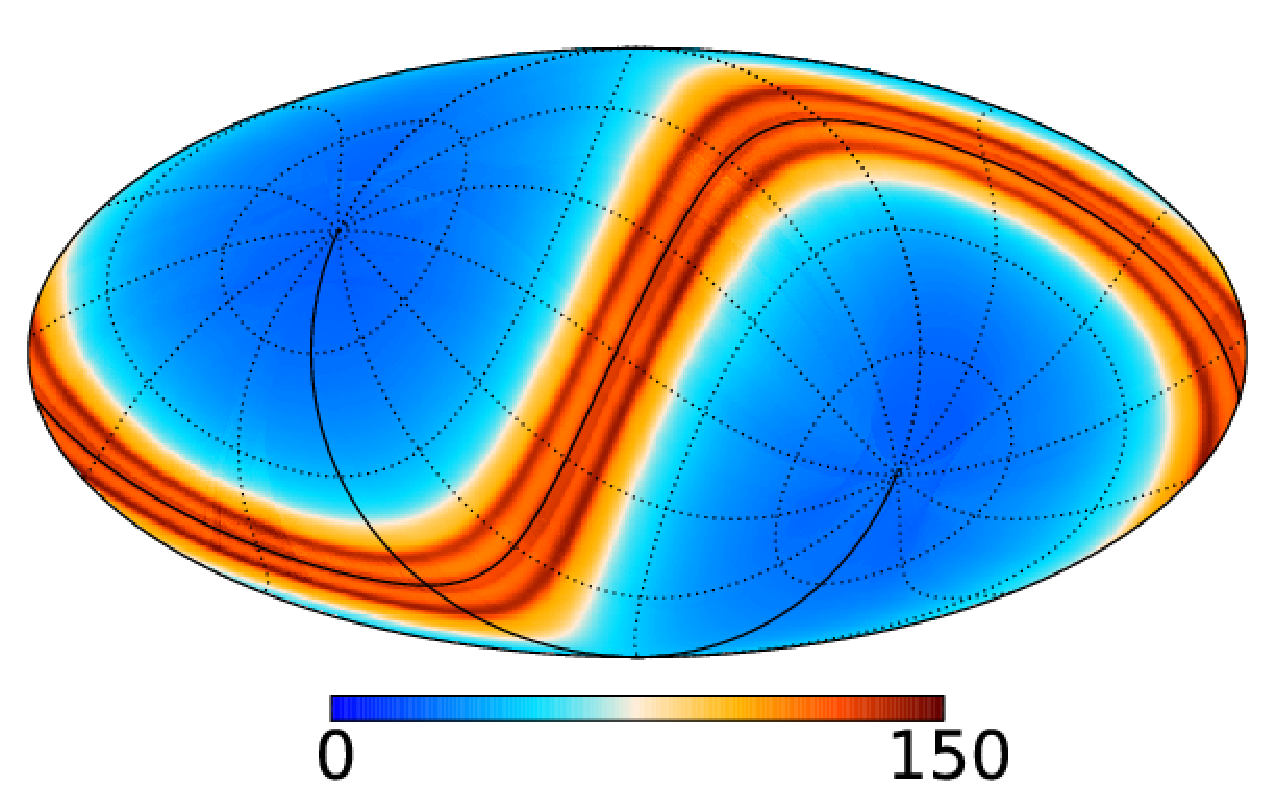}\includegraphics[clip,width=0.33\textwidth]{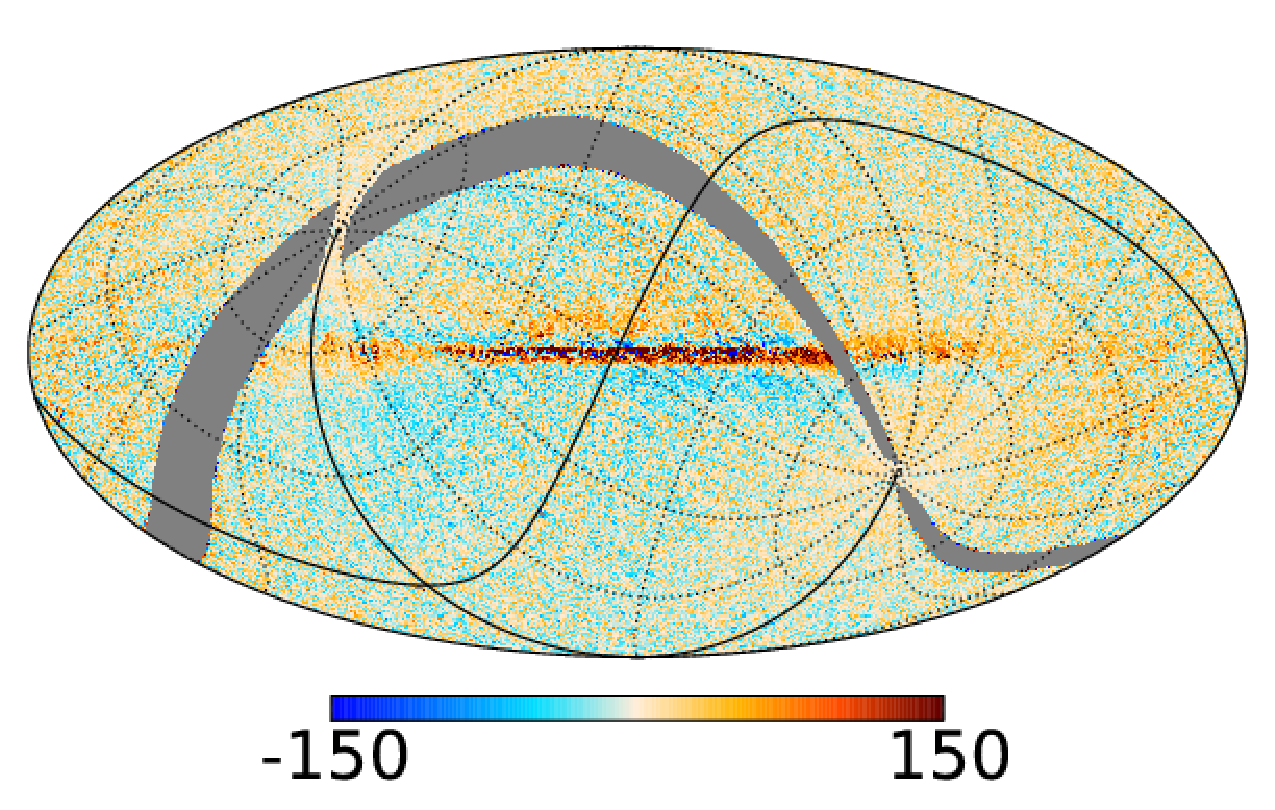}
\par\end{centering}

\begin{centering}
\includegraphics[clip,width=0.33\textwidth]{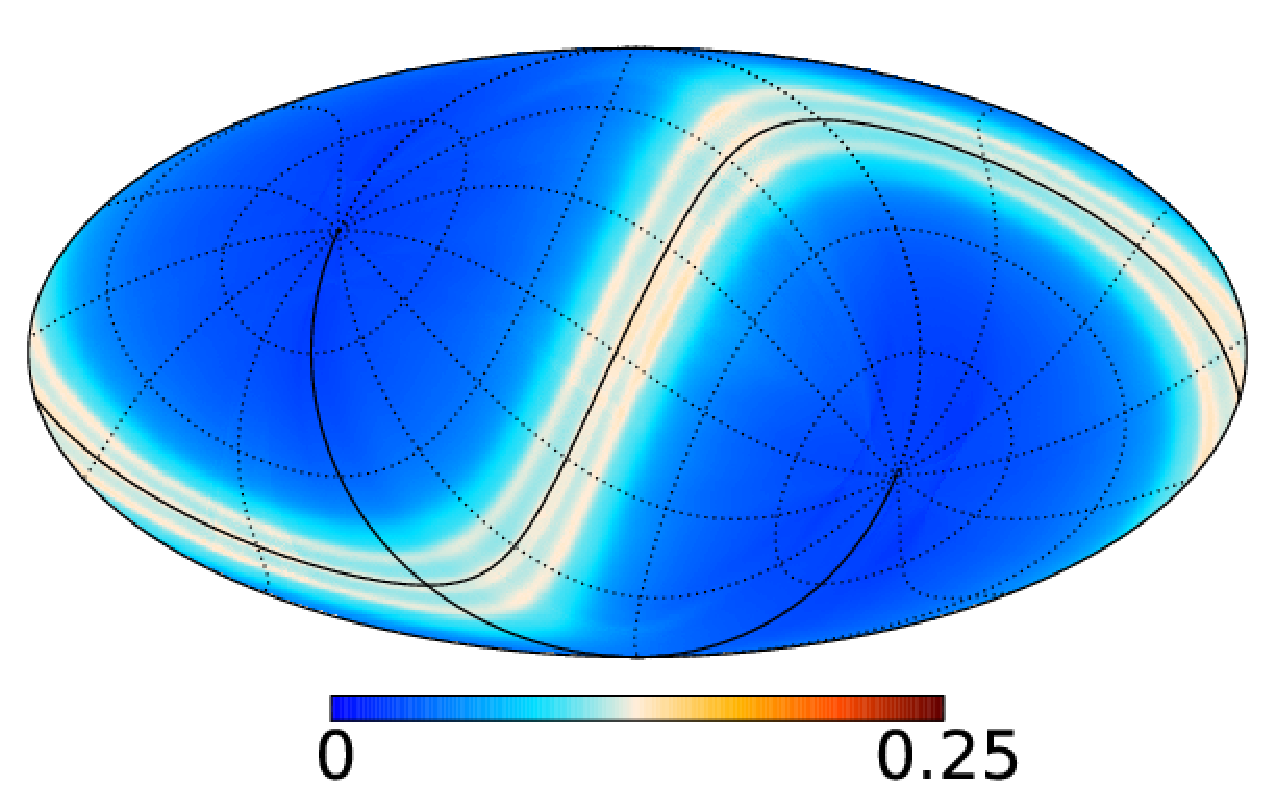}\includegraphics[clip,width=0.33\textwidth]{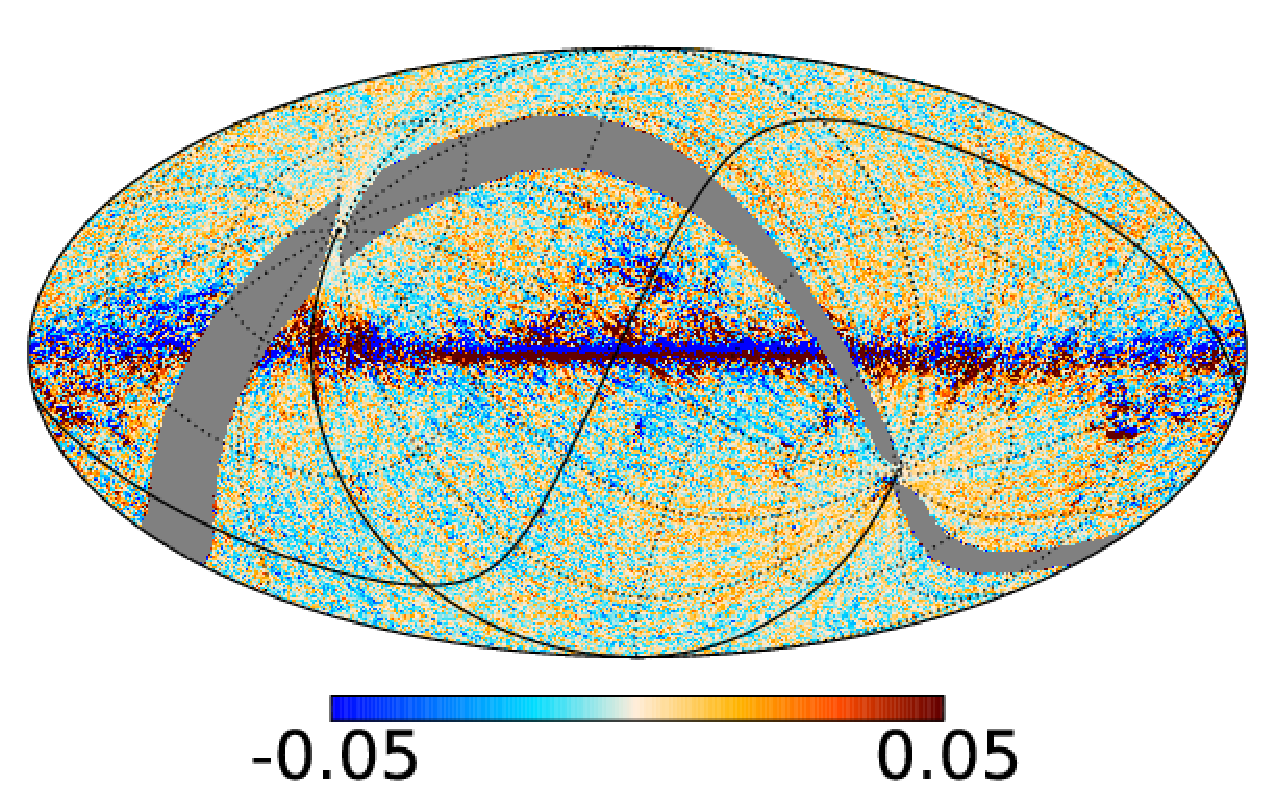}
\par\end{centering}

\begin{centering}
\includegraphics[clip,width=0.33\textwidth]{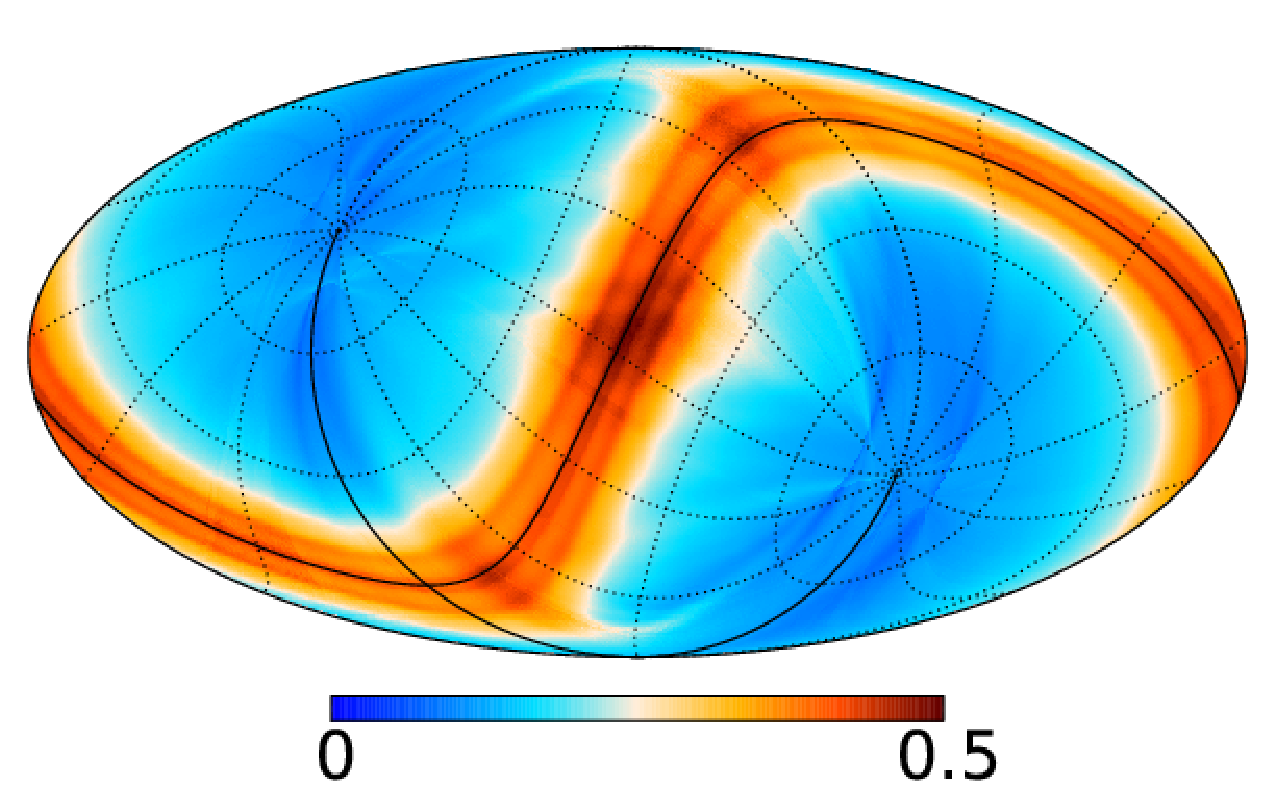}\includegraphics[clip,width=0.33\textwidth]{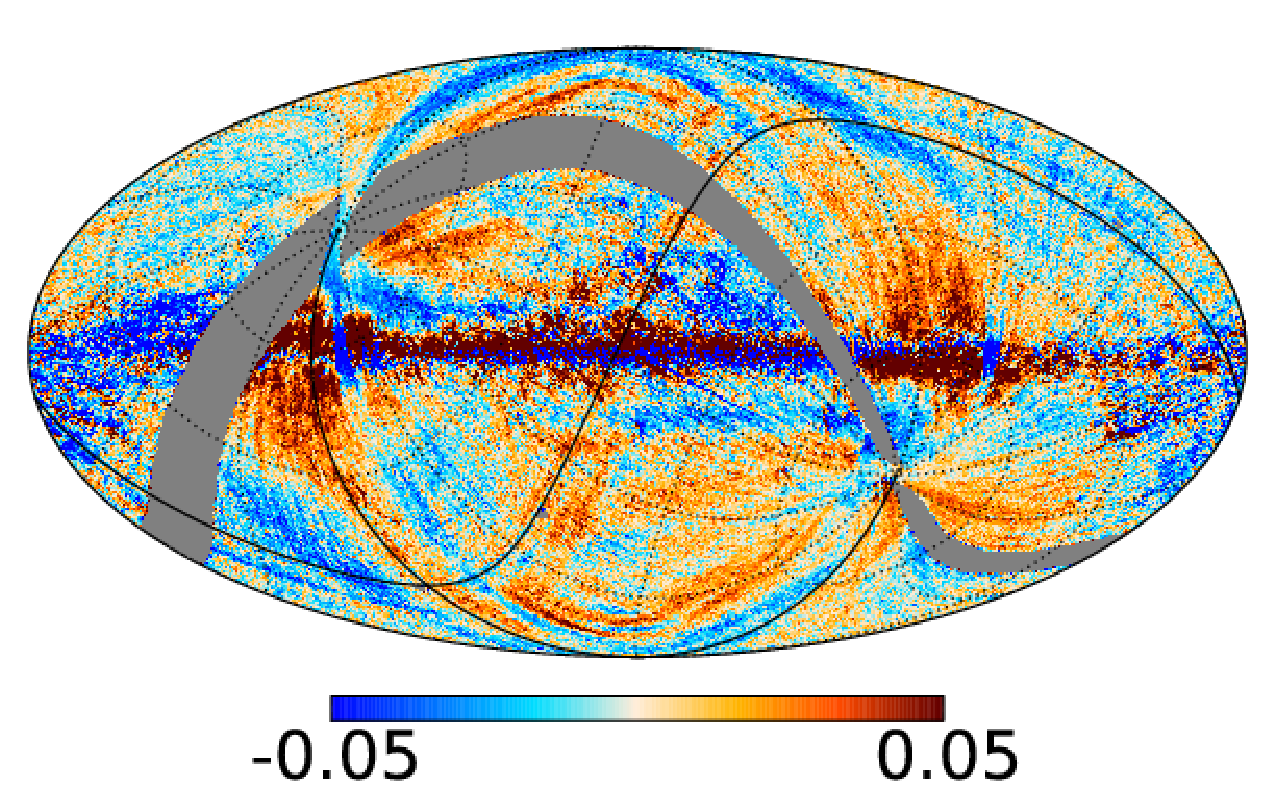}
\par\end{centering}

\centering{}\protect\caption{\label{fig:zodCor} The zodiacal light contribution in HFI channels,
with 100\,GHz to 857\,GHz from top to bottom. Units are $\mu\mathrm{K_{CMB}}$
for 100--353\,GHz\ and MJy\,sr$^{-1}$ for 545 and 857\,GHz. \emph{Left
column}: Estimate of the correction. \emph{Right column}: The Survey
1$-$Survey 2 difference map after correction, which can be compared
with the bottom row of Figs.~\ref{fig:Imaps-100} to \ref{fig:Imaps-857}.
While there is no visible difference in the CMB channels, note the
improvement to the 353--857\,GHz maps. }
\end{figure*}

Most of the signals \Planck\  observes on the sky are fixed to the
celestial sphere. There are, however, two exceptions: zodiacal light
emission (ZLE), and far sidelobe (FSL) contamination. Because \Planck\ 
surveys the sky twice per year with a ``cycloidal'' scanning strategy
\citep{dupac2005}, each time a given location on the (distant) sky
is observed, a different column of nearby dust in our Solar System
is sampled, leading to a slightly different ZLE (see \citealt{planck2013-pip88}).
The signature of zodiacal emission is indicated by the ``S''-shaped
band along the Ecliptic plane seen in the 857\,GHz difference between
maps made six months apart (bottom row of Fig.~\ref{fig:Imaps-857}),
when \Planck\  is located at antipodal points in its orbit and therefore
looking through very different columns of zodiacal dust. Figure~\ref{fig:zodCor}
shows the sky pattern of the ZLE model in each of the HFI bands.

In addition, there is a small, asymmetric, off-axis beam response
that also leads to additional time variability. This is seen as the
arcs near the north and south Galactic poles in Fig.~\ref{fig:Imaps-857}.
Both of these are most pronounced in the highest-frequency channels.
We remove these signals so that further Galactic and extragalactic
studies will not suffer from interplanetary dust emission.

The removal process is described in more depth in \citet{planck2013-pip88}.
It consists of fitting survey maps from each horn to the \emph{COBE}
zodiacal emission model \citep{kelsall1998} to find the emissivities
at the HFI wavelengths, plus templates for the Galactic and dipole
signal seen through a model of the far sidelobes. With the amplitudes
for each template from this fit, we again turn to the \emph{COBE}
and far-sidelobe models to reconstruct the implied zodiacal emission
at all times during the survey for each individual detector. We convert
the values to $\mu\mathrm{K}_{\mathrm{CMB}}$, and remove these signals
before the data are combined into maps.

\begin{figure}[th]
\centering{}\includegraphics[clip,width=1\columnwidth]{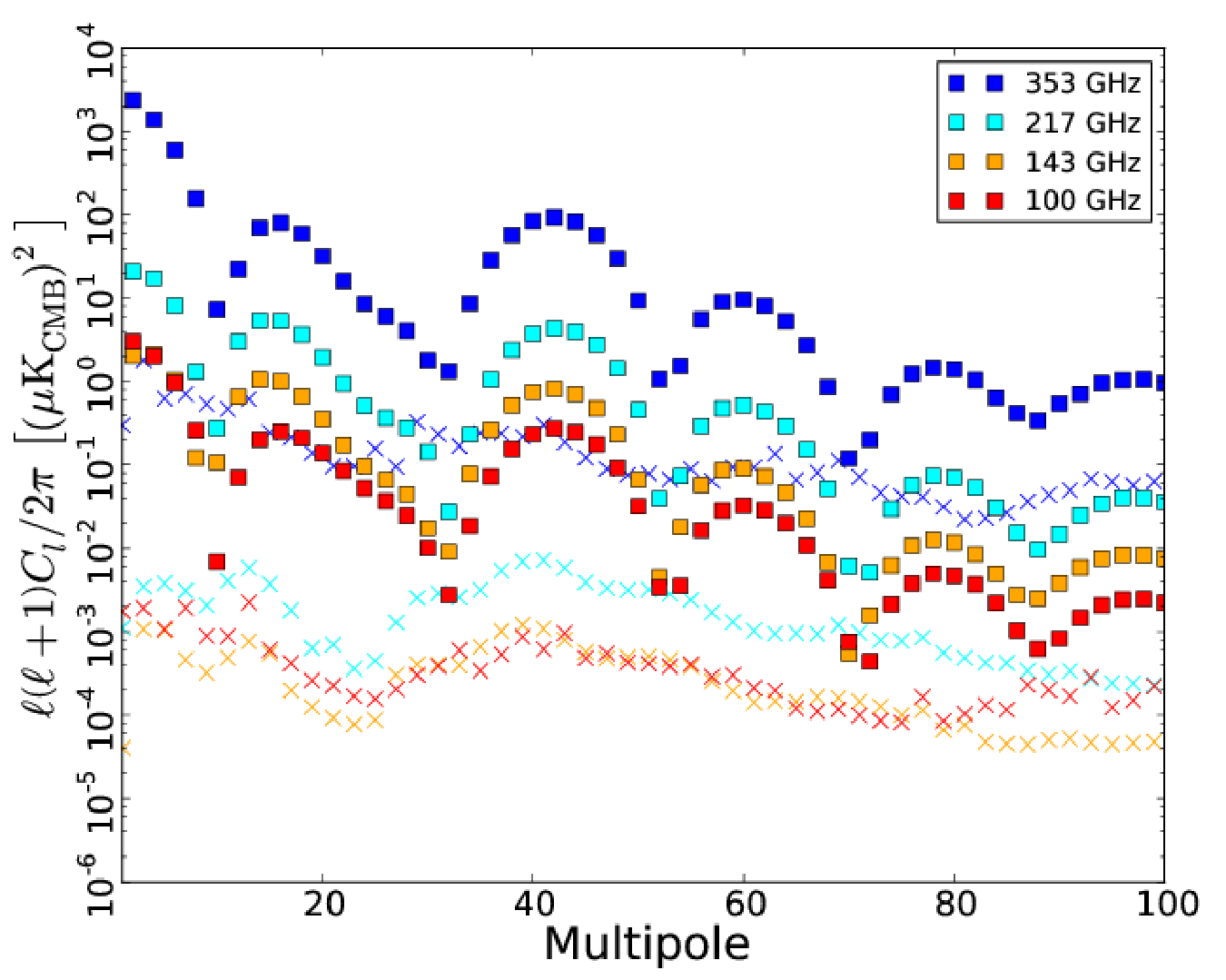}\protect\caption{\label{fig:zodCl}Angular power spectrum of the zodiacal light contribution
from each channel. Even multipoles are shown with filled boxes, odd
multiples with crosses. The systematic difference reflects the approximate
north/south symmetry of the zodiacal contribution. Spectra of other
channels are given in \citet{planck2013-pip88}.}
\end{figure}
The fit coefficients for the zodiacal emission are presented in \citet{planck2013-pip88},
as are representative maps of the amount of signal removed from each
map, and the power spectra of these maps. At 143\,GHz, the zodiacal
emission consists essentially of a band of emission between Ecliptic
latitudes of roughly $\pm15\deg$, reaching a maximum of 5\,$\mu\mathrm{K}_{\mathrm{CMB}}$.
The power spectrum at 143\,GHz is small: it appears only at multipoles
less than roughly 75, and expressed as $\left(2\pi\right)^{-1}\ell\left(\ell+1\right)C_{\ell}$
is everywhere less than $2.5\,\mu\mathrm{K_{CMB}^{2}}$ for all multipoles
(and even smaller if multipoles are binned).

Maps are made available both with and without ZLE removal. In addition,
as there is a detection of the Galaxy seen through the far sidelobes
in some horns at 857 and 545\,GHz, these signals are removed as well,
for these frequencies only, along with ZLE removal. Figure~\ref{fig:zodCor}
shows the sky map of the signal removed, and Fig.~\ref{fig:zodCl}
shows its power spectrum at 100--$353\GHz$, separately for odd and
even multipoles, reflecting the approximate north/south symmetry of
the pattern. The correction is much less uncertain at the three highest
frequencies, but the low-frequency spectra allow us to gauge the rough
level of the contribution to the low-$\ell$ power spectra in uncorrected
maps. At $\ell \ge 50$ the ZLE contribution in uncorrected maps is
smaller than $1\:\mu\mathrm{K_{\mathrm{CMB}}^{2}}$ and can thus be
neglected. \textcolor{black}{But in order to extract the CMB, we found
that one cannot use only a partial correction of the ZLE at the frequencies
where it is well determined, i.e., at 353\,GHz and higher frequencies,
since lower-amplitude ``S''-shaped residuals (in the CMB map) are
found when the corrected maps are used. Nevertheless, overall, we
discovered that the best CMB cleaning was achieved by using the full
multi-component CMB and foreground separation process itself on maps
not otherwise explicitly corrected for the ZLE. This is addressed
in more detail in \citet{planck2013-p06}, but the main reason can
be seen right away: the relative weights of the maps that are used
to generate a CMB map are found to be the same when one set of HFI
maps or the other is used, and for the }\texttt{NILC}\textcolor{black}{{}
component-separation technique described in \citealp{planck2013-p06}
are given by: 30\,GHz: $-0.01$; 40\,GHz: $-0.04$; 70\,GHz: $-0.16$;
100\,GHz: $-0.22$; 143\,GHz: 1.2; 217\,GHz: 0.33; 353\,GHz: $-0.11$;
545\,GHz: $3\times10^{-4}$; and 857\,GHz: $1.8\times10^{-5}$ .
The typical peak-to-peak amplitude of the zodiacal correction is:
70\,GHz: $6\,\mu\mathrm{K}^{2}$; 100\,GHz: $6\,\mu\mathrm{K}^{2}$;
143\,GHz: $5\,\mu\mathrm{K}^{2}$; 217\,GHz: $15\,\mu\mathrm{K}^{2}$;
353\,GHz: $150\,\mu\mathrm{K}^{2}$; 545\,GHz: $2400\,\mu\mathrm{K}^{2}$;
and 857\,GHz: $2\times10^{-5}\,\mu\mathrm{K}^{2}$. One sees that
with the above weights the 353--857\,GHz contribution roughly cancels
the 143--217\,GHz one, while lower frequency channels have both a
smaller zodiacal contribution and a smaller weight in determining
the CMB map. }

\subsection{CO correction\label{sec:Correct-CO} }

Some emission from the rotational transition lines of CO is present
in the HFI bands. It is especially significant in the 100, 217, and
353\,GHz channels, due to the 115, 230, and 345\,GHz CO transitions.
This emission comes largely from the Galactic interstellar medium,
mainly located at low and intermediate Galactic latitudes ($\left|b\right|<20\,^{\circ}$).
Because of the wide spectral coverage of \Planck\  \citep{planck2013-p03d},
we are able to produce velocity-integrated CO line maps for the first
three \rev{transitions} $J$=1$\rightarrow$0, 2$\rightarrow$1, and 3$\rightarrow$2.
A complete description of the methods used, the validation tests,
and the CO products are given in \citet{planck2013-p03a}. These all-sky
CO maps can be used for astrophysical studies of the interstellar
medium%
\begin{comment}
 (e.g., \textcolor{red}{{[}REF IN DRAFT FORM planck2013-pip58 CO{]}})
\end{comment}
{} or in component separation methods.

Given the fraction of sky used for the estimation of cosmological
parameters ($f_{\mathrm{sky}}\simlt0.5$) and the fact that CO is
significant over less than $1\,\%$ of the sky for these high Galactic
latitudes, simply masking the CO-emitting regions is the course of
action taken in cosmological studies with \Planck.

\begin{figure}[th]
\centering{}\includegraphics[clip,width=0.95\columnwidth]{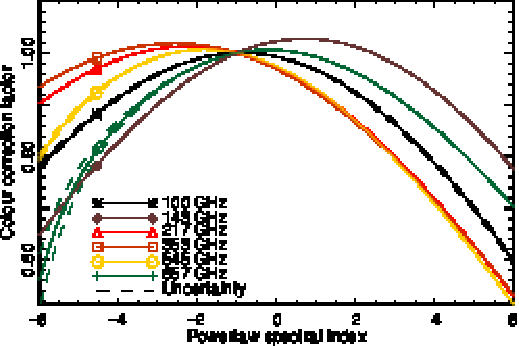}\protect\caption{\label{fig:colour_corrections} Band-averaged colour correction for
various power-law frequency spectra, normalized to spectral index
$-1$. The dashed lines provide estimates of the uncertainty of these
corrections. }
\end{figure}

\subsection{Spectral bands and colour corrections\label{sec:Correct-color}}

The HFI detector spectral response was determined through a series
of ground-based tests conducted with the actual HFI focal plane in
a simulated space environment, i.e., evacuated and cryogenic, prior
to launch. During these tests, the HFI focal plane was coupled to
a broadband radiation source observed through a continuously scanned
polarizing Fourier transform spectrometer (FTS). The individual HFI
detector responses to the modulated FTS signal, i.e., the observed
interferograms, were recorded alongside that of a reference bolometer
located inside an integrating sphere within the experimental setup.
Multiple scans were averaged together to increase the signal-to-noise
ratio of the resultant spectra and to allow an estimate of the spectral
uncertainty. Standard FTS data processing techniques were employed
to obtain the observed spectra of each HFI detector and that of the
reference bolometer; the ratios of the detector spectra against the
reference bolometer spectrum provide the relative response of each
HFI detector. This method incorporates the detector throughput (i.e.,
étendue $A\Omega$, see \citealt{planck2013-p03d}) in the spectral
response as the reference bolometer accepts a much wider field of
view, i.e., approximately $2\pi$\,sr over all frequencies. The HFI
detector spectral transmission data, and their frequency band-average
transmission spectra, are described more fully in \citet{planck2013-p03d}
and \citet{planck2013-p28}, where a description of the band-averaging
scheme is also provided.

\begin{figure*}[th]
\centering{}\includegraphics[bb=0bp 0bp 508bp 337bp,width=0.97\textwidth]{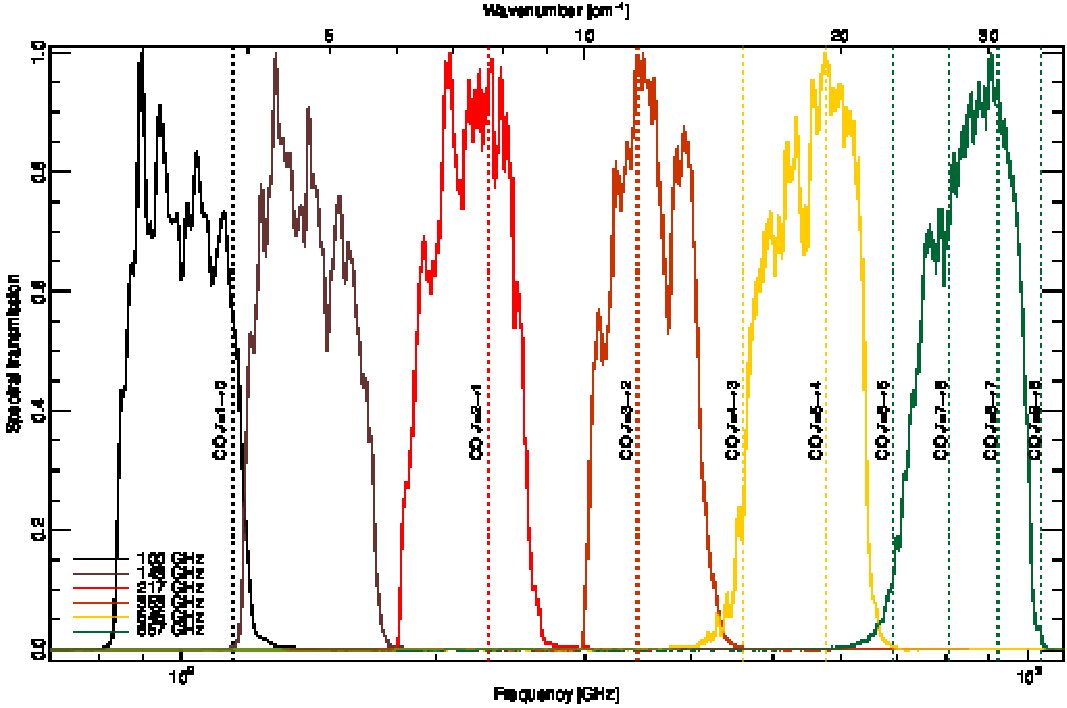}
\protect\caption{Band-averaged transmission spectra for all HFI frequency bands. The
locations of the various relevant CO transitions are marked with vertical
lines.\label{fig:HFIspec} }
\end{figure*}

As discussed in Sect.~\ref{sec:Abs-calib}, HFI is calibrated using
the CMB dipole for the four 100--353\,GHz bands, expressed in units
of $\mathrm{K_{\mathrm{CMB}}}$, and using planets for the two 545--857\,GHz
bands, expressed in units of W\,m$^{-2}$\,sr$^{-1}$\,Hz$^{-1}$
(equivalently, MJy\,sr$^{-1}$). This unit conversion is defined
to translate data between units of $\mathrm{K_{\mathrm{CMB}}}$ and
MJy\,sr$^{-1}$, the latter for spectra with a profile consistent
with a spectral index of $-1$, following the \emph{IRAS} convention
\citep{IRASExSup}. Colour-correction coefficients are provided for
a variety of spectral indices (see Fig.~\ref{fig:colour_corrections})
and other common spectral profiles including modified blackbodies
and planetary spectra. It is thus important to ensure that data processing,
e.g.,\ component separation, is conducted in a consistent manner
in units appropriate for the frequency band. The HFI policy is to
operate in units of MJy\,sr$^{-1}$, as this is valid for all of
the HFI frequency bands whereas $\mathrm{K_{\mathrm{CMB}}}$ is only
valid for 100--353\,GHz. The $\mathrm{K_{\mathrm{CMB}}}$ to MJy\,sr$^{-1}$
unit conversion coefficients are therefore given by the following
ratios 
\begin{equation}
U(\mathrm{K}_{\mathrm{CMB}}\textrm{ to }\mathrm{MJy\, sr^{-1}})={\displaystyle \frac{{\displaystyle \int\!\! d\nu~\tau(\nu)b_{\nu}'}}{{\displaystyle \int\!\! d\nu~\tau(\nu)(\nu_{\mathrm{c}}/\nu)}}\times10^{20}\;\left[\frac{\mbox{MJy\,}\mbox{sr}{}^{-1}}{\mathrm{K_{\mathrm{CMB}}}}\right]\;,}\label{eq:KCMB_MJysr}
\end{equation}
where $\tau(\nu)$ is the relevant spectral transmission curve, $b_{\nu}'=dB_{\nu}/dT|_{T_{\mathrm{CMB}}}$
is the Planck function derivative, and $\nu_{\mathrm{c}}$ is the
nominal band centre frequency.

Other unit conversions are also provided, to $y_{\mathrm{SZ}}$ (the
Sunyaev-Zeldovich Compton parameter, discussed in \citealp{planck2013-p05b}),
K\,km\,s$^{-1}$ (CO transmission; see \citealt{planck2013-p03a}),
and K$_{\mbox{\tiny{RJ}}}$ (brightness temperature). It is important
to note that brightness temperature is not restricted to a Rayleigh-Jeans
spectral profile, but is a convention for an alternate expression
of an arbitrary flux density in temperature units \citep[e.g., ][]{1986RybickiLightman}.

The HFI unit conversion and colour correction philosophy expresses
equivalent flux density at a predefined nominal frequency depending
on the spectral profile in question, rather than maintaining a set
signal amplitude at a varying effective frequency, which would also
depend on the spectral profile in question. This is primarily done
to ease the requirements on component separation, since having all
components expressed at the same effective frequency for all detectors
within a frequency band is much preferred over dealing with a variety
of different effective frequencies for each detector, and each component. 

More details on the unit conversion factors are provided in \citet{planck2013-p03d},
while tables of conversion factors for \Planck\  detectors (only
HFI channels for some spectral shapes, e.g., CO lines) are provided
in \citet{planck2013-p28}. Figure~\ref{fig:HFIspec} illustrates
the band-averaged spectra for each of the HFI bands; similar, and
related, plots for individual HFI detectors are provided in \citet{planck2013-p28}.

A series of IDL scripts, called the \texttt{UcCC} routines, is available
to allow users of \Planck\ data to determine unit conversion and
colour correction coefficients, and uncertainties, for user-specified
profiles, in addition to those nominally provided in the public \IMO.
Further details on these scripts are provided in \citet{planck2013-p28}. 

\begin{figure*}[t]
\begin{centering}
\includegraphics[width=1\columnwidth]{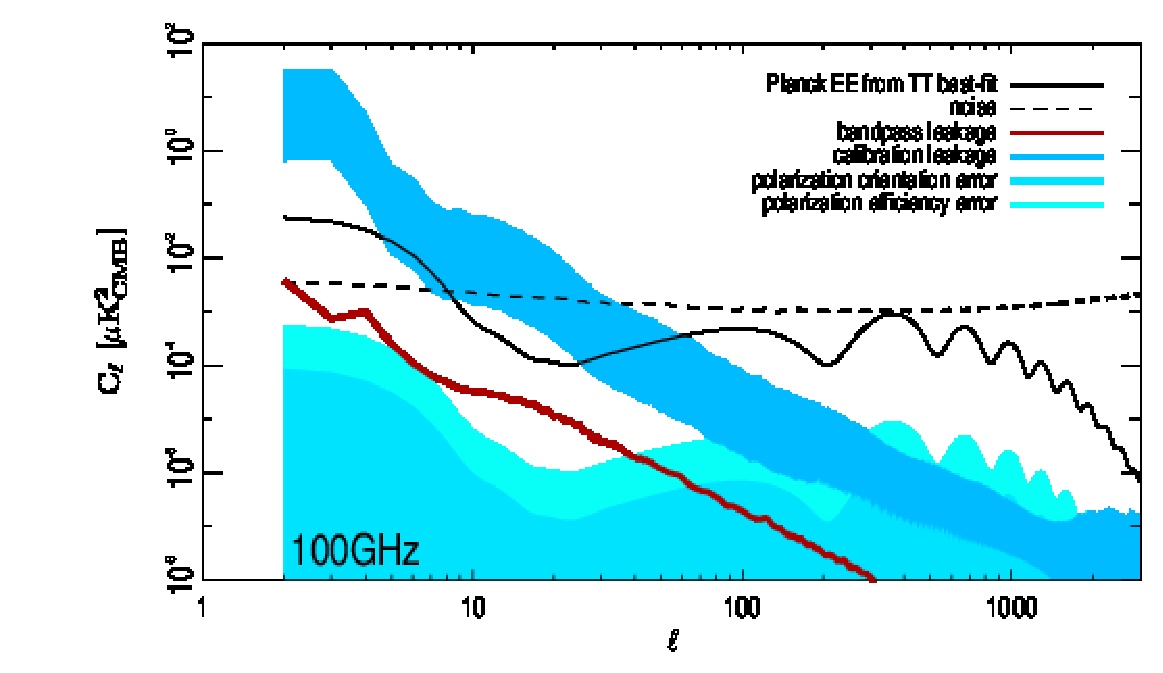}\includegraphics[width=1\columnwidth]{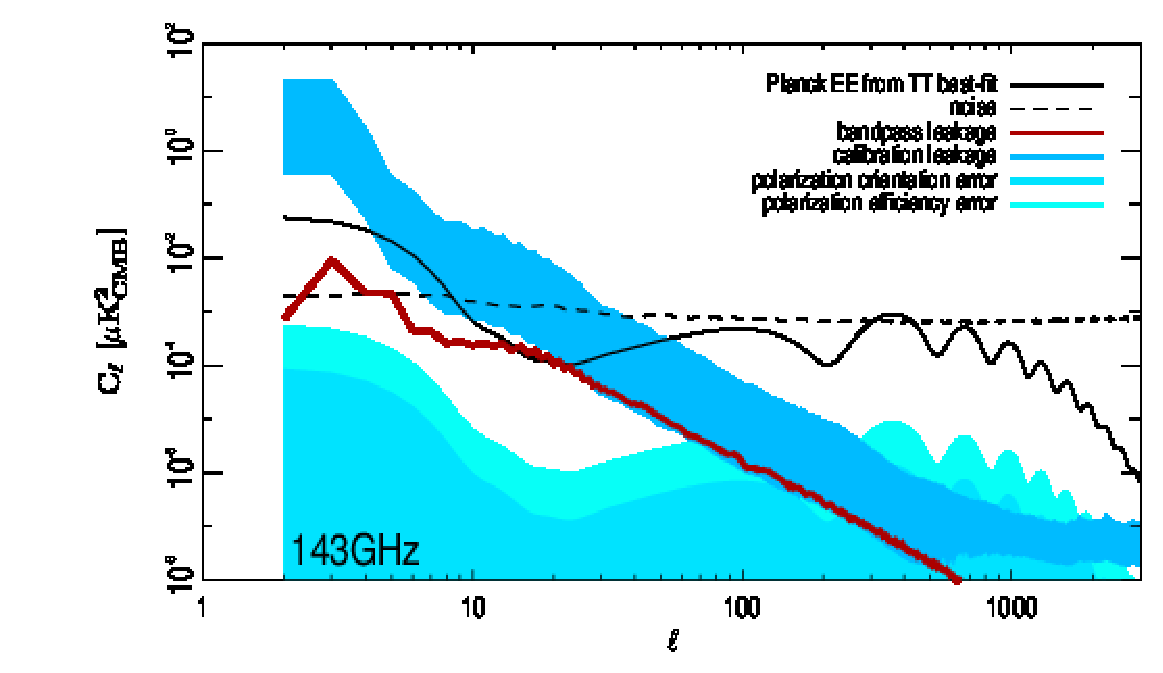}
\par\end{centering}

\begin{centering}
\includegraphics[width=1\columnwidth]{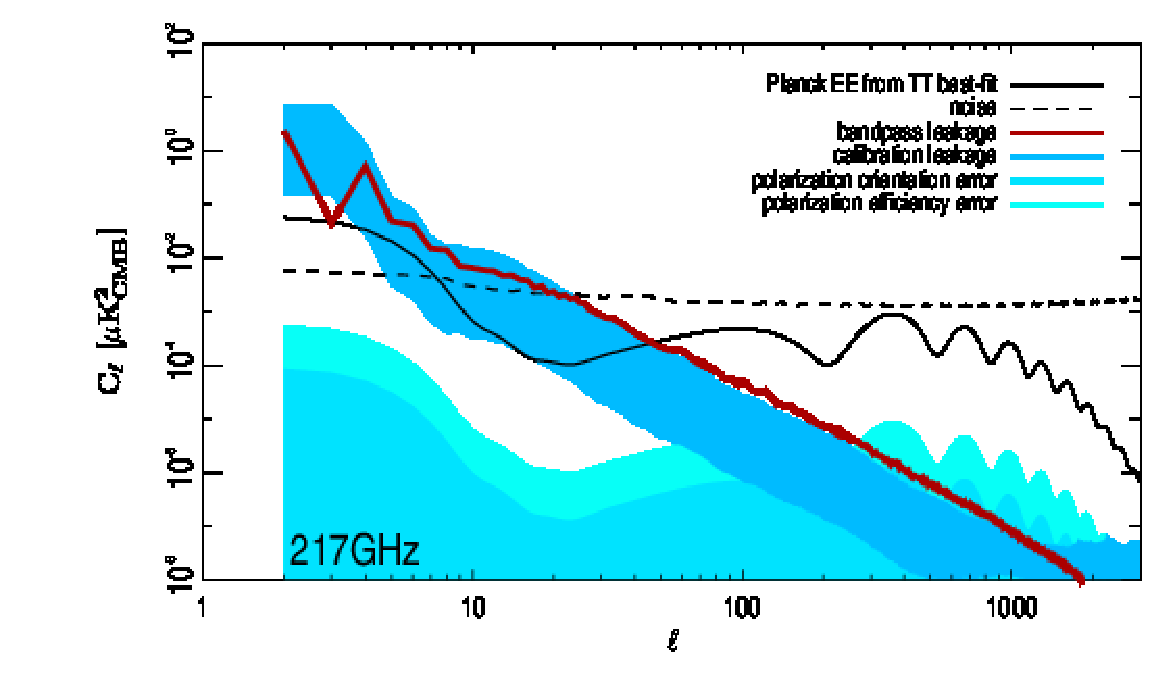}\includegraphics[width=1\columnwidth]{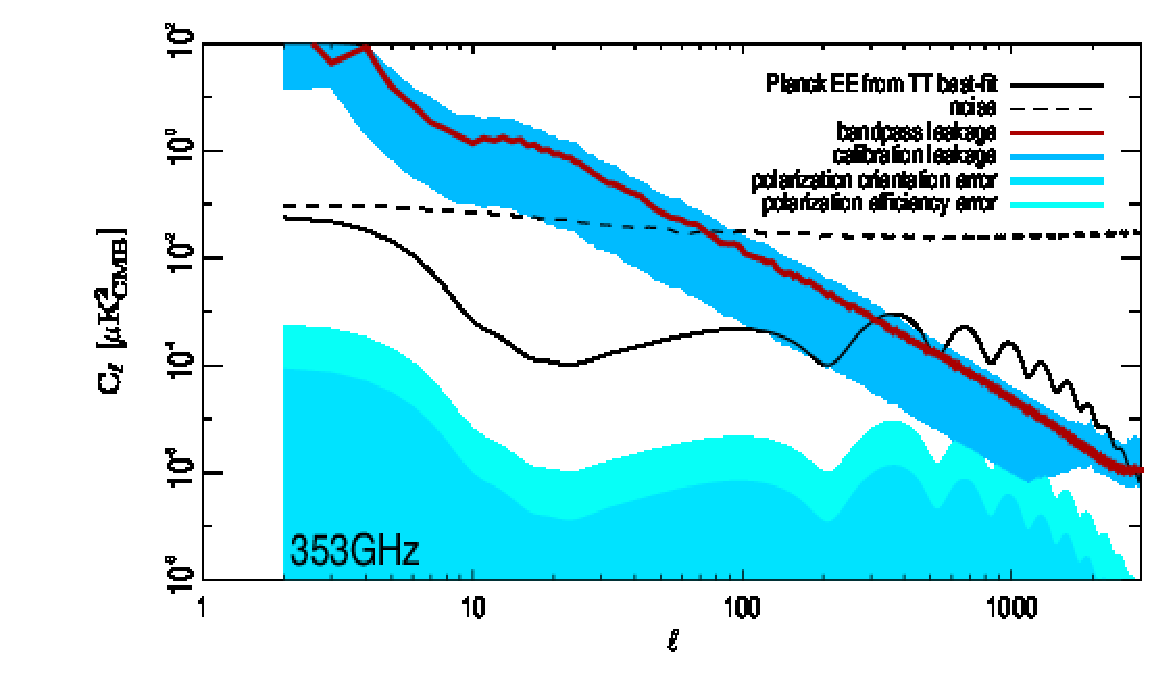}
\par\end{centering}

\protect\caption{Uncertainties on polarized power spectra due to residual systematics
in the HFI polarization maps, compared to the $EE$ spectrum predicted
for the best-fit model from \Planck\ temperature data.\label{fig:polar_sys}
}
\end{figure*}

\subsection{Polarization\label{sec:Polar}}

Together with the intensity maps, we derive polarization maps for
each HFI frequency from 100\,GHz to 353\,GHz \citep[cf. ][]{planck2013-p03f}.
The $Q$ and $U$ Stokes parameter maps \rev{have a nearly white noise spectrum and are dominated by Galactic dust emission and CMB with a signal-to-noise ratio of roughly 100 for $\ell<100$.} Nevertheless,
at the present state of the reconstruction, polarization maps are
dominated at large scales by systematic effects. As a consequence,
they are not included in the present release. However, these data
will be used in forthcoming analyses of foreground polarization, and
hence we describe the major systematic effects remaining in the polarization
data. 

The \Planck\ scanning strategy does not allow recovery of the polarization
signal for each detector independently. We use a combination of at
least four detectors oriented at 45\deg\ in order to reconstruct
the $Q$ and $U$ Stokes parameters. Any miscalibration between the
data sets will thus lead to a mix between Stokes parameters and, in
particular, a leakage from intensity to polarization. As discussed
in \citet{planck2013-p03f}, absolute calibration is derived from
the solar dipole and gain variations due to ADC nonlinearities are
corrected to a few tenths of a percent in the time domain. Final intercalibration
uncertainties between HFI bolometers are of the order of 0.4\,\%
or less for the three lowest HFI frequencies (0.39, 0.28, and 0.21\,\%
respectively at 100, 143, and 217\,GHz) and rise to 1.35\,\% at
353\,GHz. The induced leakage is dominated by the solar dipole intensity,
which produces large-scale features in the polarization maps. We have
estimated the leakage due to calibration mismatch using simulations.
Figure~\ref{fig:polar_sys} shows that the leakage is dominant over
the CMB signal up to $\ell=100$ in the $EE$ angular power spectrum,
as predicted by the best-fit cosmological parameters measured from
\Planck\ temperature data \citep{planck2013-p11}.

In addition to the calibration mismatch, differences in bolometer
spectral transmissions will also introduce a leakage from dust emission
intensity into polarization. Using the spectral transmission as measured
on the ground \citep{planck2013-p03d}, we estimated the level of
leakage introduced by bandpass mismatch, as shown in Fig.~\ref{fig:polar_sys}.

Polarization angles and efficiency have been assessed during ground
measurements. Values and uncertainties are reported in \citet{rosset2010}.
We derive uncertainties on angular power spectra using simulations
for both polarization angle and efficiency independently.

Figure~\ref{fig:polar_sys} shows the level of specific sources of
error in $EE$ polarization angular power spectra compared to the
CMB anisotropies and the noise level (estimated from half-ring differences)
computed on 90\,\% of the sky, excluding the Galactic plane. In the
polarized HFI bands (100, 143, 217, and 353\,GHz) systematic errors
dominate over the CMB signal at low multipoles ($\ell<100$) due to
miscalibration leakage and bandpass mismatch leakage. Polarization
angle and polarization efficiency uncertainties are second-order effects.
For Galactic emission polarization studies, the signal is more than
an order of magnitude more intense. At higher multipoles ($\ell\simgt200$),
HFI polarization maps are noise-dominated but there is no evidence
for correlated noise from channel to channel.

\section{Data consistency and validation\label{sec:VALIDATION}}

\subsection{Consistency with a severe selection of data}

Another check based on real data consists of comparison with data
in which the pointing periods have been much more severely censored.
As mentioned in Sect.~\ref{sec:Toi-Qualification}, the \rev{normally}
discarded pointing periods represent less than 1\,\% of the total
integration time for the full mission and 0.2\,\% for the nominal
mission, the difference being due to \rev{some} solar flare events
\rev{and the HFI end-of-life operations } during the fifth survey. 

\rev{NB: the next 2 paragraphs and the figure were revised.}

\begin{figure}[h]
\centering{}\includegraphics[clip,width=0.95\columnwidth]{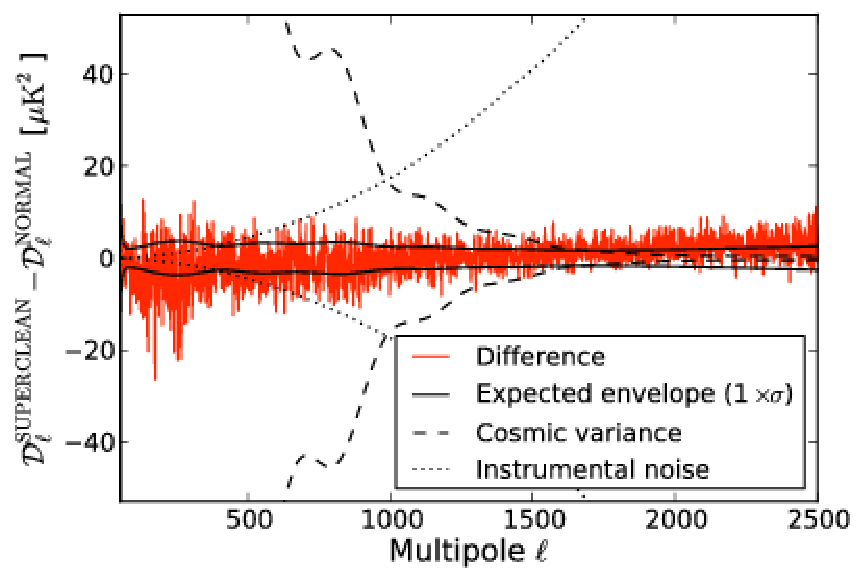}\protect\caption{\label{fig:difference_143_no_bins} The difference of the pseudo-spectra
for the normal and the severe selections (in red) for the 143~GHz
case. This difference is negligible with respect to both the cosmic
variance and the instrumental noise (computed from FIRST$\times$LAST
half-ring differences). No binning is applied.}
\end{figure}

Following much more severe criteria, 30~\% of the data of the full
mission are rejected. For example, very conservative thresholds on
the stationarity of the noise are applied (see the details in Appendix~\ref{sec:SUPERCLEAN}).
We then create a map at each frequency with this selection and compute
the pseudo-spectra of both the normal and the severe selections. We
show in Fig.~\ref{fig:difference_143_no_bins} the difference between
those two pseudo-spectra (red curve) and its expected 1\,$\sigma$
envelope, as well as the cosmic variance and the instrumental noise.
This difference computed at each multipole is effectively centred
around zero and shows a very low absolute value compared to cosmic
variance or instrumental noise (even in the {[}2--500{]} $\ell$-range
where it remains below a few percent).

Some departures from ideality can be seen with the severe selection,
e.g., a slight tilt in the difference \revv{spectra},
and a higher dispersion than expected, especially at low $\ell$,
but these features are also present in the checks we performed. These
comparisons with respect to the expected dispersion are detailed in
Appendix~\ref{sec:SUPERCLEAN}. \revv{This brings evidence that the level of non-stationarity of the noise during the nominal mission is negligible, and that the tilt induced by the data selection is more likely due to the variation in the structure of the hit map than to a variation in the level of a residual systematic effect}. 

\subsection{Difference map consistency tests\label{sec:Consistency-diff-map} }

\begin{figure*}[t]
\centering{}\includegraphics[width=1\columnwidth]{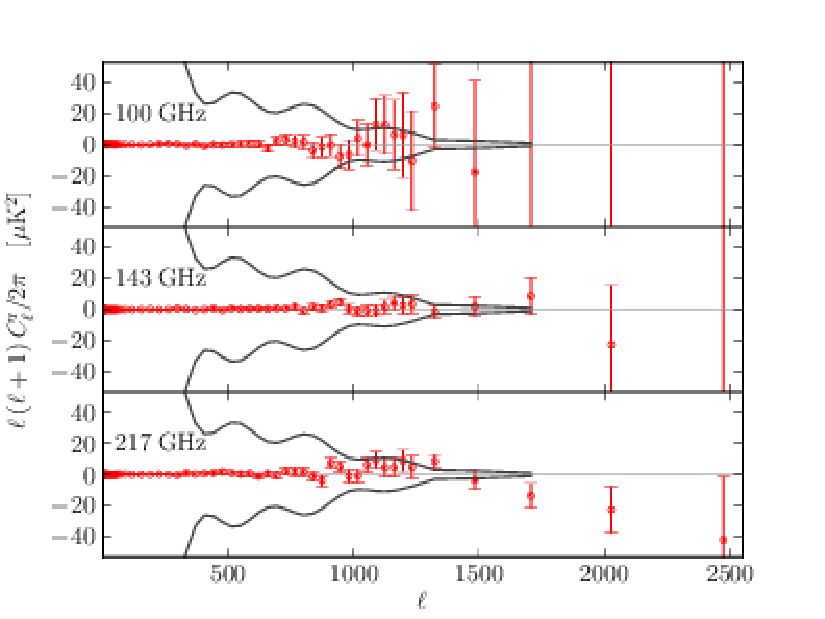}\includegraphics[width=1\columnwidth]{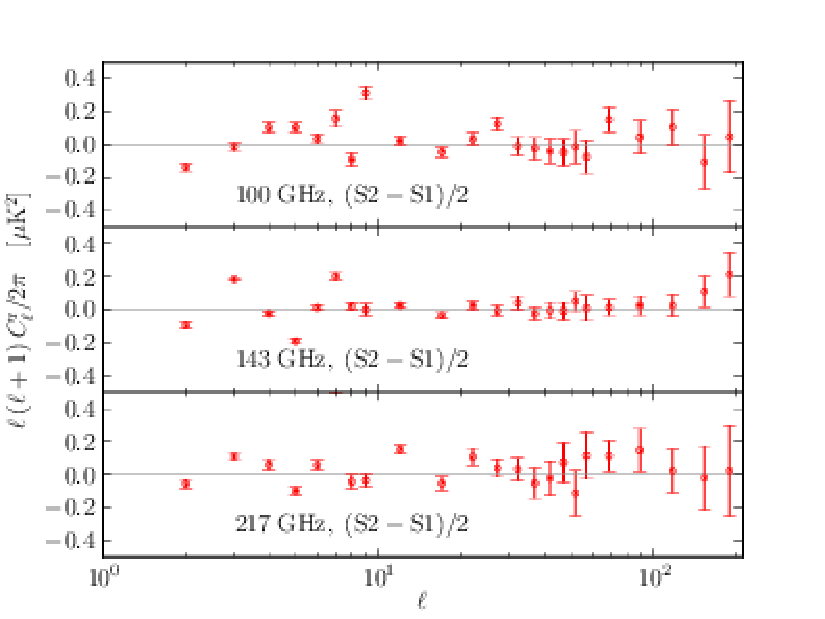}\protect\caption{\label{fig:survey_diff}(Survey 2 $-$ Survey 1)$/2$ consistency
test result from the ds1 $\times$ ds2 cross-spectra at 100, 143,
and 217\,\GHz, with the $f_{\mathrm{{sky}}}\simeq0.30$ mask derived
from that used in the primary cosmological analysis (see text). \emph{Left}:
The residual signal in each bandpower up to the highest multipole
used in the likelihood code \citep{planck2013-p08}. For clarity,
the bins used in the likelihood analysis are grouped in sets of four
for $\ell>60$ and sets of eight for $\ell>1250$. We show in each
bin the variance of the $C_{\ell}$ distribution from the simulations.
Note that the amplitude of the residual signal is under the binned
sample variance envelope expected for the $f_{\mathrm{{sky}}}\simeq0.30$
mask (shown in black) up to $\ell\sim1000$ at all three frequencies.
\emph{Right}: Zoom-in on the low-$\ell$ part of the spectrum, showing
that although the difference test failures are highly significant
at these scales, their amplitudes at all three CMB frequencies are
less than about 0.5\,$\mathrm{\mu K^{2}}$ for all multipoles up
to $\ell=200$, which is a tiny fraction of the binned sample variance
at these scales.}
\end{figure*}

We perform a series of difference map consistency tests to evaluate
the contribution of residual systematic effects to the angular power
spectrum of the temperature anisotropies. We generate difference maps
of halves of the nominal mission data in which the signal is expected
to fully subtract out in the absence of systematic effects. The resulting
maps are then subtracted from systematics-free \revv{simulated} ``Yardstick''
realizations (Appendix~\ref{sec:YARDSTICK}) of the same differences
propagated through the same mapmaking pipeline as that used for data
processing. The amplitude of the power of the remaining signal in
each likelihood bin can thus be used as an estimate of the level of
the residual systematic contamination in this bin. This estimate is
derived from cross-spectra between different detector sets to allow
for a direct comparison with the spectra used as inputs by the likelihood
code \citep{planck2013-p08}.

Many differences of halves of the data can be expected to be compatible
with an absence of signal. Given \Planck's scanning strategy, we
expect the bulk of the remaining systematic contamination to be captured
in survey difference maps. In particular, transfer function errors,
far-sidelobe pickup, gain instability, pointing drifts, and residual
glitches will leave an imprint in these maps. Half-focal-plane difference
maps will be particularly sensitive to relative gain errors, while
also providing some sensitivity to sidelobe pickup. Finally, difference
maps obtained by subtracting PSB-only from SWB-only sky maps probe
any unexpected behavioural difference (e.g., beam mismatch) between
polarization-sensitive bolometers and unpolarized dectectors, and
provide a check of the $(I,Q,U)$ decomposition applied to the data
from the former. Bandpass difference misestimation between the two
detector technologies will also leave a signal in these maps.

\begin{figure}[t]
\centering{}\includegraphics[width=1\columnwidth]{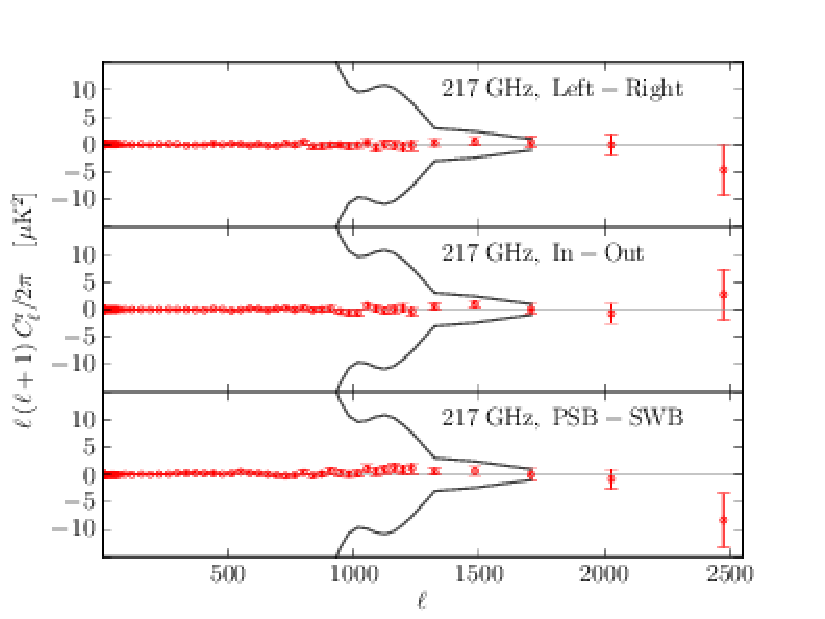}\protect\caption{\label{fig:217_other_diff}Examples of map difference consistency
tests at 217\GHz\ with the $f_{\mathrm{{sky}}}\simeq0.30$ mask,
derived from that used in the primary cosmological analysis (see the
text). The range of multipoles shown and the binning used are identical
to those described in Fig.~\ref{fig:survey_diff}. We again show
in each bin the variance of the $C_{\ell}$ distribution from the
simulations, and in black the binned sample variance for our mask.
``Left'' and ``Right'' refer to the two halves of the focal plane,
whereas ``In'' and ``Out'' refer to detectors toward the centre
and the periphery of the focal plane. Unlike in the survey difference
test shown in Fig.~\ref{fig:survey_diff}, no significant null test
failure is detected here. The detector combinations used in the example
plots above are $(5-8)\times(1-4)$, $(7-8)\times(3-4)$, and $\mathrm{(ds1}-3)\times\mathrm{(ds2}-4)$
for, respectively, the (Left$-$Right), (In$-$Out), and (PSB$-$SWB)
tests. As in Fig.~\ref{fig:survey_diff}, these power spectra are
divided by four to allow for a direct comparison of the amplitude
of the residuals to that of the components in the nominal mission
power spectrum. }
\end{figure}

Since the survey difference test is the most stringent difference
map consistency test at our disposal, we focus the discussion in this
section on the corresponding results. We did not observe null test
failures in any of our other tests from the list above. The nominal
mission survey difference test results are shown in Fig.~\ref{fig:survey_diff},
while Fig.~\ref{fig:217_other_diff} shows the outcome of a suite
of difference map consistency tests at 217\,GHz.

In Fig.~\ref{fig:survey_diff}, the angular power spectrum of the
residual signal is obtained by taking the cross-spectrum between detector
sets ds1 and ds2 at each frequency (see Table~\ref{tab:detsets})
after application of the \foreignlanguage{english}{$f_{\mathrm{{sky}}}=0.48$}
mask used in the primary cosmological analysis \citep{planck2013-p08},
combined with a mask of the regions not seen by either Survey 1 or
Survey 2 and a mask of the point sources included in the input \Planck\
Sky Model (PSM; \citealt{delabrouille2012}) maps. The resulting sky
fraction is 0.30 at 100\,GHz, and 0.29 at both 143\,GHz and 217\,GHz.
Results are then binned according to the likelihood binning. For clarity,
bins at multipoles higher than $\ell=60$ are grouped in sets of four,
whereas those at $\ell>1250$ are grouped in sets of eight. The plot
at left shows the residual signal over the whole range of multipoles
used in the likelihood analysis. At right, the same results are shown
with a low-$\ell$ $(\ell<200)$ zoom. In both cases, we report the
variance of the $C_{\ell}$ distribution in each bin as computed from
the simulations.

At all three CMB frequencies, the $\ell<200$ residuals are strikingly
small, never exceeding about $0.5\,\mu\mathrm{{K}^{2}}$. Excluding
a single bin in each case, the residuals actually remain below or
at the level of $0.2\,\mu\mathrm{{K}^{2}}$ over this range of multipoles.
Although these non-zero differences are detected with very high statistical
significance, they are many orders of magnitude smaller than the binned
sample variance (shown in black in Fig.~\ref{fig:survey_diff}) at
these scales. In particular, they cannot affect the cosmological analysis.
This stays true all the way up to $\ell\sim1000$, at which point
residuals become higher than the binned sample variance at 100 and
217\,GHz. At 143\,GHz, this does not happen until $\ell\sim1500$.

In the multipole range from 1000 to 2500, although the amplitude of
the residuals is almost always greater than the binned sample variance,
the variance in the simulation results is significantly larger than
in the $\ell<1000$ regime. As a result, the 100 and 143\,GHz residuals
for that range of multipoles are fully compatible with zero. However,
the 217\,GHz residuals in the same multipole range are not, as can
be directly inferred from the plot, where an apparently significant
oscillatory feature starts at $\ell=1000$. 

This survey-difference test has been performed at each frequency for
all combinations of the input maps used in the likelihood analysis
(two at 100\,GHz, five at 143\,GHz, and six at 217\,GHz), and for
two survey differences: Survey~1 $-$ Survey~2 and Survey~1 $-$
Survey~3. In addition to the 217-ds1 $\times$ 217-ds2 cross-spectrum
shown in Fig.~\ref{fig:survey_diff}, only two other cross-spectra
fail this test, namely 217-1 $\times$ 217-ds2 and 217-1 $\times$
217-3. \rev{We have evidence that these failures result in part from a systematic feature at} $\ell\sim1800$  \rev{mitigated by additional data flagging to reduce the effect of electromagnetic interferences from the 4\,K cooler drive and the read-out electronics (see Sect.~\ref{sec:4Kline-removal}).} In
the likelihood analysis \citep{planck2013-p08}, we have checked \revv{(cf. Fig.24 and 25) that the inclusion, or not, of these three cross-spectra has no discernible influence on the determination of all basic CDM cosmological parameters (but for a slight broadening of the posterior distribution, while the main foreground parameter change is a  shift of the point source amplitude at 217 GHz).}

\revv{Finally, to best address any remaining concerns about systematic errors resulting from data selection and differences in the first-half and second-half spectra, we present additional, high precision, tests of systematics performed since the March 2013 data release on the 217 GHz \Planck\ channel. Indeed, this channel has the highest angular resolution within the "CMB frequency" range 100 - 217\,GHz used for most of the cosmological results. As a consequence, it has the least degree of redundancy of the CMB frequencies, since one cannot cross-check the 217\GHz\ spectrum with the lower frequency spectra at high multipoles ($\ell \simgt 2000$). Appendix~\ref{sec:fullmission} presents tests of the fidelity of the \Planck\ power spectra using additional data from the full mission (29 months, 4.8 sky surveys, with the same processing than for the nominal mission). By using the full mission it is possible to perform more extensive and higher signal-to-noise tests than those described by \cite{2013arXiv1312.3313S}, which were based exclusively on the 2013 nominal mission data release. Reassuringly, the outcome lends further support to the claimed  robustness and accuracy of the cosmological results based on the current processing of HFI data. }

\subsubsection{ADC nonlinearity impact\label{sec:ADC-non-linearity}}

\begin{figure}[t]
\begin{centering}
\includegraphics[clip,width=0.8\columnwidth]{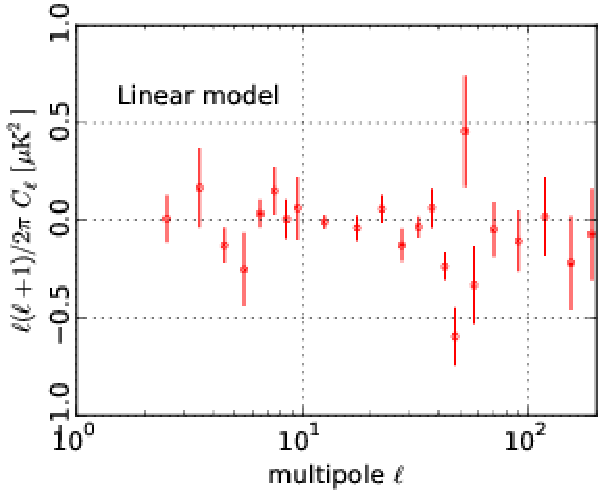}
\par\end{centering}

\begin{centering}
\includegraphics[clip,width=0.8\columnwidth]{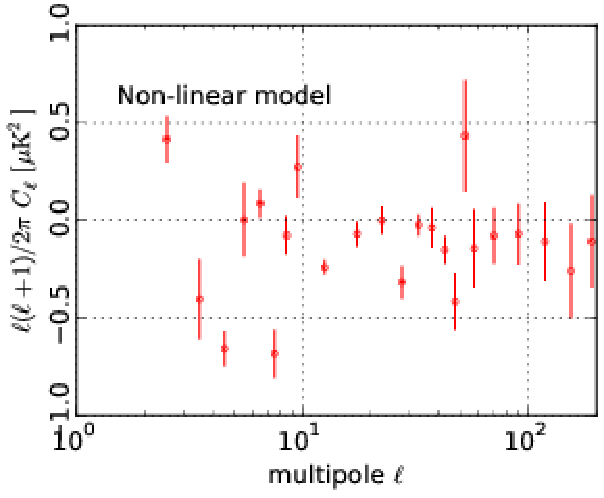}
\par\end{centering}

\centering{}\includegraphics[clip,width=0.8\columnwidth]{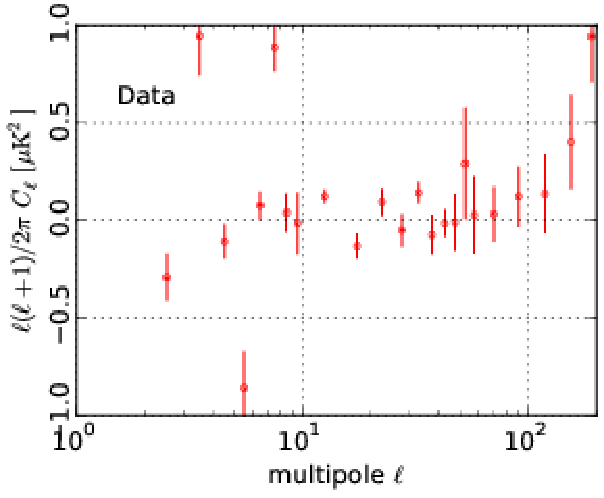}\protect\caption{\label{fig:ADC_vs_Yardstick}Simulated effect of the ADC nonlinearity
on the Survey 1$-$Survey 2 difference at 143\,GHz. \emph{Top}: Difference
test of a single realization of the full pipeline, assuming a linear
response of the ADC. \emph{Middle}: the same, but with our current
best model of the nonlinearity response of the ADC. This can be compared
with the bottom panel, which shows the result of the test applied
to the delivered data in exactly the same conditions. This demonstrates
that indeed the effect of ADC nonlinearity can explain most of the
coherent deviation seen at $\ell\simlt100$\textbf{. }}
\end{figure}

We check the impact of including the effect of ADC nonlinearity through
a comparison of data and simulations of the two PSB pairs at 143\,GHz.
\rev{Note that this effect had been suspected to be detectable, but is partially degenerate with time-dependent gain variations in the detectors. Beyond correcting for such gain variations, the effect was not accounted for in the nominal mission products,} nor
included in the corresponding Yardstick simulation, since the relevant
modelling information was only obtained after the end of the HFI cryogenic
phase through dedicated data gathering during the warm phase. Figure~\ref{fig:ADC_vs_Yardstick}
shows that the inclusion of this effect leads to similar $\ell\simlt100$
deviations from the Yardstick compared to those observed in the survey
difference consistency test of the 143\,GHz data, which suggests
this is one of the main limiting factors for measuring the low-$\ell$
polarization with the current processing of the data. The inclusion
of this effect makes little difference at higher $\ell$.

\subsubsection{Noise estimation bias from half-ring maps \label{sec:Noise-Estimation-Bias}}

\begin{figure}[t]
\begin{centering}
\includegraphics[width=1\columnwidth]{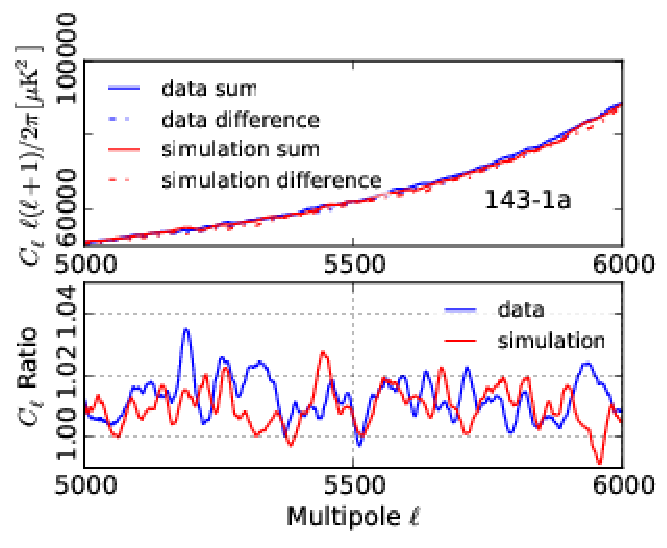}
\par\end{centering}

\centering{}\protect\caption{\label{fig:HRD_bias}Comparison of the high-$\ell$ part of the power
spectra of the sum and difference of half-ring maps for the 143-1a
detector, both for flight data and simulations where glitches were
introduced and deglitched with the actual processing pipeline. The
comparison shows that the simulations provide a good description of
this effect (which is not included in the Yardstick simulations).
\emph{Top}: The sum and difference spectra. \emph{Bottom}: Ratios
of the sum and difference spectra for both data and simulations, which
differ by about 2\,\%. }
\end{figure}

A detailed comparison of the power spectra of the sum and difference
of half-ring maps shows a constant offset of a few percent at high
multipoles. At these scales, there is no longer any signal remaining
in the sum map, due to the beam cut-off (see Fig.~\ref{fig:HRD_bias}
for an example), and we would expect that the difference map spectrum
would give a precise determination of the high-$\ell$ tail of the
sum map, while it is actually slightly below. This offset shows that
a noise estimate from half-ring maps difference is a slightly biased
indicator of noise in the sum maps. This results from the deglitching
algorithm. 

In the first step, the pipeline estimates the signal by averaging
the timelines into rings. This signal estimate is removed in order
to aid detection of glitches by comparison with the noise level. Each
positive sample greater than three times the noise level is flagged
as a glitch. One side effect of this thresholding is that the positive
tail of the timeline noise distribution is now clipped at $3\sigma$.
Note that the signal estimate within a ring is of course still noisy,
so that its removal from the timeline before deglitching subtracts
this residual noise component periodically (coherently at all harmonics
of the spin frequency), in effect periodically modulating (in both
half-rings) the flagging and clipping level of the noise distribution.
In the difference of half-ring maps, this common contamination disappears,
while it is present in the sum map, leading to this slight bias. 

Figure~\ref{fig:HRD_bias} shows that the inclusion of the deglitching
step in simulations accounts quantitatively for the difference in
the power spectra of the sum and difference of half-ring maps. By
fitting the ratio of the spectra of sum and difference maps (masked
to retain 37\,\% of the sky) at very high $\ell$ (around $\ell\sim7000$---not
shown in the figure) one finds the bias levels which are given in
Table~\ref{tab:bias_correction}.

\begin{table}[th]
\protect\caption{Percentage increase of the power spectrum of the difference of half-ring
maps needed to \rev{correct the bias of} the noise level estimate
in frequency maps.\textcolor{red}{{} \label{tab:bias_correction}}}
%\begingroup
%\newdimen\tblskip \tblskip=5pt
%\nointerlineskip
%\vskip -3mm
%\footnotesize
%\setbox\tablebox=\vbox{
%   \newdimen\digitwidth 
%   \setbox0=\hbox{\rm 0} 
%   \digitwidth=\wd0 
%   \catcode`*=\active 
%   \def*{\kern\digitwidth}
%%
%   \newdimen\signwidth 
%   \setbox0=\hbox{+} 
%   \signwidth=\wd0 
%   \catcode`!=\active 
%   \def!{\kern\signwidth}
%%
%\halign{\hfil#\hfil\tabskip=2em& % frequency
%\hfil#\hfil& % 100
%\hfil#\hfil& % 143
%\hfil#\hfil& % 217
%\hfil#\hfil\/\tabskip=0pt\cr  % 357           % Template goes here.
%\noalign{\doubleline}
%Frequency [GHz]& 100& 143& 217& 353\cr % Table headings go here.
%\noalign{\vskip 3pt\hrule\vskip 5pt}
%                                               % Body of table goes here.
%&$0.53\pm 0.08$\,\%&  $0.61\pm 0.08$\,\%&  $0.75\pm 0.09$\,\%& $0.44\pm 0.09$\,\%\cr 
%\noalign{\vskip 5pt\hrule\vskip 3pt}}}
%%\endPlancktable                   % ends one-column \halign
%\endPlancktablewide                 % ends two-column \halign
%%\tablenote a Footnote a.\par
%%\tablenote b Footnote b.\par
%\endgroup
%

\begingroup
\newdimen\tblskip \tblskip=5pt
\nointerlineskip
\vskip -3mm
\footnotesize
\setbox\tablebox=\vbox{
   \newdimen\digitwidth 
   \setbox0=\hbox{\rm 0} 
   \digitwidth=\wd0 
   \catcode`*=\active 
   \def*{\kern\digitwidth}
   \newdimen\signwidth 
   \setbox0=\hbox{+} 
   \signwidth=\wd0 
   \catcode`!=\active 
   \def!{\kern\signwidth}
\halign{\hfil#\hfil\tabskip=2em& % frequency
\hfil#\hfil\/\tabskip=0pt\cr  % percentage increase           % Template goes here.
\noalign{\doubleline}
Frequency [GHz]& Fractional change [\%]\cr % Table headings go here.
\noalign{\vskip 3pt\hrule\vskip 5pt}
                                               % Body of table goes here.
100& $0.53\pm 0.08$\cr
143& $0.61\pm 0.08$\cr
217& $0.75\pm 0.09$\cr
353& $0.44\pm 0.09$\cr 
\noalign{\vskip 5pt\hrule\vskip 3pt}}}
\endPlancktable                   % ends one-column \halign
%\endPlancktablewide                 % ends two-column \halign
%\tablenote a Footnote a.\par
%\tablenote b Footnote b.\par
\endgroup
\end{table}

\subsection{Power spectrum consistency tests\label{sec:Concistency-power-spectra}
}

\begin{figure*}[!t]
\begin{centering}
\includegraphics[clip,width=1\columnwidth]{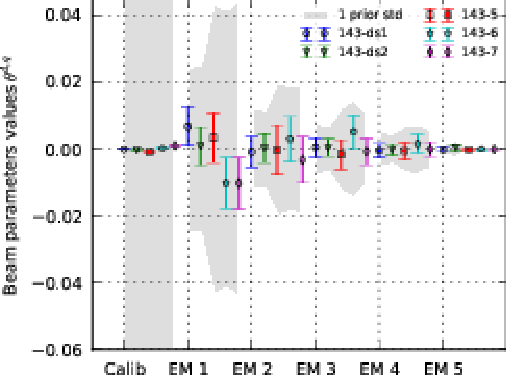}
\includegraphics[clip,width=1\columnwidth]{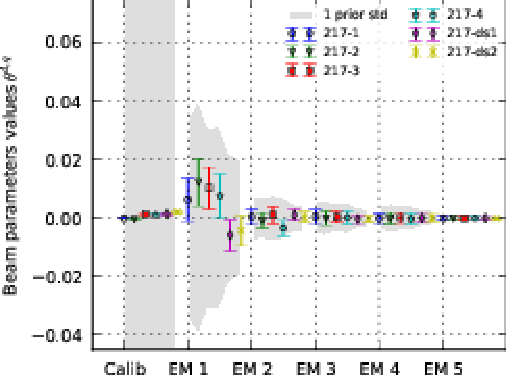}
\par\end{centering}

\begin{centering}
\includegraphics[clip,width=1\columnwidth]{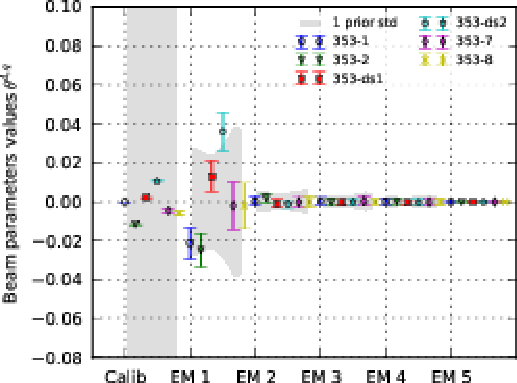}
\includegraphics[clip,width=1\columnwidth]{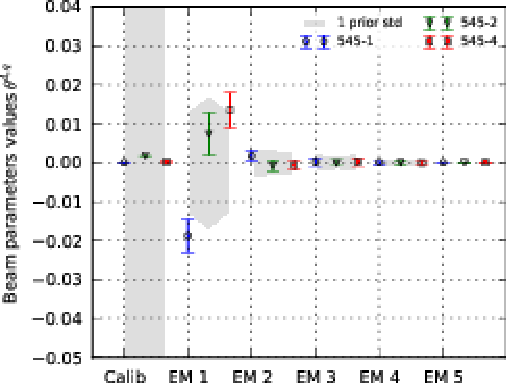}
\par\end{centering}

\begin{centering}
\includegraphics[clip,width=1\columnwidth]{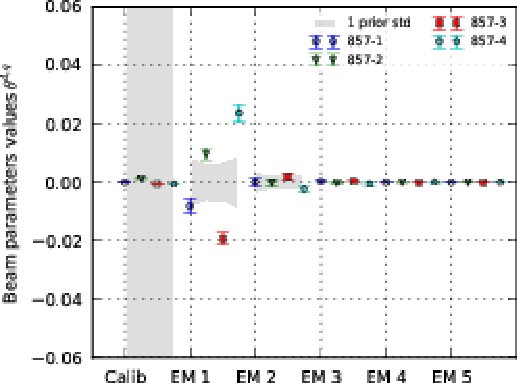}\includegraphics[clip,width=1\columnwidth]{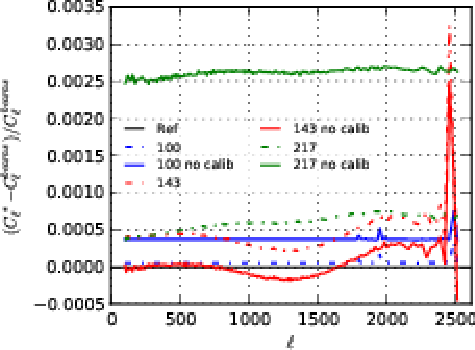}
\par\end{centering}

\centering{}\protect\caption{Calibration (``Calib'') and beam uncertainty mode (``EM'') determination
for all detectors (or detector sets) within each frequency band. The
grey areas show the $1\sigma$ priors on each beam parameter (without
assuming a prior for gain calibration; the first detector in each
series is used as a fixed reference). The error bars show the $1\sigma$
posteriors. The bottom right\textbf{ }shows the variations of the
power spectra in the primary CMB channels when applying the recalibration
and beam uncertainty corrections described in the text. Altogether,
this leads to variations smaller than 0.4\,\% at all $\ell \simlt 2500$.
Note that the shape variations with $\ell$ are much smaller (a fractional
variation less than about $5\times 10^{-4}$).\label{fig:intraFreqCalibBeam}
}
\end{figure*}

A powerful check of the data follows from assessing the signal consistency
between detectors at the same frequency given the prior information
on the effective beam uncertainties and calibration. At each frequency,
an estimate of an adapted power spectrum model is processed with the
\texttt{SMICA} likelihood \citep{planck2013-p08}. At 143 and 217\GHz\ 
where the CMB is dominant, the spectrum on the measured sky is modelled
as $\vec{a}\vec{a}^\dagger+\vec{n}_\ell$, a sum of the the free signal
power, $C_\ell$, plus a free independent noise spectrum, $\vec{n}_\ell$,
where the known vector $\vec{a}$ accounts for the fixed CMB emissivity
at each frequency (bold-faced vectors have elements at each frequency).
At higher frequencies, dust emission begins to dominate the CMB, and
some foreground modelling is necessary. Thus the CMB power is fixed
to the angular spectrum determined at 143\GHz, the noise is kept
free and a free foreground is added, with colour correction determined
from a greybody spectral shape integrated though the measured spectral
response of each detector {[}40\,\% of the sky is used to estimate
the binned empirical power spectrum with \spice\  \citep{Szapudi2001}{]}.
Furthermore, beam errors are modelled as a linear combination of templates,
thanks to a beam error-mode analysis \citep{planck2013-p03c}. Then,
five beam parameters along with a calibration are estimated. The full
model is written as

\begin{equation}
\vec{R}_{\ell}=\vec{B}_{\ell}\left(\vec{a}\vec{a}^{\dagger}c_{\ell}^{143}+\vec{b}\vec{b}^{\dagger}g_{\ell}\right)+\vec{n}_{\ell}\label{eq:BPAR}
\end{equation}
where $\vec{R}_{\ell}$ is the modelled spectrum, $\vec{B}_{\ell}$
the free beam model, $\vec{b}$ the free dust emmisivity at each frequency,
$g_{\ell}$ the free dust power, and $\vec{n}_{\ell}$ the free noise
power. The beam model for detector $d$ is written $B_{\ell}[d]=\exp(\sum_{q}\theta^{d,q}T_{\ell}^{d,q})$,
where $\theta$ are the beam-eigenmode amplitude parameters and $T$
the templates. Due to the intrinsic ambiguity between spectral power
and beams, beam simulations are used to determine the variances of
a Gaussian prior on these parameters. Finally, at 100\GHz, with only
two detector sets, three parameters per $\ell$ (one signal and two
noise powers) allow an estimate of the empirical spectrum, but relative
calibration and beam parameters estimation are then unfeasible.

Figure~\ref{fig:intraFreqCalibBeam} shows the result of such an
analysis. At 143 and 217\GHz, where the CMB is dominant and used
as calibrator, the only adjustments necessary are a gain recalibration
by a very small amount (0.2\,\% at most) and the addition of a fraction
of a $\sigma$ of the first beam uncertainty eigenmode. At 545 and
857\GHz, where the CMB is subdominant and planets are used as calibrators,
the required recalibration remains just as small (about 0.2\,\%),
but larger beam adjustments are necessary. The 353\GHz\ calibration
is the one requiring the largest changes, about 1\,\%. At 217\GHz,
and to a lesser extent 353\GHz, the SWB and PSB beam window functions
are at opposite sides of the ``optimal'' one. In conclusion, by
properly modelling foregrounds at high frequency, cross-checks on
signal anisotropy with 40\,\% of the sky shows that maps are well
calibrated (relatively) within a frequency band.

\begin{figure}[t]
\begin{centering}
\includegraphics[clip,width=1\columnwidth]{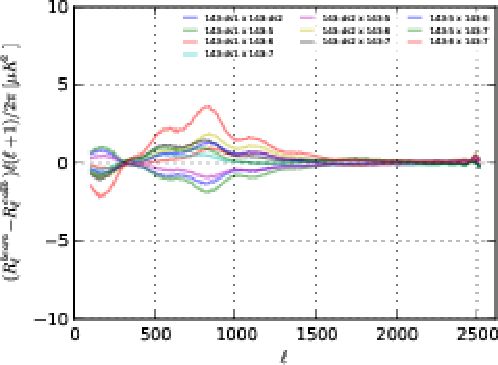}
\par\end{centering}

\begin{centering}
\includegraphics[clip,width=1\columnwidth]{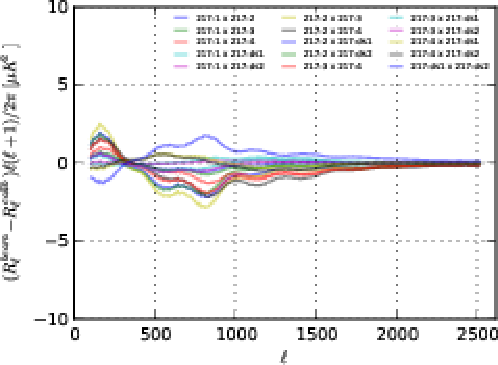}
\par\end{centering}

\centering{}\protect\caption{Derived overall beam correction of each detector set which minimizes
their mismatch, given the model used (Eq.~\ref{eq:BPAR}). This is
plotted for 143 and 217\,GHz, where there are sufficient data to
simultaneously estimate beam parameters along with signal and noise
power spectra. These corrections peak around $\ell\sim200\ \mathrm{and}\ 800$
and are at most of the order of 3\,$\mu\mathrm{K}^{2}$. \label{fig:intraFreqDesetBeams}
}
\end{figure}

Figure~\ref{fig:intraFreqDesetBeams} shows the magnitude of the
beam corrections determined by maximizing the internal consistency
for the most sensitive CMB channels of \Planck. The largest corrections
would be about $2\,\mu\mathrm{K^{2}}$ around $\ell\sim100$ and about
$4\,\mu\mathrm{K^{2}}$ around the third peak of the CMB spectrum
($\ell\sim800$). For the delivered frequency maps, which combine
all detectors at the same frequency, the effect is even weaker, and
for the likelihood analysis, this is fully accounted for (these beam
eigenvalues are either \rev{directly estimated} or marginalized over,
including in addition their expected cross-correlations as discussed
in \citealt{planck2013-p03c} and \citealt{planck2013-p08}).

\subsection{Consistency of the calibration across frequencies}

In the previous section, we looked at the consistency of detector
spectra within a given frequency band. A similar analysis of the spectra
used in the high-$\ell$ part of \Planck\ likelihood \citet{planck2013-p08}
shows that the best relative recalibration (for the mask considered
and accounting for the first beam eigenmode), between all 13 detectors
maps in the 100, 143 and 217\,GHz channels, would be less than 0.2\,\%.
This level has an insignificant effect on the constraints on cosmological
parameters. 

\begin{figure}[t]
\begin{centering}
\includegraphics[clip,width=1\columnwidth]{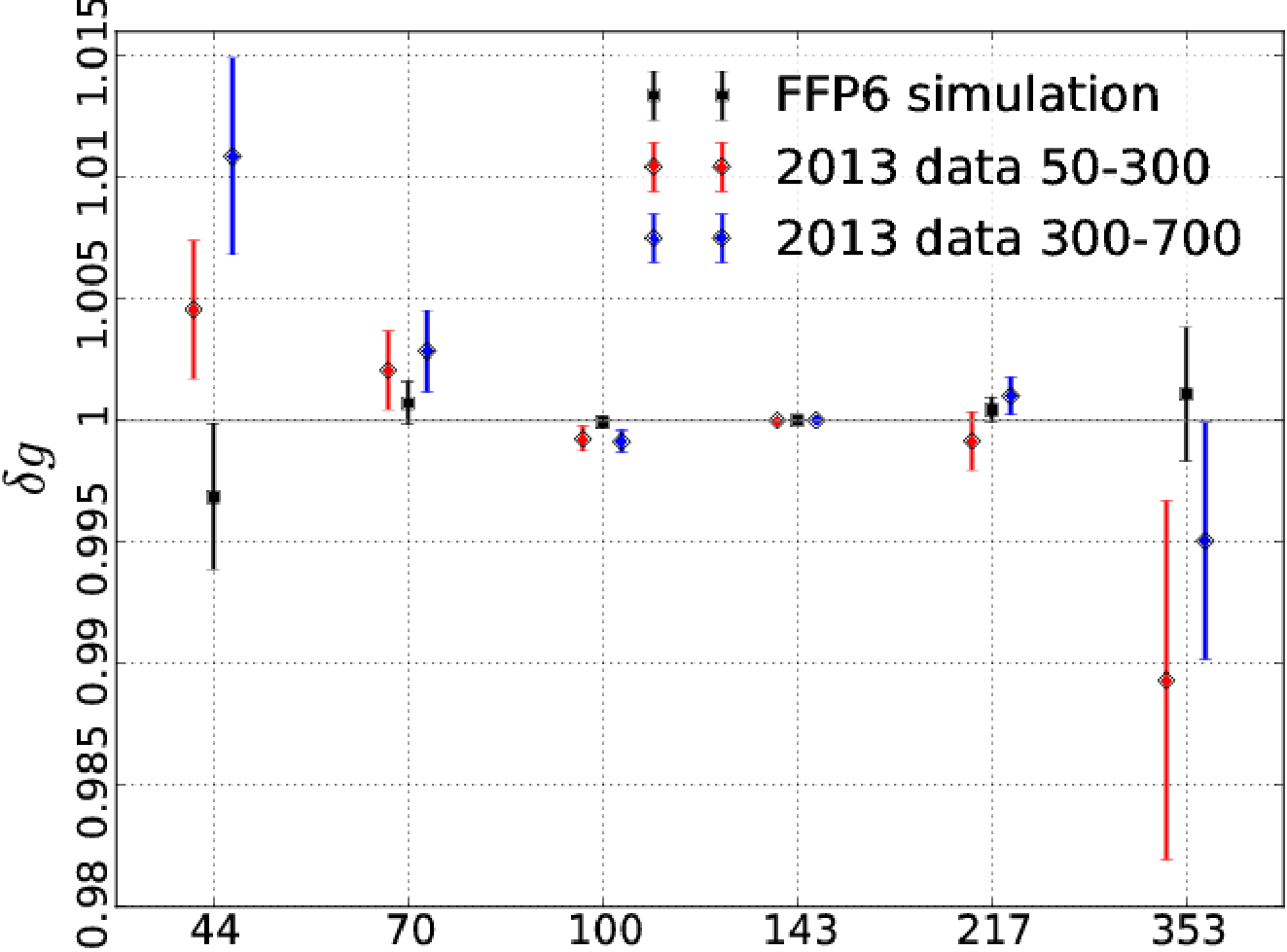}
\par\end{centering}

\centering{}\protect\caption{Recalibration factor maximizing the CMB consistency in in simulations
(black) and in the data, considering different multipole ranges (red
and blue), at each \Planck\ frequency in GHz (on the horizontal axis).
\label{fig:interFreq-SMICA} }
\end{figure}

Finally, we can further broaden our consistency checks by using the
\texttt{SMICA} component separation method \rev{on both LFI and HFI data}
from 44 to 353\,GHz, in effect intercalibrating on the common CMB
anisotropies themselves, taking 143\,GHz as the reference (see Sect.~7.3.3
of \citealp{planck2013-p03f} and \citealp{planck2013-p06}). Figure~\ref{fig:interFreq-SMICA}
compares the results of such an analysis%
\footnote{The analysis is actually carried out over 40\,\% of the sky, for
various $\ell$ ranges; the results discussed here show the average
in the multipole range 100--679 for the simulations. %
} on a full focal-plane (FFP6) simulation \citep{planck2013-p28} and
on the 2013 maps, using a five-component foreground model. The simulations
accurately validate the approach, in a realistic setting (albeit with
decreasing accuracy at 44 and 353\,GHz). For flight data, the multipole
range over which the calibration factor is averaged has been split
in two, a relatively low-$\ell$ range (50--300), more appropriate
to include the noisier 44 and 70\GHz\ LFI channels, and a higher-$\ell$
range, (300--700), more appropriate for HFI data. The figure confirms
that the 100--217\GHz\ map calibrations are consistent to 0.2\,\%,
as discussed earlier. The 353\,GHz results would suggest a slightly
higher calibration than the one we finally adopted, although the \texttt{SMICA}
analysis does not account for the Rayleigh scattering correction \citep[see][]{planck2013-p03f}
and we may also reach the limit of a five-component foreground model.
This analysis also suggests a residual calibration difference between
LFI at 70\GHz\  and HFI at 100\GHz\  of roughly 0.3\,\%. 

Further checks involving comparison with the LFI and \emph{WMAP} data
are discussed in \citet{planck2013-p01a}.

\subsection{Further validation\label{sec:Other-papers-validation}}

Other tests are described in various other papers: 
\begin{itemize}
\item \citet{planck2013-p01a} compares the maps from the HFI and LFI instruments
on \Planck\ and from the \emph{WMAP} satellite. We demonstrate excellent
consistency amongst the three sets of maps produced with different
instruments over a a factor of thirty in wavelength, and especially
over the narrow range of 70--100\GHz\ covered by all three instruments,
illustrating the excellent control of systematic errors we have achieved
with \Planck.
\item \citet{planck2013-p06} discusses several methods for separating the
various astrophysical and cosmological data. Methods with very different
assumptions about the physics of the different components, and about
the nature of the noise in the \Planck\ maps, are able to converge
on equivalent maps for different components, most importantly of the
CMB itself.
\item \citet{planck2013-p08} discusses the angular power spectrum, $C_\ell$,
computed from \Planck\ data, as well as the likelihood function that
embodies its full probability distribution. The likelihood is computed
by taking into account explicit models of foreground contamination
and of systematic effects, such as beam errors and the relative calibration
of different detectors. The resultant spectra explicitly illustrate
consistency across those bands with the highest signal-to-noise ratio
for the cosmological spectra at 100, 143, and 217\GHz, at the level
of power spectra and of cosmological parameters. As discussed as well
in \ref{sec:Consistency-diff-map}, the likelihood is robust to the
inclusion (or otherwise) of even the 217\GHz\  detectors that show
some mild non-zero features in the spectra of their difference maps. 
\end{itemize}

\section{Summary of product characteristics and conclusions \label{sec:SUMMARY} }

The data provided to the \Planck\ collaboration \rev{and the community}
by the HFI DPC, in addition to the \IMO, consisted of the following
(letters refers to entries in the final column of Table~\ref{tab:detsets}).
\begin{enumerate}
\item C: beam-corrected power spectra for each CMB detector set used in
the \Planck\  likelihood code, together with a description of the
relevant beam uncertainties;
\item N: frequency maps for the nominal mission duration, not corrected
for ZLE/FSL;
\item B: effective beams and uncertainties (for N);
\item F: spectral bands and uncertainties;
\item H: hit count maps (for N);
\item HR: half-ring maps, HR1 and HR2, two maps made from the first and
the second half of each pointing period of the nominal mission;
\item S: survey maps, S1 and S2, made from the data of Survey 1 only and
Survey 2 only; and
\item Z: zodiacal light emission and far sidelobe corrected frequency maps
and their survey maps (Z-N, Z-S1, Z-S2).
\end{enumerate}
Table~\ref{tab:detsets} summarizes the various detector sets from
which we made these products, while Table~\ref{tab:summary} summarizes
the map characteristics needed to make use of the maps. Further details
and comments are given below.

\begin{table*}[t]
\centering{}\protect\caption{HFI Nominal Maps --- Main Characteristics. \label{tab:summary}}
\begingroup
\newdimen\tblskip \tblskip=5pt
\nointerlineskip
\vskip -3mm
\footnotesize
\setbox\tablebox=\vbox{
   \newdimen\digitwidth 
   \setbox0=\hbox{\rm 0} 
   \digitwidth=\wd0 
   \catcode`*=\active 
   \def*{\kern\digitwidth}
   \newdimen\signwidth 
   \setbox0=\hbox{+} 
   \signwidth=\wd0 
   \catcode`!=\active 
   \def!{\kern\signwidth}
\halign{\hbox to 2.5in{#\leaderfil}\tabskip 2.2em&
%%#\hfil\leaderfil\tabskip=1em&
%% \hfil#\hfil& % Units
\hfil#\hfil& % 100
\hfil#\hfil& % 143
\hfil#\hfil& % 217
\hfil#\hfil& % 353
\hfil#\hfil& % 545
\hfil#\hfil& % 857
\hfil#\hfil\/\tabskip=0pt\cr % Notes              
% Template goes here.  %%%% PLEASE LEAVE NO SPACE BEFORE "&" (it does affect format)
\noalign{\doubleline}
\omit\hfil Quantity \hfil&                                                    &           &           &           &           &           & Notes\cr
%---------------------------------------------------------
\noalign{\vskip 3pt\hrule\vskip 5pt}                                      
Reference frequency $\nu$ [GHz]           &      100&        143&        217&        353&        545&        857& \tablefootmark{a1}\cr
Number of bolometers                      &        8&         11&         12&         12&          3&          4& \tablefootmark{a2}\cr
%---------------------------------------------------------
%\noalign{\vskip 3pt\hrule\vskip 5pt}                                      
%Detector noise - S [$\mu$K\,s$^{1/2}$]          &         &         49&       *66&         205&           &           & \tablefootmark{a3}\cr
%Detector noise - P [$\mu$K\,s$^{1/2}$]          &      132&         65&       101&         397&           &           & \tablefootmark{a3}\cr
%Detector noise - S [kJy sr$^{-1}$\,s$^{1/2}$]   &         &           &          &            &         52&         56& \tablefootmark{a3}\cr
%---------------------------------------------------------
\noalign{\vskip 3pt\hrule\vskip 3pt}  
Effective beam solid angle $\Omega$ [arcmin$^2$]      &   105.78&      59.95&      28.45&      26.71&      26.53&      24.24& \tablefootmark{b1}\cr
Error in solid angle $\sigma_\Omega$ [arcmin$^2$]     &   **0.55&      *0.08&      *0.03&      *0.02&      *0.03&      *0.03& \tablefootmark{b2}\cr
Spatial variation ({rms}) $\Delta\Omega$ [arcmin$^2$] & **0.31&      *0.25&      *0.27&      *0.25&      *0.34&      *0.19& \tablefootmark{b3}\cr
Effective beam FWHM$_1$ [arcmin]                      &   **9.66&      *7.27&      *5.01&      *4.86&      *4.84&      *4.63& \tablefootmark{b4}\cr
Effective beam FWHM$_2$ [arcmin]                      &   **9.65&      *7.25&      *4.99&      *4.82&      *4.68&      *4.33& \tablefootmark{b5}\cr
Effective beam ellipticity $\epsilon$                 &   *1.186&      1.036&      1.177&      1.147&      1.161&      1.393& \tablefootmark{b6}\cr
Variation ({rms}) of the ellipticity $\Delta\epsilon$ &*0.023&      0.009&      0.030&      0.028&      0.036&      0.076& \tablefootmark{b7}\cr
%---------------------------------------------------------
\noalign{\vskip 3pt\hrule\vskip 5pt}                                      
Sensitivity per beam solid angle [$\mu$K]       &       10&          6&        12&         39&           &           & \tablefootmark{c1}\cr
\phantom{Sensitivity per beam solid angle}
                                 [kJy sr$^{-1}$]&         &           &          &           &         13&         14& \tablefootmark{c1}\cr
 Sensitivity                       [$\mu$K\,deg]&      1.8&        0.8&       1.0&        3.5&           &           & \tablefootmark{c2}\cr
 \phantom{Sensitivity}      [kJy sr$^{-1}$\,deg]&         &           &          &           &        1.1&        1.1& \tablefootmark{c2}\cr
%---------------------------------------------------------
\noalign{\vskip 3pt\hrule\vskip 3pt}  
Relative calibration accuracy [\%]& $\simlt 0.2$& -& $\simlt 0.2$& $\simlt 1**$& $\simlt *5$& $\simlt *5$& \tablefootmark{d1}\cr
Absolute calibration accuracy [\%]& $\simlt 0.5$& $\simlt 0.5$& $\simlt 0.5$& $\simlt 1.2$& $\simlt 10$& $\simlt 10$& \tablefootmark{d2}\cr
%---------------------------------------------------------
\noalign{\vskip 3pt\hrule\vskip 3pt}  
Galactic zero-level offset        [MJy\,sr$^{-1}$]&   0.0047&     0.0136&     0.0384&     0.0885&     0.1065&     0.1470& \tablefootmark{e1}\cr
Galactic zero-level uncertainty   [MJy\,sr$^{-1}$]&   0.0008&     0.0010&     0.0024&     0.0067&     0.0165&     0.0147& \tablefootmark{e2}\cr
CIB monopole prediction           [MJy\,sr$^{-1}$]&   0.0030&     0.0079&     0.033*&     0.13**&     0.35**&     0.64**& \tablefootmark{e3}\cr
%---------------------------------------------------------
\noalign{\vskip 5pt\hrule\vskip 3pt}}}
\endPlancktablewide% ends two-column \halign
\raggedright
\tablenote \textit{a1} {~Channel map reference frequency, and channel identifier.}\par
\tablenote \textit{a2} {~Number of bolometers whose data was used in producing the channel map.}\par
%\tablenote \textit{a3} {~Estimate of the total detector noise, i.e., the {\emph rms} noise at the map level for one second of integration time per bolometer. S \& P respectively stand for unpolarized (spider web) and polarized detectors. See Table~\ref{tab:Total-noise}.}\par
%---------------------------------------------------------
\tablenote \textit{b1} {~Mean value over detectors at the same frequency. See Sect.~\ref{sec:DETECTOR-BEAMS}.}\par
\tablenote \textit{b2} {~As given by simulations. }\par
\tablenote \textit{b3} {~Variation ({rms}) of the solid angle across the sky. }\par
\tablenote \textit{b4} {~FWHM of the Gaussian whose solid angle is equivalent to that of the effective beams.}\par
\tablenote \textit{b4} {~FWHM of the mean best-fit Gaussian.}\par
\tablenote \textit{b6} {~Ratio of the major to minor axis of the best-fit Gaussian averaged over the full sky. }\par
\tablenote \textit{b7} {~Variability ({rms}) on the sky. }\par
%---------------------------------------------------------
\tablenote \textit{c1} {~Estimate of the noise per beam solid angle given in \textit{b1}. See Sect.~\ref{sub:Maps-overview}.}\par
\tablenote \textit{c2} {~Estimate of the noise scaled to $1\deg$ assuming that the noise is white.}\par
%---------------------------------------------------------
\tablenote \textit{d1} {~Relative calibration accuracy between frequency channels. See Sect.~\ref{sec:MAP-MAKING}.}\par
\tablenote \textit{d2} {~Calibration uncertainty including the estimated uncertainty of the calibrating source.}\par
%---------------------------------------------------------
\tablenote \textit{e1} {~Offset to remove at each frequency to set the Galactic zero level. The values quoted correspond to $\nu I_\nu=\mathrm{const}$. At the four lowest frequencies, the conversion factor is about 244, 371, 483, and  287\,MJy\,sr$^{-1}$\,K$^{-1}_{\mathrm{CMB}}$.}\par
\tablenote \textit{e2} {~Overall error on the map zero point, for a constant $\nu I_\nu$ spectrum.}\par
\tablenote \textit{e3} {~According to the \cite{bethermin2012} model, whose uncertainty is estimated to be at the 20\,\% level. (Also for constant $\nu I_\nu$).}\par
\endgroup
\end{table*}

\subsubsection*{Angular Response}

Effective beams provide the response of a map pixel to the sky (lines
\emph{b1}--\emph{b7} of Table~\ref{tab:summary}). They are based
on a determination of the scanning beam from planet scans (Sect.~\ref{sub:Scanning-beams}),
which include the effect of the optical beam, the electronic detection
chain, and the TOI processing pipeline. Scanning beams are an important
intermediate product, although they are usually not relevant for astrophysical
applications. The effective beam further accounts for the combined
effect of the scanning strategy and additional data processing. 

Different applications need different levels of accuracy and detail,
from the mean full-width at half-maximum (FWHM) of a symmetrical Gaussian
description (line \emph{b4}) to an actual point spread function at
each specific map location. Beam solid angles are given in line \emph{b1}.
Mean ellipticities are given in line \emph{b6}. Lines \emph{b2}, \emph{b3},
and \emph{b7} provide information on the angular response uncertainties
and variation across the sky. Our complete uncertainty budget, including
covariances of the beam eigenvalues between detectors, in a form which
is usable for the power spectra analyses of \citet{planck2013-p08},
is provided in \citet{planck2013-p03c}.

\subsubsection*{Sensitivity}

The numbers given in line \emph{c1} of Table~\ref{tab:summary} indicate
the rms contribution of the noise per beam solid angle, for the median
integration time, as scaled from maps at the full resolution (pixels
of 1.7 arcmin). The numbers given in line \emph{c2} of the table also
convert this into an rms of the noise in pixels of $1\deg$ on a side,
again assuming that the rms varies inversely proportional to the pixel
scale, as if it were white noise. While convenient to compare the
sensitivity of different experiments, this number may be misleading,
since it is actually an overestimate of the real noise in the map
at the $1\deg$ scale. 

It is interesting to note that the sensitivity of the map delivered
is comparable to pre-flight predictions \citep{planck2005-bluebook},
despite the flagging of a significant fraction of data. This decrease
of integration time per pixel is approximately compensated by the
better-than-required sensitivity of the individual bolometers. A more
detailed description of the noise is made available as ``half-ring''
maps whose differences offer a quite accurate view of the small scale
noise, varying by up to half a percent.

\subsubsection*{Photometric Accuracy}

The photometric calibration (Sect.~\ref{sec:Abs-calib}) of the 100
to 353\GHz\  channels relies on the solar dipole. Comparing the common
CMB component at these frequencies shows that the relative accuracy
between these channels is better than 0.2\,\% between 100 and 217\,GHz,
and at the percent level for the 353\,GHz channel. For the two highest
frequencies, this relative accuracy is at the five percent level (lines
\emph{d1}--\emph{d2}). We have also estimated the zero level offset
of the maps which needs to be accounted for in Galactic and cosmic
infrared background studies (lines \emph{e1}--\emph{e3}). 

\subsubsection*{Spectral response and conversions}\label{sec:spectral-response}

The accuracy of the HFI spectral response characterization (Sect.~\ref{sec:Correct-color})
is validated using a variety of HFI in-flight observations. Comparisons
of Zodiacal light observations (Sect.~\ref{sec:Correct-FSL=000026Zodi})
with the bandpass data reveal an out-of-band signal rejection of better
than $10^{8}$. The bandpass-based unit conversions and colour corrections
were compared against those derived using sky-only data for sources
including Sunyaev-Zeldovich (SZ) clusters, dust emission, and CO emission.
The SZ observations show agreement with the bandpass data within the
quoted uncertainty, as do the 100, 217, and 353\,GHz dust coefficient
comparisons. The 143\,GHz dust coefficient comparison shows differences
between the sky and bandpass calibration methods employed; however,
the dust component is not dominant at 143\,GHz so this discrepancy,
although important to understand, does not significantly impact the
results. The CO-based dust colour correction coefficients show good
agreement with their bandpass-based counterparts, while there are
some differences between the bandpass-based and sky-based CO unit
conversion coefficients. Investigating these differences continues,
and will improve the understanding of the HFI instrument and data;
this will in turn lead to better data analysis and improved knowledge
of systematic and calibration uncertainties. Details are provided
elsewhere, primarily in \citet{planck2013-p03d} and \citet{planck2013-p28}.

\subsubsection*{Data limitations, validation, and checks for systematic errors}

The most powerful top-down tests are the statistics of differences
between sky maps. The power spectra of these difference maps have
been computed in the same way as those used in the \Planck\ CMB likelihood
(which are based on the nominal maps). We have applied a large number
of such tests, differencing maps made from various detector groups
(in/out, left/right), and sky surveys. The sky survey differences
maximize the ability to detect systematic effects potentially affecting
a specific detector or detector set; they have been applied to all
likelihood inputs. We have found three potentially significant departures
at large multipoles, for cross-spectra involving 217\,GHz detectors.
A jackknife test described in the CMB power spectrum and likelihood
paper \citep{planck2013-p08} verified that their removal does not
affect significantly the cosmological constraints . 

We also found small, microkelvin-level, residuals at low multipoles,
which although very significant statistically, are insignificant for
a cosmological analysis of the temperature. We have verified through
simulations that most of this effect is due to the limitations of
the partial correction made for the ADC non-linearity. This limits
the accuracy of our dipole calibration to about 0.2\,\% in the CMB
channels, and prevents us from using polarization information at low
multipoles. However, the data collected at the end of the cryogenic
chain will allow a more satisfactory correction for ensuing \Planck\ data
releases. 

\subsubsection*{Conclusions}

The six intensity maps and ancillary information presented above constitute
an unprecedented source of information in this frequency range, which
is in line with our pre-flight expectations. The level of understanding
reached while analysing this data release bodes well for the next
release of \Planck\ data, which will focus on polarization on all
scales, and use nearly twice as much data.
\begin{acknowledgements}
The development of \Planck\ has been supported by: ESA; CNES and
CNRS/INSU-IN2P3-INP (France); ASI, CNR, and INAF (Italy); NASA and
DoE (USA); STFC and UKSA (UK); CSIC, MICINN, JA and RES (Spain); Tekes,
AoF and CSC (Finland); DLR and MPG (Germany); CSA (Canada); DTU Space
(Denmark); SER/SSO (Switzerland); RCN (Norway); SFI (Ireland); FCT/MCTES
(Portugal); and PRACE (EU). A description of the \Planck\ Collaboration
and a list of its members, including the technical or scientific activities
in which they have been involved, can be found at \url{http://www.sciops.esa.int/index.php?project=planck&page=Planck_Collaboration}. 
\end{acknowledgements}
\nocite{planck2013-p01, planck2013-p02, planck2013-p02a, planck2013-p02d, planck2013-p02b, planck2013-p03, planck2013-p03c, planck2013-p03f, planck2013-p03d, planck2013-p03e, planck2013-p01a, planck2013-p06, planck2013-p03a, planck2013-pip88, planck2013-p08, planck2013-p11, planck2013-p12, planck2013-p13, planck2013-p14, planck2013-p15, planck2013-p05b, planck2013-p17, planck2013-p09, planck2013-p09a, planck2013-p20, planck2013-p19, planck2013-pipaberration, planck2013-p05, planck2013-p05a, planck2013-pip56, planck2013-p06b, planck2013-p28}

\bibliographystyle{aa}
\bibliography{HFIDPC_bib,Planck_bib}

\appendix

\section{Yardstick simulations\label{sec:YARDSTICK}}

The Yardstick simulation pipeline has been developed for two main
purposes: to quantify the level of residual systematic effects in
the maps produced by the HFI DPC; and to validate our correction (or
lack thereof) of the data for the systematic effects that we are able
to model. The former is achieved by comparing combinations of the
data from which the sky signal is expected to vanish with an ensemble
of signal and noise realizations of the same combinations of data
subsets propagated through the same DPC processing as the data. A
mismatch between the two, e.g., in angular power spectrum space, reveals
a residual systematic effect in the data. These tests are described
in Sect.~\ref{sec:Consistency-diff-map}. In addition, validating
the correction of a given systematic effect requires the implementation
of both a model of the effect and its correction (the latter is of
course unnecessary if the aim of a test is to justify the absence
of a correction). In Sect.~\ref{sec:ADC-non-linearity}, the validation
of the non-correction in the data of the ADC non-linearity is an example
of such a test. The Yardstick pipeline has been designed in a modular
way to allow for both kinds of tests.

For all simulations, the sky signal is taken from the \textit{\Planck\ }Sky
Model \citep[PSM; ][]{delabrouille2012}. This parametric model allows
the generation of all-sky temperature and polarization maps of the
CMB, the SZ effects, and diffuse Galactic emission (in particular
synchrotron, free-free, and thermal dust) with a resolution of a few
arcminutes at all \textit{\Planck\ }frequencies. The PSM also includes
an extensive point-source catalogue, as well as spinning dust, CO
line, and $\ion{H}{ii}$ region models. From the point of view of
data validation, a particularly useful feature of the PSM is its ability
to generate random realizations constrained to match observational
data within their uncertainties. Although these realizations cannot
be fully independent (with the exception of those of the CMB), they
do allow for the propagation of some sky modelling uncertainty through
the DPC pipeline. The PSM also handles convolution with the spectral
response of the HFI bolometers. 

The Yardstick pipeline uses the $a_{\ell m}$ spherical harmonic coefficients
from the PSM as inputs. After generating the pointing timelines directly
from \textit{\Planck}'s attitude history file, it feeds both to the
LevelS-Core pipeline, software developed jointly by the HFI and LFI
teams and documented in the HFI data processing early paper \citep{planck2011-1.7}.
LevelS-Core returns the beam-convolved signal TOI for each bolometer,
which is fed back into the Yardstick pipeline. Noise TOI are generated
by a dedicated module and include both white noise and $1/f$ components.
The high-frequency component is taken to be the detector noise described
in \citet{planck2013-p03f}, while the $1/f$ noise is modelled following
\citet{planck2011-1.5}. The overall amplitude of the noise is given
on a ring-by-ring basis by half-ring map null tests, further discussed
in Sect.~\ref{sec:Noise-Estimation-Bias}. The noise model does not
include correlations between detectors.

Instrumental effects can be added to the simulated data, each by the
addition of a dedicated module to the pipeline. We have used this
ability to improve our understanding of pointing and beam issues,
as well as to test various aspects of glitch and 4\,K line removal,
among others. Section~\ref{sec:ADC-non-linearity} shows an example
of the use of the Yardstick simulation facility to evaluate the impact
on null-test residuals of the non-correction of the ADC non-linearity.
When evaluating the quality of the correction of a systematic effect,
we simply apply the relevant TOI processing prior to proceeding with
mapmaking. The impact of the transfer function convolution and deconvolution
is in any case accounted for.

Finally, TOI are propagated through the DPC mapmaking and calibration
pipelines following the procedures described in this paper, including
gain correction and application of the various flags \citep{planck2013-p03f}.
This allows for a one-to-one comparison of the simulated and data
maps. Each simulation leads to the generation of per-detector and
per-detector-set maps for each survey, and for the nominal mission
on which the results in this release are based.

\begin{figure*}[t]
\begin{centering}
\includegraphics[width=1\columnwidth]{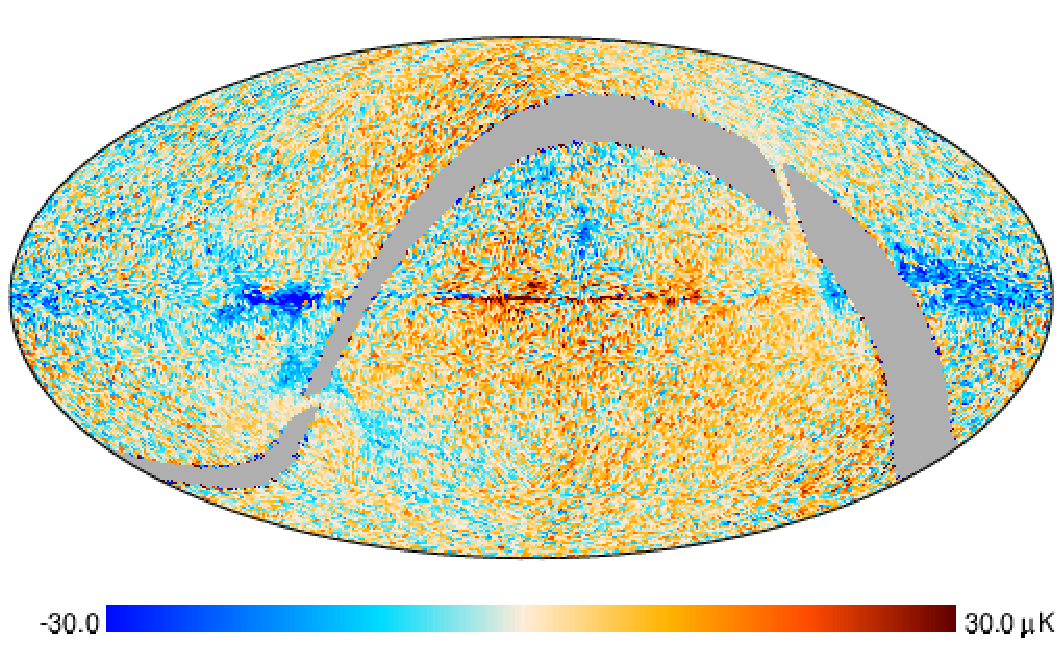}\includegraphics[width=1\columnwidth]{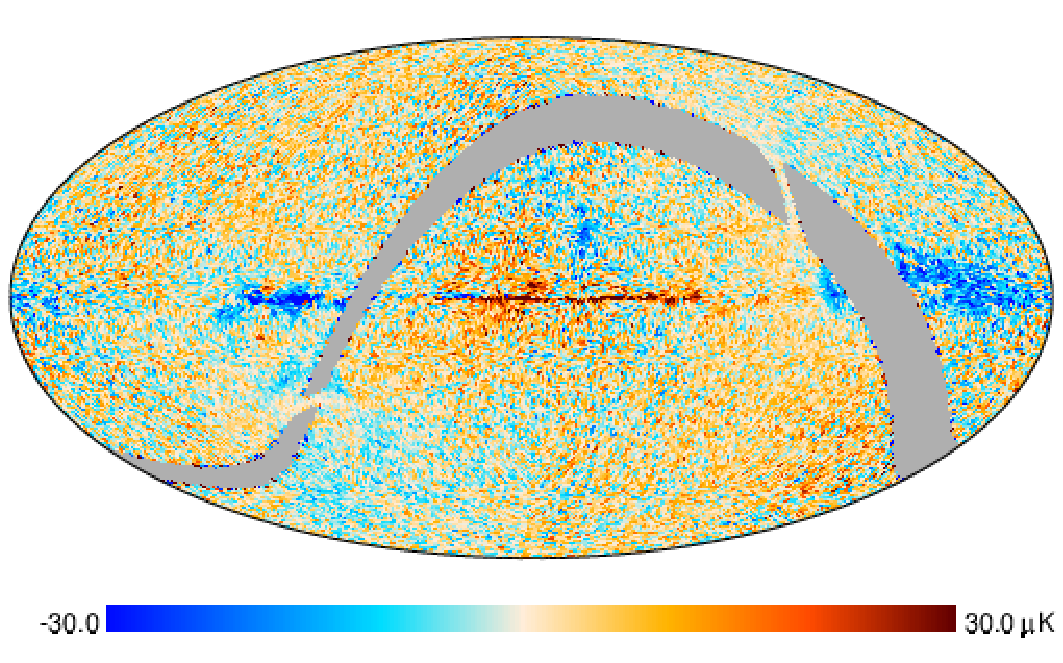}
\par\end{centering}

\begin{centering}
\includegraphics[width=1\columnwidth]{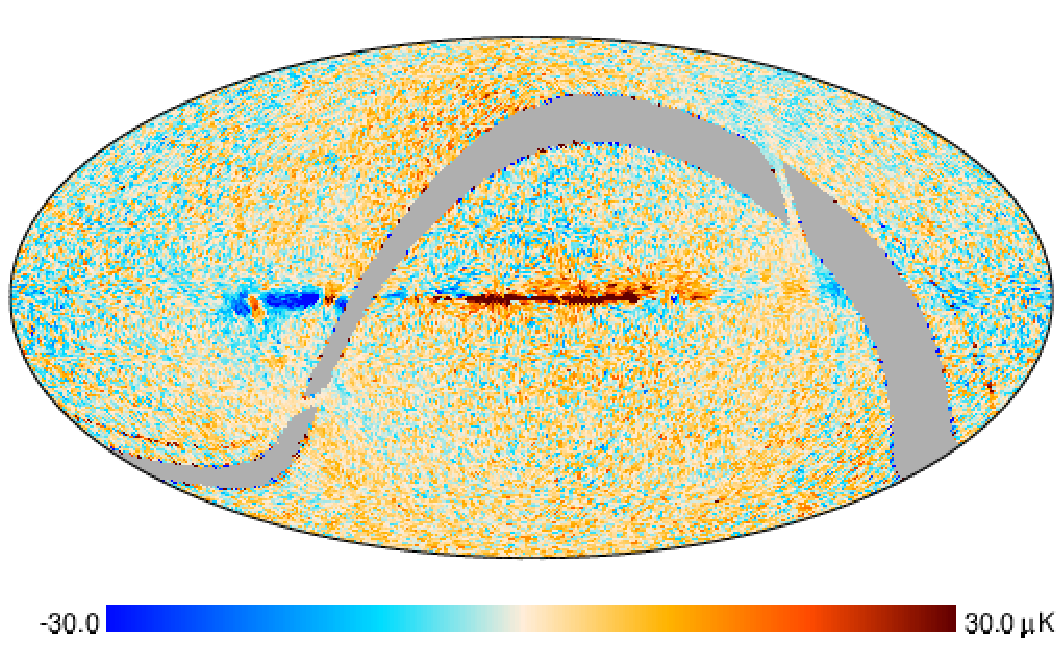}\includegraphics[width=1\columnwidth]{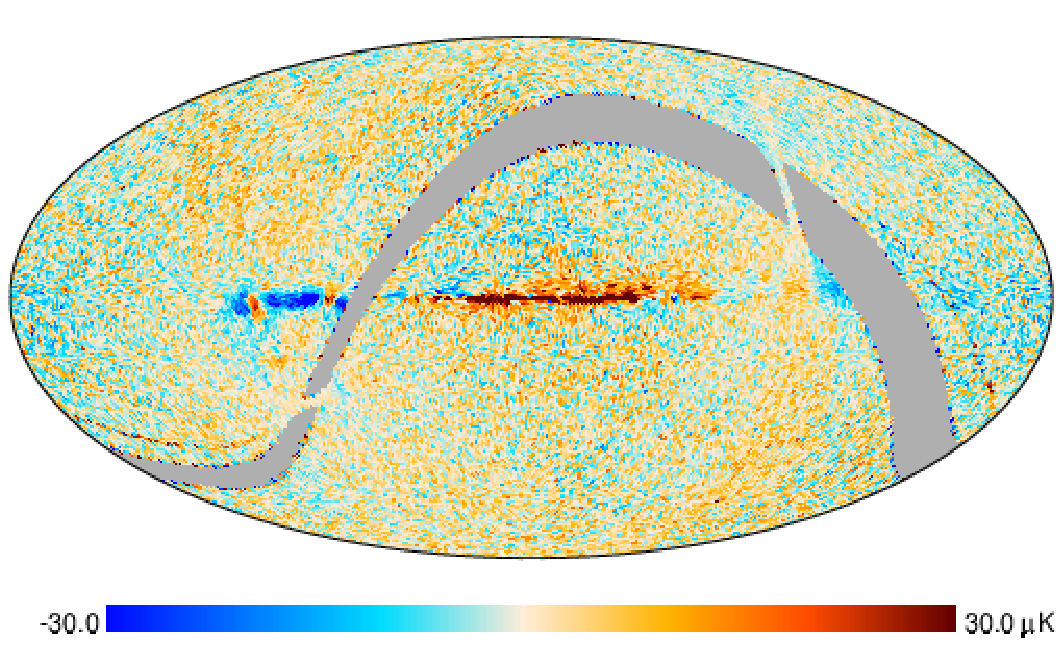}
\par\end{centering}

\centering{}\protect\caption{\emph{Top}: data difference maps for the 143-1a bolometer. \emph{Bottom}:
differences from a simulation including the effect of ADC non-linearity
and 4\,K lines.\emph{ Left column}: Survey 1$-$Survey 2 differences
for maps made without destriping. \emph{Right column}: Survey 1$-$Survey
2 differences after the destriping used to create the delivered data.
The match is sufficiently good that one can use such simulations to
estimate the combined effect of the non-linearity of the ADC on the
delivered data.\textcolor{red}{{} }\label{fig:simADC} }
\end{figure*}

\section{Comparison with a severe selection of data. \\ 
%\rev{NB: This annex has been completely revisited} 
\label{sec:SUPERCLEAN}}

This appendix describes
the selection of pointing periods by more severe criteria and the
impact at the map and angular power spectrum levels.

\subsection{Data selection}

Nominal TOI processing is used; only the selection of rings (i.e.,
data obtained during a stable pointing period) is changed and described
here. The selected rings include only those showing sufficiently stationary
noise, within strict bounds. To keep enough data, the full mission
is used for this test, not only data from the nominal 15.5~months.
We list here the applied criteria.
\begin{itemize}
\item \textbf{Normally discarded rings.} Of course, rings that are already
discarded remain so.
\item \textbf{4\,K cooler lines. }Resonant rings are the ones for which
a 4\,K cooler line (Sect.~\ref{sec:4Kline-removal}) coincides with
a harmonic of the spin frequency. The 4\,K line correction comes
from an interpolation that could introduce some problems. The 10\,Hz
line is the most troublesome, so we discard all the 10\,Hz 4\,K
line resonant rings. This criterion is common to all detectors.
\item \textbf{Baseline jump correction. }Rings where a jump correction (Sect.~\ref{sec:Jump-corrections})
was performed are discarded.
\item \textbf{Bursts in the noise.} Any anomaly in the noise level (as measured
by the total noise NET) can be discarded. From Sect.~\ref{sec:Toi-Qualification},
these are due to either a baseline drift or a small jump. The basic
threshold used here is $\pm3\,\%$ of the median noise per ring.
\item \textbf{Two-level noise. }As described in \citet{planck2013-p28},\textcolor{red}{{}
}%
\begin{comment}
\textcolor{red}{{[}AJ: needs to be described or referenced{]}}
\end{comment}
the following bolometers exhibit a larger than usual noise level in
a specific range of rings indicated between brackets:

\begin{itemize}
\item 353-2 --- {[}4994,\emph{ end}{]};
\item 353-3a --- {[}5811, 9057{]} and {[}18890,\emph{ end}{]};
\item 143-3a --- {[}9410,\emph{ end}{]}; and
\item 353-5b --- {[}2572, \emph{end}{]}.
\end{itemize}

The corresponding ranges are entirely discarded.

\item \textbf{Integration time per ring.} It is possible that the performance
of the deglitching procedure might differ for rings with a significantly
smaller integration time than average. Hence we discard rings with
an integration time less than 35 minutes. \emph{This criterion is
common to all detectors.}
\item \textbf{Planets. }Rings including a big planet (i.e., flagged as in
Sect.~\ref{sec:Input-flags}) can perturb the mapmaking solution
if long tails are unseen. \emph{This criterion is common for detectors
in the same row of the focal plane.}
\item \textbf{Medium and small RTS.} All rings known to be affected by RTS
(Sect.~\ref{sec:Toi-Qualification}) are discarded.
\item \textbf{Fifth survey. }Survey 5 is fully discarded because of solar
flares, unusual integration time and end-of-life operations. \emph{This
criterion is common to all detectors. }\textcolor{red}{}%
\begin{comment}
\textcolor{red}{{[}AJ: or do we just want to say above that we use
data from the first four surveys?{]}. LilianS: Not exactly as we use
the 5th survey for random discarding ring process}
\end{comment}

\item \textbf{PSB. }If one of the bolometers of a PSB pair is flagged for
a ring, then both bolometers are discarded for that ring. 
\end{itemize}

\subsection{Resulting ring statistics}

Each criterion has a different impact on both global statistics and
those of individual bolometers.

\begin{figure*}[th]
\centering{}\includegraphics[clip,width=1\textwidth]{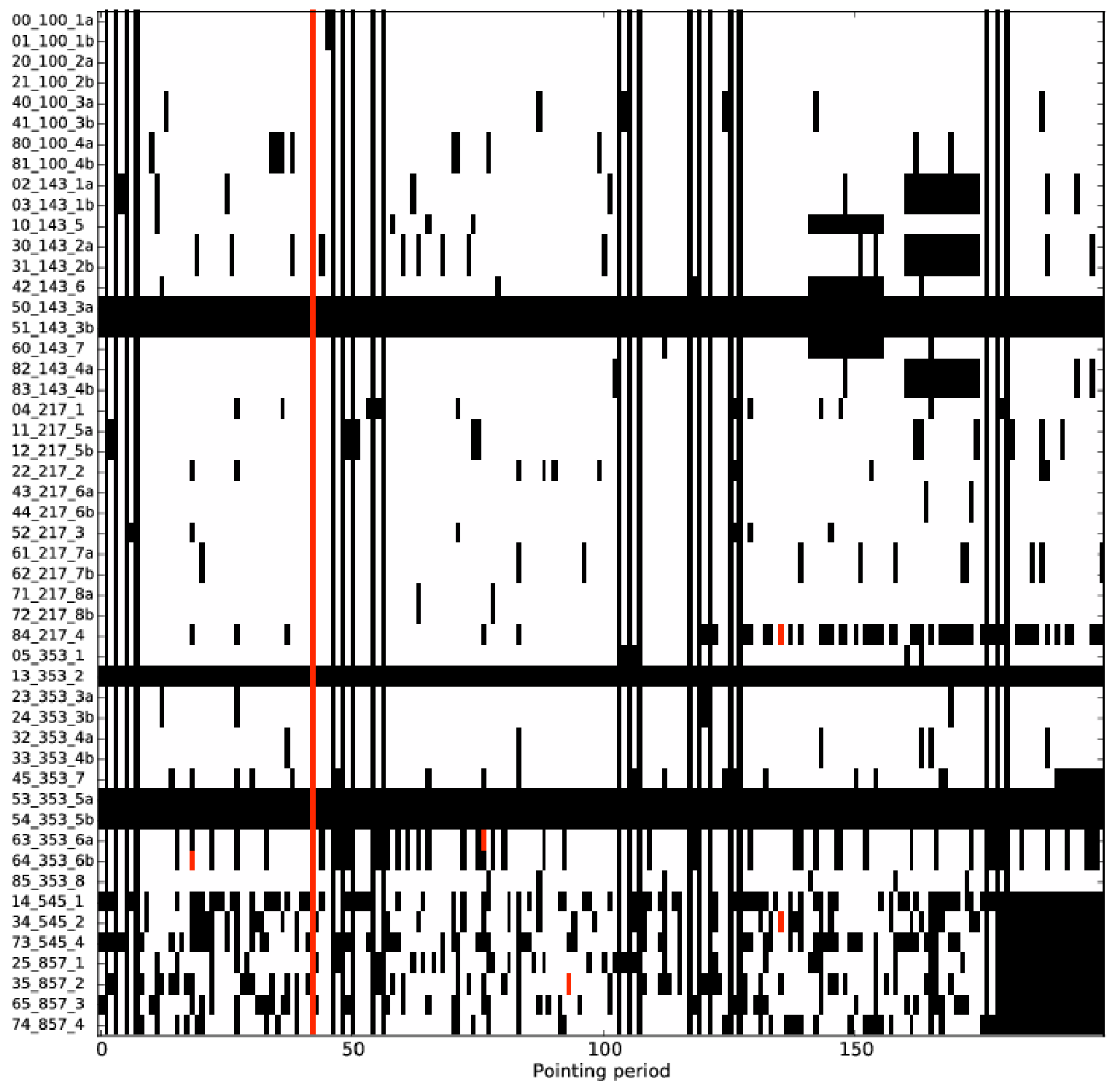}\protect\caption{\label{fig:superclean_OR}Example of normally discarded rings (red)
and the severe selection (black) for 201 consecutive pointing periods
(the abscissa has an arbitrary offset). The vertical axis gives the
individual HFI detector (the first two digits describe the organization
of the HFI readout electronics as described in \citealt{planck2013-p28}).}
\end{figure*}

A logical-\texttt{OR} of all of the criteria is shown in Fig.~\ref{fig:superclean_OR}.
Black pixels indicate that the corresponding rings of the corresponding
bolometer are not used for the severe subset, whereas the red ones
are those that were already discarded with the old criteria. Several
features are notable, such as differences between the highest and
lowest frequency channels, and the fact that some criteria affect
all the bolometers.

The global impact of this ring selection is that about 41\,\% of
the rings are discarded for the full mission, about 35\,\% for the
100--353\,GHz channels, and 45\,\% for 545--857\,GHz. With the
normal criteria, only about 1\,\% of the rings are discarded for
the full mission (0.2\,\% in the nominal mission). 

In order to distinguish the impact of a simple decrease of the observation
time and the impact of the use of the severe selection, we check the
difference between discarding 35\,\% of the rings with the criteria
above, and discarding 35\,\% of the rings randomly for the 143\,GHz
channel.

\subsection{Impact of the severe ring selection on map making}

\begin{figure*}[th]
\centering{}\includegraphics[clip,width=1\textwidth]{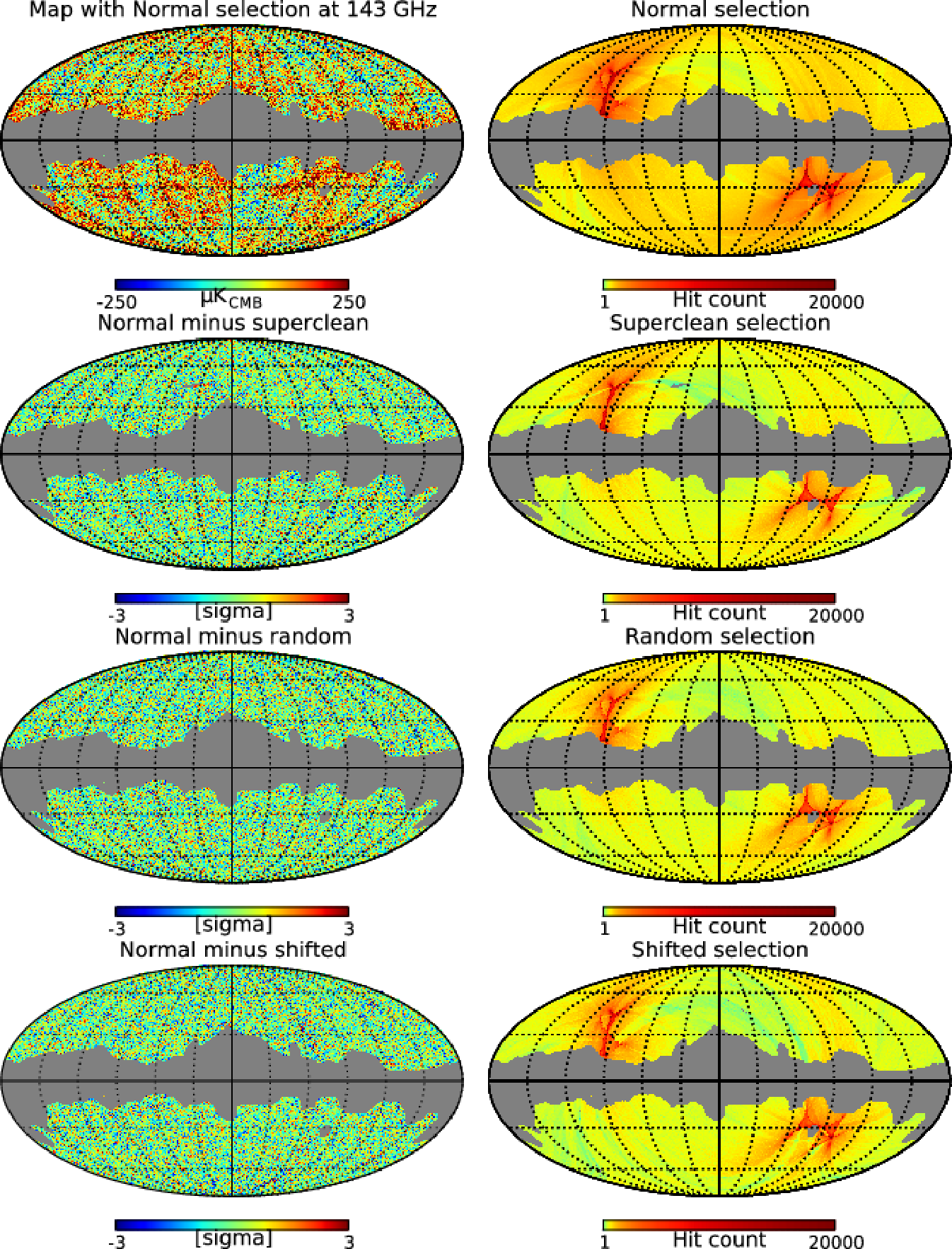}\protect\caption{\label{fig:maps_143}\emph{ Top}: 143\,GHz intensity (\emph{left)}
and corresponding hit-count map (\emph{right}).\emph{ Lower panels}:
difference (\emph{left}) between a given selection and normal maps,
whitened with the noise calculated from the corresponding hit-count
map (\emph{right}), in units of the standard deviation. For the severe
selection, striping of the hit numbers is more visible.}
\end{figure*}

We use the standard pipeline to create maps for each frequency. Along
with the global reduction of the number of samples, the severe ring
selection will create stripes on the hit-maps, due to the temporal
structure of some criteria. The sanity check with the randomly selected
rings will have only the impact of the reduction of the hit number.
This can be seen in the right column of Fig.\,\ref{fig:maps_143}:
the number of hits in the severe maps is not as homogeneous as in
the normal map or the random map but rather looks like the shifted
selection map. On the left side, the top panel represents the masked
intensity of the normal map, whereas the lower panels correspond to
the difference with the severe, random, and shifted maps, respectively.
The differences have been whitened with a variance proportional to
the inverse of the following effective hit number: 
\[
n_{\mathrm{eff}}=(1-g)n_{\mathrm{normal}}\,,
\]
where $n$ is the number of hits of the labelled map and the average
ratio of the hit numbers in the severe map to the hit numbers in the
normal map is $g=0.65$ for 100--353\,GHz and $g=0.55$ for 545--857\,GHz.

\subsection{Impact of the severe ring selection on spectra}

We then compare the spectra for the severe ring selection and the
normal ring selection. We used cross-spectra of offset-corrected half-ring
maps, FIRST$\times$LAST. The offset correction is done in order to
avoid discrepancies at low-$\ell$ due to mask leakage. To avoid contamination
from the Galaxy, the masks applied to compute these offsets depend
on the frequency: $f_\mathrm{sky}$= [70,70,40,40,20,20]\,\% for {[}100,143,217,353,545,857{]}\,GHz.
Notice that, as the same sky mask is used with the normal and severe
maps, the error bars do not include the cosmic variance. 

Actually, in the Gaussian case, the expected variance of the difference
is induced by noise, and can be computed as:
\begin{equation}
\sigma_{\mathrm{expected}}^{2}\left(C_{\ell}^{\mathrm{S}}-C_{\ell}^{\mathrm{N}}\right)=N_{\ell}\frac{2C_{\ell}^{\mathrm{N}}\frac{1-g}{g}+N_{\ell}\frac{1-g^{2}}{g^{2}}}{f_{\mathrm{sky}}(2\ell+1)}\,,\label{eq:ExpectedVarianceSuperClean}
\end{equation}
where $C_{\ell}^\mathrm{S}$ is the pseudo-spectrum for the severe
selection, $C_{\ell}^\mathrm{N}$ is the pseudo-spectrum for the normal
selection, $N_{\ell}$ is the noise pseudo-spectrum of the normal
map. This formula holds if the noise of the individual half-ring maps
have the same pseudo-spectrum and are totally uncorrelated, and if
the only difference between the normal and the severe selections is
the number of hits in them.

\begin{figure}[th]
\begin{centering}
\includegraphics[clip,width=1\columnwidth]{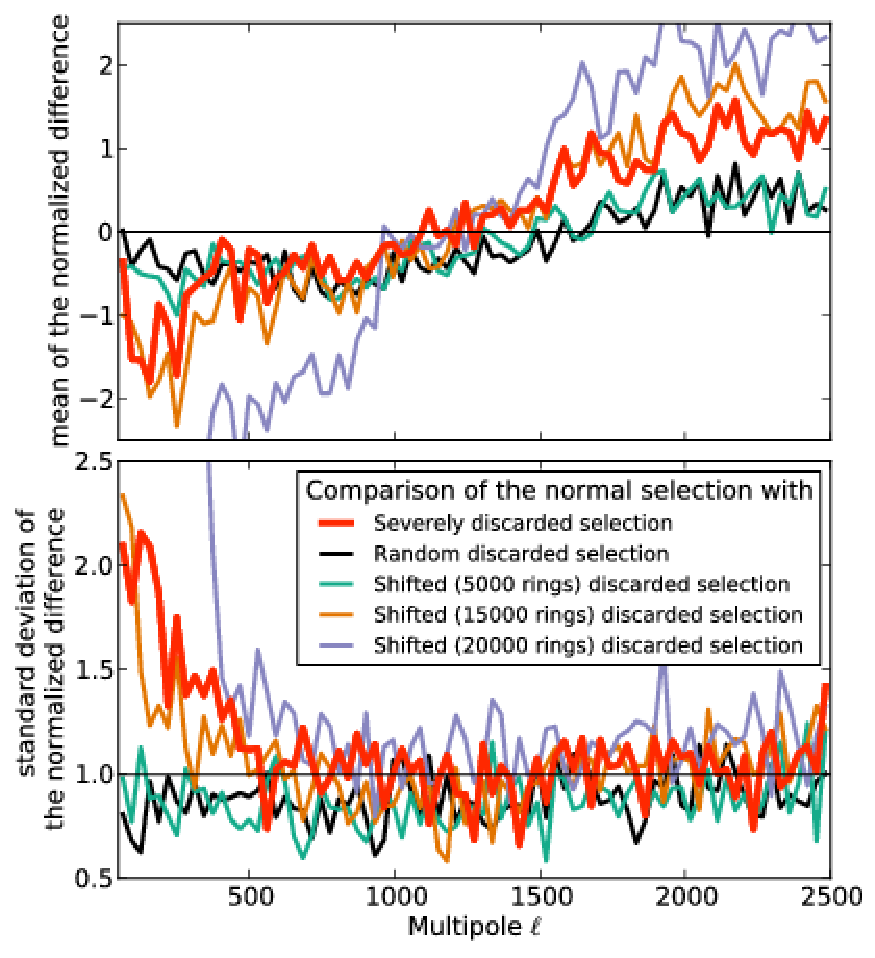}
\par\end{centering}

\protect\caption{\label{fig:binned_31_difference_143_mean_std}Binned
  difference between the severe and the normal pseudo-spectra \revv{at
    143~GHz} in units of the expected standard deviation. If our assumptions
  are correct, the residual noise has a Gaussian distribution centred on zero
  with a variance of one.  The top and bottom panels represents the mean and
  standard deviation of the difference in bins of 31 samples, obtained from
  the data in Fig.~\ref{fig:difference_143_no_bins}. The black points
  represent a sanity check where the rings have been discarded randomly, and
  the blue, orange and green lines represent three sanity checks where the
  selection of discarded rings has been shifted with respect \revv{to} the severe
  one. }
\end{figure}

This formula does not take into account the exact scanning strategy
and its associated geometrical effects. We plot in Fig.~\ref{fig:binned_31_difference_143_mean_std}
the difference of the two pseudo-spectra, as in Fig.~\ref{fig:difference_143_no_bins},
but normalized by the standard deviation expected with Eq.~\ref{eq:ExpectedVarianceSuperClean}.
Consequently, we would expect in the ideal case the mean to be 0 and
the standard deviation to be 1. 

%
% Figure is put back in on revision 2
\begin{figure*}[th]
  \centering{}\includegraphics[width=1\textwidth]{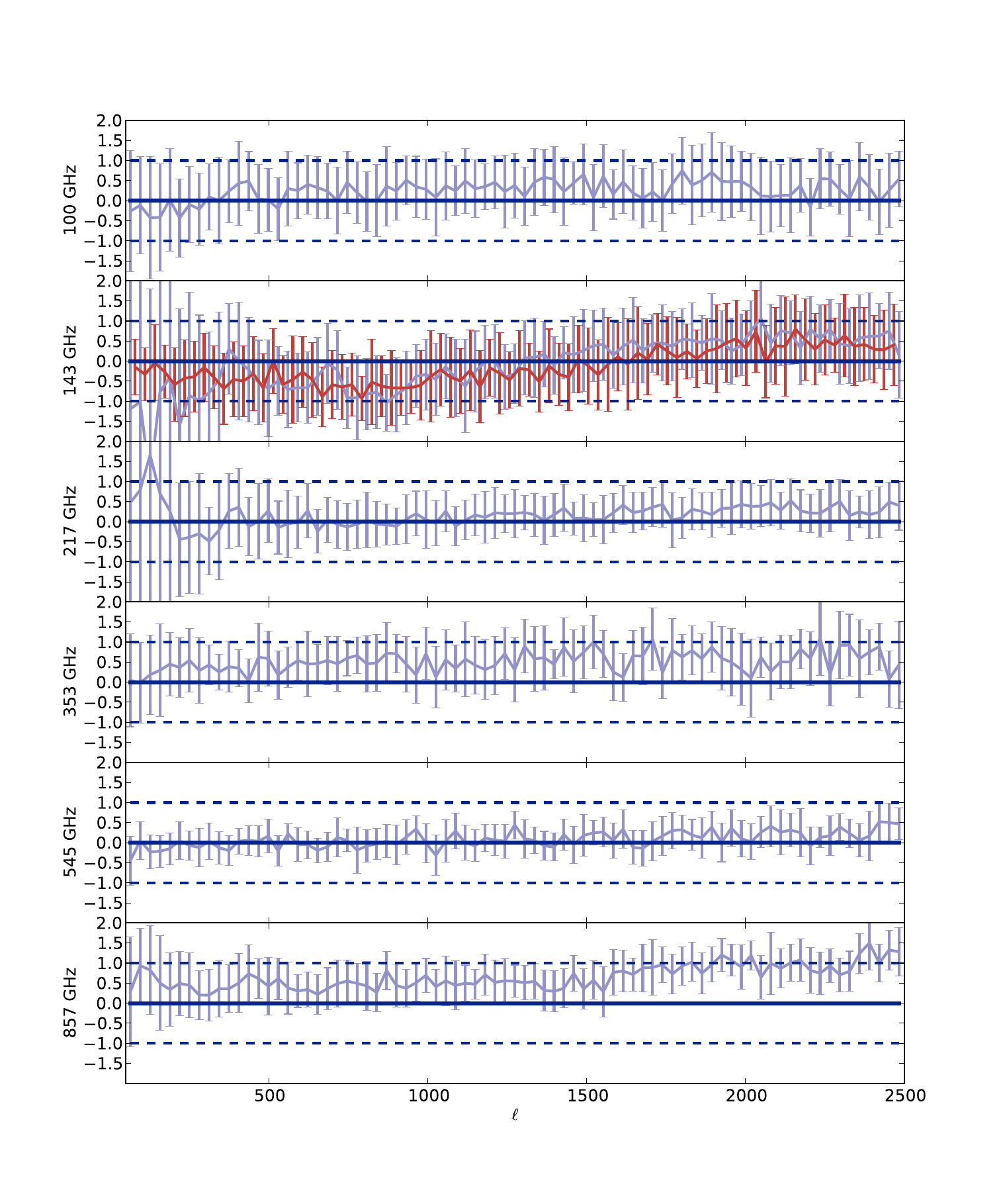}\caption{\label{fig:binned_difference_all_freqs}For
    each of the HFI channels, the difference of the normal spectrum and the
    spectrum obtained with a more severe selection of rings, is shown in units
    of the expected standard deviation. If our assumptions are correct, the
    residual noise is supposed to have a Gaussian distribution centred on zero
    with a dispersion of one. The red points for the 143\,GHz correspond to a
    sanity check with a random discarding of rings as discussed in
    Fig.~\ref{fig:binned_31_difference_143_mean_std}, (the points here are
    band-averaged in bins of width $\Delta\ell\sim 31$).}
\end{figure*}

\begin{figure}[th]
\begin{centering}
\includegraphics[clip,width=1\columnwidth]{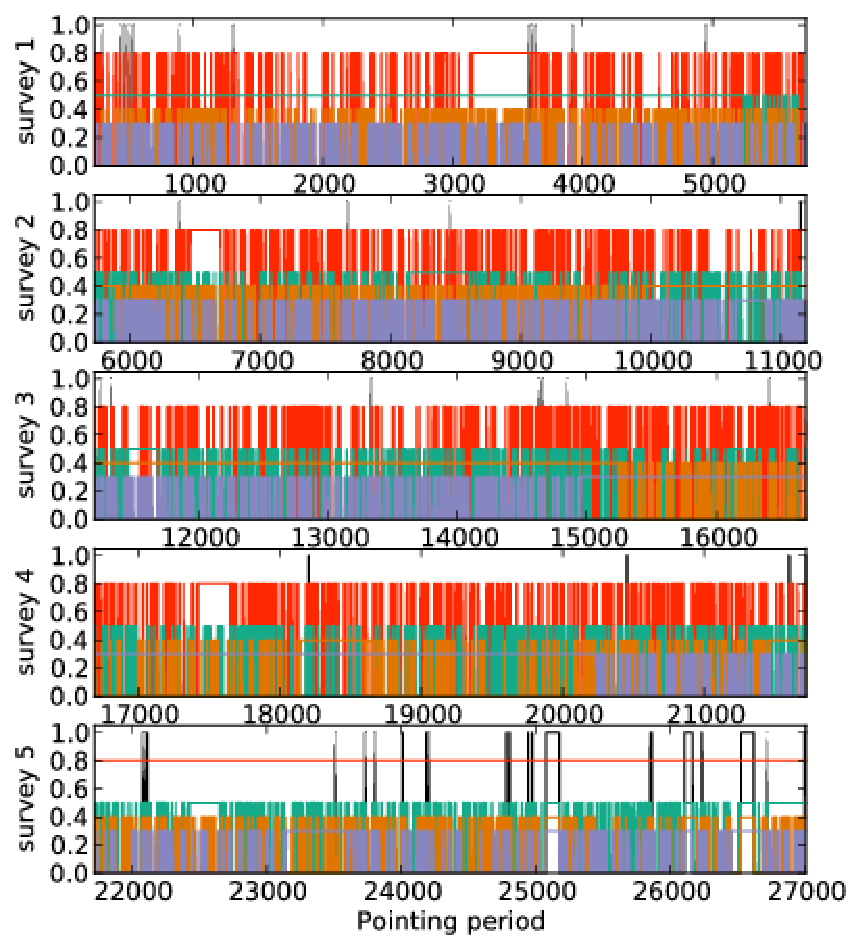}
\par\end{centering}

\protect\caption{\label{fig:example_ROI_shifted}Selection of rings for one
  143~GHz bolometer, for the normal selection (in black, value 1.0 if
  discarded), the severe selection (in red, 0.8 if discarded), three shifted
  selections (in blue, orange and green, respectively 0.3, 0.4, 0.5 if
  discarded).  Each panel represents a survey. The color code for the shifted
  selections is the same as in
  Fig.~\ref{fig:binned_31_difference_143_mean_std}.}
\end{figure}

The red curve is obviously not compatible with this ideal case. To
check if this effect is due to the hit counts on the map or to the
content of the rings, two sanity checks consisting of discarding the
same number of pointing periods than the severe selection have been
performed: 
\begin{itemize}
\item a selection of randomly discarded pointing periods,
\item a selection of discarded pointing periods shifted with respect to
the severe one, by adding a constant to the list of discarded rings,
modulo the total number of periods. The shifts are large to assess
the impact of the fifth survey. Of course, the standard discarded
ring list is still taken into account. We show in Fig.~\ref{fig:example_ROI_shifted}
three examples of these selections which can be compared to the normal
and the severe ones.
\end{itemize}

All the selections have the same number of discarded rings, but a
different impact on the scanning strategy. The difference between
the three realizations is driven by the absence or the presence of
large stripes on the hit count maps, and depends on the shift.

As can be seen in Fig.~\ref{fig:maps_143}, the random selection
has fewer stripes than the shifted one in the hit count map, so we
expect smaller deviations at low $\ell$ for this selection. This
is due to the large chunk of consecutive discarded rings in the severe
selection. One can see in Fig.~\ref{fig:binned_31_difference_143_mean_std}
that the effects affecting the severe selection (the tilt in the mean,
and the increase of dispersion at low $\ell$) are also present 
\revv{at different levels in the} sanity checks.

The difference between the severe and the normal selection cannot be cast into
just a global coefficient $g=1-0.35$, but the sanity check with shifted
selections reproduce the observed bias well 
\revv{especially when the shift is
  by 15000 rings which corresponds to about one year. Therefore the small
  features and the global slope of the difference between the normal and
  severe spectra can be attributed to the structure of the hit maps (directly
  linked to the scanning strategy) rather than 
  to the content of the signal maps.}

\revv{Figure~\ref{fig:binned_difference_all_freqs} shows those normalized
differences for all of the channels. The difference is plotted in
units of $\sigma_\mathrm{expected}(C_{\ell}^{S} - C_{\ell}^{N})$,
so that the expected error bars are $0 \pm 1$. This is the case from
$\ell=500$ to $\ell = 2500$ for all frequencies.
The rise at 143 and 217 GHz is as discussed in Fig.~\ref{fig:binned_31_difference_143_mean_std}}

The maps are thus fully compatible, and no effect of non-stationarity
of the noise in the timelines (for instance) has been detected. The
normal selection can be used safely to extract science from the frequency
maps.

\section{\label{sec:fullmission}\revv{Additional tests at 217\,GHz from full mission survey data}}

We denote the combined detector maps produced for Sky Survey~1, Sky Survey~2, etc., as S1, S2, S3, S4. Here we consider additional
tests allowed using the full S3 and S4 surveys, i.e., using some
data collected after the end of the nominal mission (and processed
with the same pipeline). The 2013 likelihood is based on cross-spectra
between detector-set maps (or detsets) to eliminate possible
biases from inaccurate instrument noise determinations. At 217\,GHz,
we use six temperature maps from the available detsets over the nominal
mission duration. One could have computed cross-survey spectra, for
example $\hbox{S1} \times \hbox{S2}$; however, one then pays a substantial 
penalty in signal-to-noise ratio, since half the data are effectively discarded. Furthermore,
since each individual survey does not cover the complete sky, S1 and
S2 contain non-identical missing areas of sky, leading to irregularly
shaped sky masks. In the \Planck\ likelihood code released in 2013,
we cross-correlated detset maps (excluding auto-spectra). This has
the advantage of retaining almost all of the information in the \Planck\
data, but has the disadvantage of susceptibility to biases caused
by correlated systematics between detectors observing the sky at the
same time.  One therefore has to make a choice between analysing cross-survey
spectra, with associated loss of signal-to-noise ratio, or analysing
detset cross-spectra with potentially greater susceptibility to systematics.
Of course, if systematics can be controlled to a sufficiently low
level, then the detset approach is the more powerful. 

From each of the 217\,GHz survey maps, we compute the following cross-spectra $\hbox{S1}\times \hbox{S2}$, $\hbox{S3}\times \hbox{S4}$, $0.5(\hbox{S1}\times \hbox{S2} + \hbox{S3}\times \hbox{S4}$) and $(\hbox{S1} + \hbox{S2})\times (\hbox{S3} + \hbox{S4})$.
The sky masks are the same as the 217\,GHz\  mask used in the 2013
Planck likelihood, with identical point source masks. The masks for
$\hbox{S1}\times \hbox{S2}$ and $\hbox{S3}\times \hbox{S4}$ differ in that we exclude missing
sky area and any pixels observed only in one sky survey. The yearly
cross-survey spectra $(\hbox{S1} + \hbox{S2})\times (\hbox{S3} + \hbox{S4})$ have higher signal-to-noise ratio than either $\hbox{S1}\times \hbox{S2}$ or $\hbox{S3} \times \hbox{S4}$, and have the added advantage that we can use almost identical masks to those used for the detset
spectra, thus eliminating cosmic variance in a comparison of spectra. 

\begin{figure}[t]
\begin{centering}
\includegraphics[width=1\columnwidth]{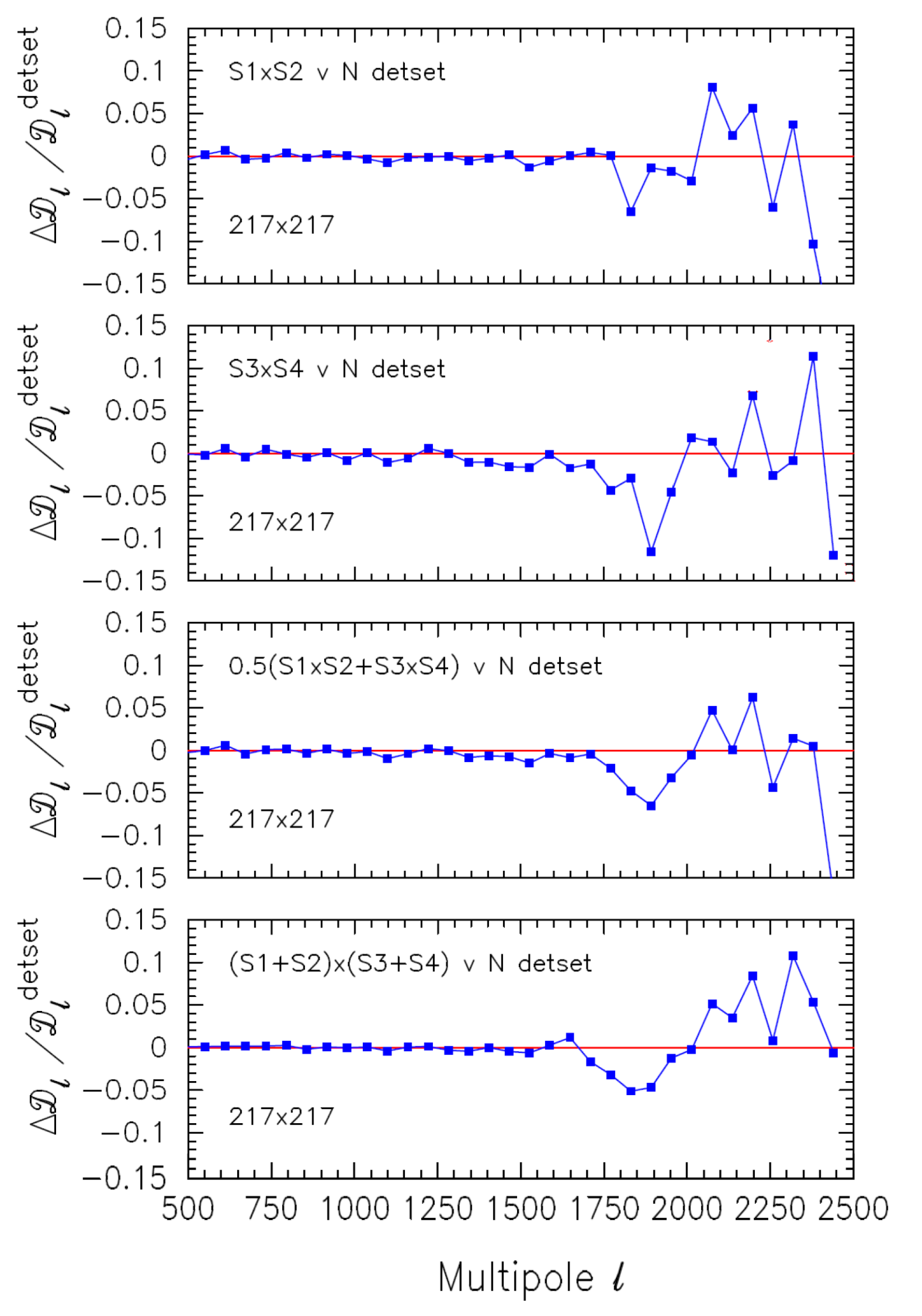}
\par\end{centering}

\centering{}\protect\caption{\label{fig:surveyspecs}Differences between the $217 \times 217$ nominal mission detset spectrum (N) and various cross-survey spectra, as labeled.   Clearly the detset and cross-survey spectra are in very good agreement. Any co-temporal systematics are extremely small.}
\label{fig:surveyspecs}
\end{figure}

The four panels in Fig.~\ref{fig:surveyspecs} show the differences
between the $217 \times 217$ nominal mission detset spectrum and
various cross-survey spectra. No corrections for Galactic dust emission
or unresolved foregrounds have been made to the spectra. Since the
masks are similar [almost identical in the case of $(\hbox{S1} + \hbox{S2})\times (\hbox{S3} + \hbox{S4})$], the spectra are highly correlated in the signal-dominated regime.
There is therefore a characteristic scale in this type of plot set
by the onset of instrument noise. Over the multipole range where the
spectra are signal-dominated, the scatter is small (much less than
the cosmic variance). In the top two plots, the scatter abruptly increases
at multipoles $\sim 1800$ because of the high noise in the $\hbox{S1}\times \hbox{S2}$
and $\hbox{S3}\times \hbox{S4}$ spectra.  The first panel shows hints of a deficit
at multipoles around 1800, and perhaps a slight excess at multipoles
of 2000. The $\hbox{S3}\times \hbox{S4}$ comparison shows further evidence of a
deficit at $\ell \sim 1800$. The yearly cross-survey spectra, which
have the highest signal-to-noise ratio of the cross-survey spectra, show
clear evidence of a deficit at $\ell \sim 1800$, and evidence for
an excess at $\ell > 2000$. As already mentioned, the dip at $\ell \sim 1800$
is caused by electromagnetic interference between the Joule-Thomson
4-K cooler electronics and the bolometer readout electronics. This
interference leads to a set of time-variable narrow lines in the time-ordered data. The data processing pipeline applies a filter to remove
these lines; however, the filtering failed to reduce the impact of
these lines to negligible levels.  Incomplete removal of the 4-K cooler
lines affects primarily the 217 $\hbox{PSB}\times\hbox{PSB}$ cross-spectrum in
Survey~1. At the time of submission of the 2013 \Planck\ papers, we had
not established clear evidence that the $\ell = 1800$ feature was
a residual of the low level data processing. The impact of this systematic
in the $217\times217$ spectrum was analysed in the revised versions
of \cite{planck2013-p11} (Appendix C.4), and was demonstrated to
have little impact on cosmological parameter determination. But it
does contributes to the weak detection of a feature in the power spectrum
reconstruction done in \cite{planck2013-p17}.

The small excess at $\ell > 2000$ is caused by low levels of correlated
noise between the 217\,GHz bolometers, and is strongest in a subset
of the $\hbox{SWB}\times\hbox{SWB}$ spectra. The amplitude of the correlated noise
implied by the fourth panel of Fig.~\ref{fig:surveyspecs} is compatible
with results discussed in \citep{planck2013-p08}, where it is found
that the impact on cosmological parameters of this effect is negligible,
with less than a $0.1\sigma$ variation on the mean posterior values. 

Finally, we constructed a full mission likelihood (denoted F) based
on the detset spectra, using sky masks identical to those used in
the publicly-distributed likelihood. The impact of systematics in
these full mission likelihoods should be substantially smaller than
in the nominal mission likelihood.  In particular, we verified that
the $\ell = 1800$ 4-K line residual is strongly diluted in the full
mission data. We find that the values of the parameters of the base
$\Lambda$CDM cosmology determined from the publicly released nominal
mission detset likelihood (N) and the full mission detset likelihood
with the same sky coverage (F) are completely compatible. For example,
$\Omega_{\mathrm{c}} h^2$ varies from $0.1199 \pm 0.0027$ (N) to
$0.1196\pm 0.0025$ (F), $\sigma_8$ is completely unchanged till
the third decimal place, and the value of $H_0$ for the nominal mission,
$67.3 \pm 1.2\,\rm{km}\, \rm{s}^{-1}\,\rm{Mpc}^{-1}$ becomes $67.6 \pm 1.1\,\rm{km}\, \rm{s}^{-1}\,\rm{Mpc}^{-1}$ for the full mission, everything else being equal.

\raggedright

\end{document}